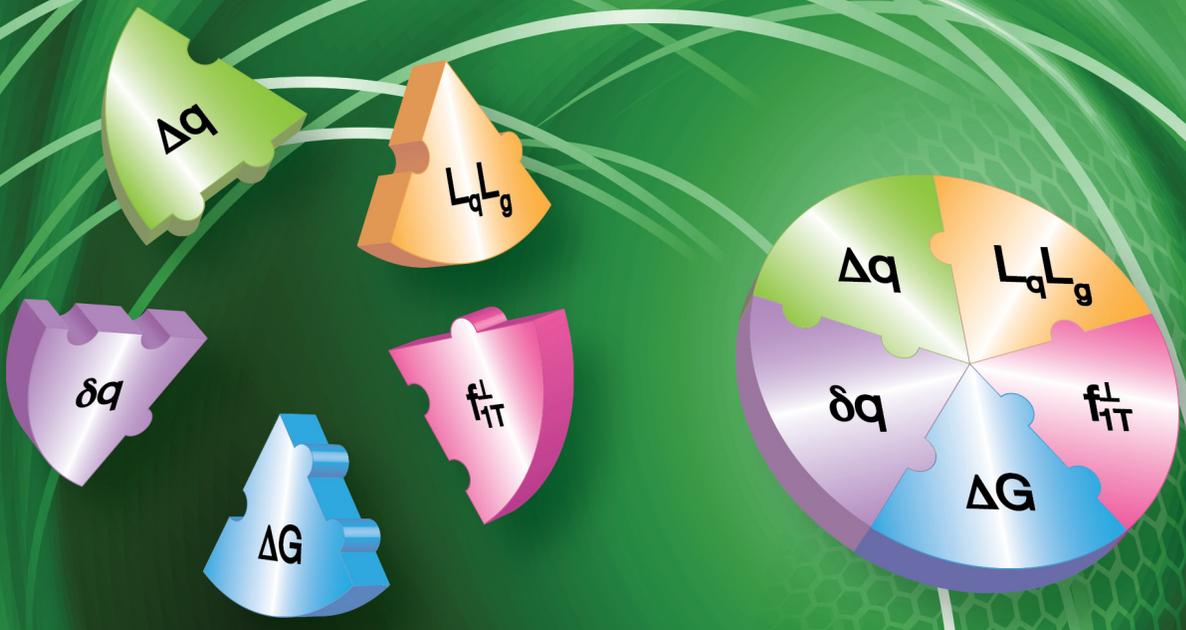

# The RHIC Cold QCD Plan for 2017 to 2023
## A Portal to the EIC

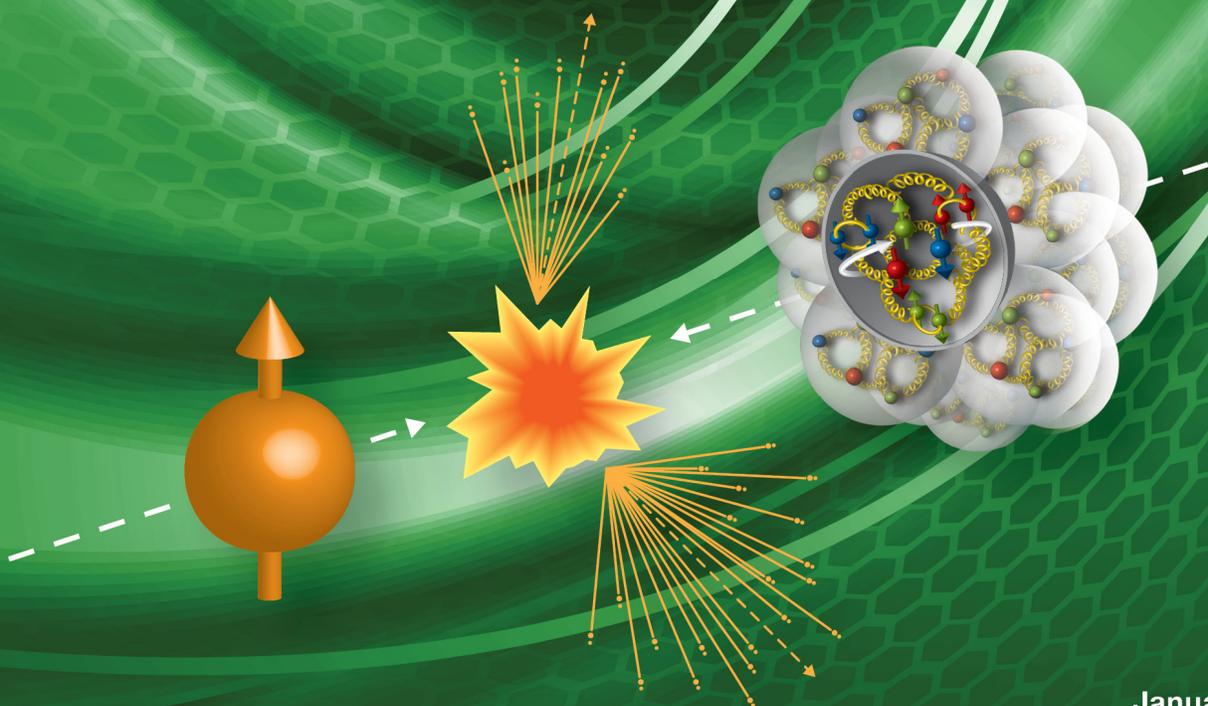

January 2016

Cover design by J. Abramowitz, BNL

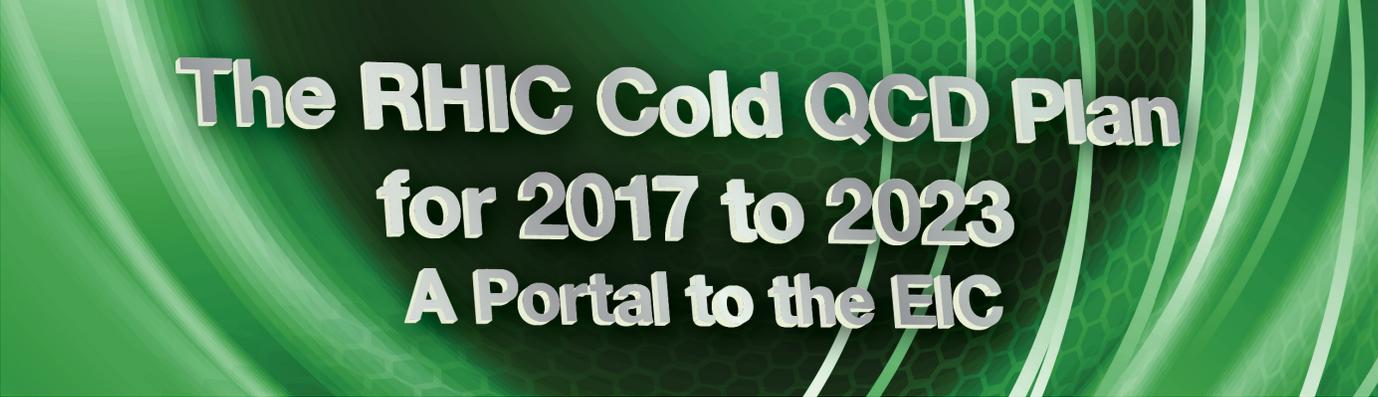

# The RHIC Cold QCD Plan for 2017 to 2023
## A Portal to the EIC

**Authors, for the RHIC SPIN Collaboration**[1]
**and the PHENIX and STAR Collaborations**

Elke-Caroline Aschenauer[2] (BNL), Christine Aidala (U. Michigan), Alexander Bazilevsky (BNL), Markus Diehl (DESY), Renee Fatemi (Kentucky U.), Carl Gagliardi (Texas A&M), Zhongbo Kang (Los Alamos), Yuri V. Kovchegov (Ohio State U.), Jamal Jalilian-Marian (Baruch U.), John Lajoie (Iowa State U.), Dennis V. Perepelitsa (BNL), Ralf Seidl (Riken), Rodolfo Sassot (Buenos Aires U.), Ernst Sichtermann (LBNL), Marco Stratmann (Univ. of Tuebingen), Stephen Trentalange (UCLA), Werner Vogelsang (Univ. of Tuebingen), Anselm Vossen (Indiana U.) and Pia Zurita (Univ. de Santiago de Compostela)

---

[1] The **RHIC Spin Collaboration** consists of the spin working groups of the RHIC collaborations**,** many theorists and members of the BNL Collider-Accelerator Department.

[2] Chair

# 1 INTRODUCTION

The exploration of the fundamental structure of strongly interacting matter has always thrived on the complementarity of lepton scattering and purely hadronic probes. As the community eagerly anticipates a future electron ion collider (EIC) in the U.S., an outstanding scientific opportunity remains to complete "must-do" measurements in p+p and p+A physics in the years preceding the EIC. As we describe in this document, these measurements will be essential to fully realize the scientific promise of the EIC by providing a comprehensive set of measurements in hadronic collisions that, when combined with data from the EIC, will establish the validity and limits of factorization and universality. The outlined program will on the one hand lay the groundwork for the EIC, both scientifically and in terms of refining the experimental requirements for the physics program at the EIC, and thus be the natural next step on the path towards an electron-ion collider. On the other hand, while much of the physics in this program is unique to proton-proton and proton-nucleus collisions and offers discovery potential on its own, when combined with data from the EIC it will provide a broad foundation to a deeper understanding of fundamental QCD.

The EIC, enthusiastically endorsed by the community in the 2015 Long Range Plan [1], is designed to study the dynamics of sea quarks and gluons in the proton and in nuclei at an unprecedented level of detail, precision and kinematic coverage (see Figure 2-5 and Figure 4-5). The importance of the measurements that we envisage in a cold QCD program at RHIC, and their synergy with those at a future EIC, rests on the following observations:

1. The separation between the intrinsic properties of hadrons and interaction dependent dynamics, formalized by the concept of factorization, is a cornerstone of QCD and largely responsible for the predictive power of the theory in many contexts. While this concept and the associated notion of universality of the quantities that describe hadron structure has been successfully tested for unpolarized and – to a lesser extent - longitudinally polarized parton densities, its experimental validation remains an unfinished task for much of what the EIC is designed to study, namely the three-dimensional structure of the proton and the physics of dense partonic systems in heavy nuclei. To establish the validity and the limits of factorization and universality, it is essential to have data from **both** lepton-ion and proton-ion collisions, with an experimental accuracy that makes quantitative comparisons meaningful.

2. Key measurements at the EIC will most likely provide the most differential and accurate constraints on the distributions that quantify the structure of the proton or of nuclei, and on their counterparts in the final state describing fragmentation of quarks and gluons into hadrons. However, RHIC measurements can probe the same functions in different processes and in a wider kinematic regime, given its significantly higher reach in collision energy. The combination of different probes and a large lever arm in momentum scales will significantly add to the impact and interpretation of data to be taken at a future EIC.

Both points are impressively validated by experience in the case of the well-known unpolarized parton distribution functions (PDFs) that describe the one-dimensional longitudinal momentum spectrum of quarks and gluons in the proton. Figure 1-2 and Table 1-1 (taken from Ref. [2]) show how a synergy of many different probes is needed in order to unravel all aspects of the unpolarized partonic structure of the proton and to test the underlying fundamental concept of universality. Experience has shown that PDF analyses without high-quality DIS data are barely possible, but that hadron-hadron collider data add essential and equally important information beyond the reach of lepton-hadron processes. We expect a very similar situation to hold with regard to measurements at the EIC and at RHIC. Figure 1-1 illustrates that the full partonic structure of nucleons and nuclei can only be unraveled if the information of different interactions and probes is combined.

| Electromagnetic Interactions | Strong Interactions | Weak Interactions |
|---|---|---|
| Photon | Gluon | Weak Bosons W & Z |
| Sensitive to electric charge | Sensitive to color charge | Sensitive to weak charge |
| → flavor | → gluon | → flavor and helicity |
| pp:     DY | jets, single hadrons, direct photon | W and Z prod'n |
| ep:     DIS, SIDIS | jets, charm | high $Q^2$ DIS |

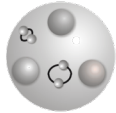  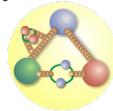  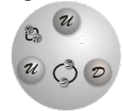

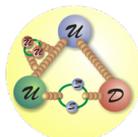

Figure 1-1: The different interactions and probes needed to unravel the partonic structure of nucleons and nuclei.

Despite significant progress both experimentally and theoretically, there remain fundamental aspects of the partonic structure of nucleons and nuclei that are still rather poorly determined, primarily because the available world data are too sparse and/or cover only a very limited kinematic region. One example is the elusive nature of the nucleon spin, another is the quest to go beyond our current, one-dimensional picture of parton densities by correlating, for instance, the information on the contribution of a parton to the spin of the nucleon with its transverse momentum and spatial position. If one extends the scope from a nucleon to nuclei, the following compelling questions, which are all at the heart of the e+A physics program at an EIC [3], immediately arise:

- Can we experimentally find evidence of a novel universal regime of non-linear QCD dynamics in nuclei?
- What is the role of saturated strong gluon fields, and what are the degrees of freedom in this high gluon density regime?
- What is the fundamental quark-gluon structure of light and heavy nuclei?
- Can a nucleus, serving as a color filter, provide novel insight into the propagation, attenuation and hadronization of colored quarks and gluons?

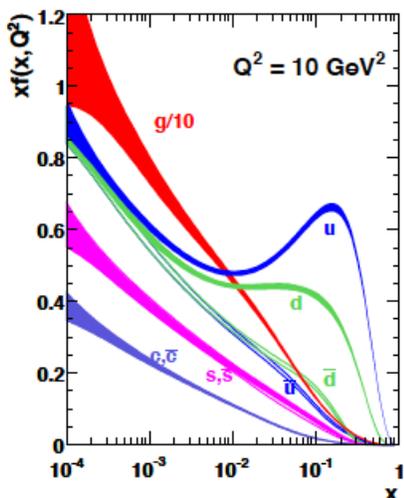

| Process | Subprocess | Partons | x range |
|---|---|---|---|
| $\ell^{\pm}\{p,n\} \to \ell^{\pm} X$ | $\gamma^* q \to q$ | $q, \bar{q}, g$ | $x \gtrsim 0.01$ |
| $\ell^{\pm} n/p \to \ell^{\pm} X$ | $\gamma^* d/u \to d/u$ | $d/u$ | $x \gtrsim 0.01$ |
| $pp \to \mu^+\mu^- X$ | $u\bar{u}, d\bar{d} \to \gamma^*$ | $\bar{q}$ | $0.015 \lesssim x \lesssim 0.35$ |
| $pn/pp \to \mu^+\mu^- X$ | $(u\bar{d})/(u\bar{u}) \to \gamma^*$ | $\bar{d}/\bar{u}$ | $0.015 \lesssim x \lesssim 0.35$ |
| $\nu(\bar{\nu}) N \to \mu^-(\mu^+) X$ | $W^* q \to q'$ | $q, \bar{q}$ | $0.01 \lesssim x \lesssim 0.5$ |
| $\nu N \to \mu^-\mu^+ X$ | $W^* s \to c$ | $s$ | $0.01 \lesssim x \lesssim 0.2$ |
| $\bar{\nu} N \to \mu^+\mu^- X$ | $W^* \bar{s} \to \bar{c}$ | $\bar{s}$ | $0.01 \lesssim x \lesssim 0.2$ |
| $e^{\pm} p \to e^{\pm} X$ | $\gamma^* q \to q$ | $g, q, \bar{q}$ | $0.0001 \lesssim x \lesssim 0.1$ |
| $e^+ p \to \bar{\nu} X$ | $W^+ \{d,s\} \to \{u,c\}$ | $d, s$ | $x \gtrsim 0.01$ |
| $e^{\pm} p \to e^{\pm} c\bar{c} X$ | $\gamma^* c \to c, \gamma^* g \to c\bar{c}$ | $c, g$ | $0.0001 \lesssim x \lesssim 0.01$ |
| $e^{\pm} p \to \text{jet} + X$ | $\gamma^* g \to q\bar{q}$ | $g$ | $0.01 \lesssim x \lesssim 0.1$ |
| $p\bar{p} \to \text{jet} + X$ | $gg, qg, qq \to 2j$ | $g, q$ | $0.01 \lesssim x \lesssim 0.5$ |
| $p\bar{p} \to (W^{\pm} \to \ell^{\pm}\nu) X$ | $ud \to W, \bar{u}\bar{d} \to W$ | $u, d, \bar{u}, \bar{d}$ | $x \gtrsim 0.05$ |
| $p\bar{p} \to (Z \to \ell^+\ell^-) X$ | $uu, dd \to Z$ | $d$ | $x \gtrsim 0.05$ |

Figure 1-2: MSTW 2008 NLO PDFs for unpolarized protons at a resolution scale $Q^2$ = 10 GeV$^2$ (taken from Ref. [2]).

Table 1-1: The main processes included in the global PDF analyses such as MSTW 2008 can be ordered in three groups: fixed-target experiments, HERA e+p collider experiments, and probes from hadron-hadron collisions (TeVatron and LHC). For each process the dominant partonic subprocesses, the primary parton species that are predominantly probed, and the approximate range of x constrained by the data are given (taken from Ref. [2]).

Measurements made in (un)polarized p+A collisions at RHIC will help to address these questions with gluon-rich complementary probes and at different momentum scales than in e+A collisions foreseen at the EIC and will serve to further focus and refine the EIC physics program. We also highlight the particular and unique strength of the RHIC p+A program as compared to p+Pb collisions at the LHC in terms of its versatility (i.e., the option of running with arbitrary nuclei), the availability of polarized proton beams, and the kinematic coverage reaching down to low $p_T$, which overlaps with the region where nuclear effects are largest.

**All projections and physics discussions are based on the following already planned data taking periods in 2017 and during the sPHENIX running periods in 2022 and 2023:**
1. **2017:** 12 weeks transversely polarized p+p at $\sqrt{s}$ = 510 GeV
   It is noted that the 2017 data-taking period will be STAR only, due to the transition from PHENIX to sPHENIX
2. **2023:** 8 weeks transversely polarized p+p at $\sqrt{s}$ = 200 GeV
3. **2023:** 8 weeks each of transversely polarized p+Au and p+Al at $\sqrt{s}$ = 200 GeV

In addition, a 20 week $\sqrt{s}$ = 500 GeV polarized p+p run, split between transverse and longitudinal polarized running is proposed based on its merits for the overall physics program laid out in this document. Analysis of the 2017 run will provide the information necessary to optimize the time-sharing between the two polarizations.

In Sections 2 to 4 we describe in detail how new data from (un)polarized p+p and p+A collisions at RHIC, summarized in Table 1-2, will serve as a gateway to the physics program at a future EIC (for further details please also [4]). Several of the discussed measurements call for improved detector capabilities at forward rapidities. Implementation strategies and first cost estimates both for STAR and sPHENIX implementations are discussed in Section 5.

The proposed program builds on the particular and unique strength of the RHIC accelerator facility compared to JLab, Compass and the LHC in terms of its versatility (i.e., the option of running with arbitrary nuclei), the availability of polarized proton beams, and wide kinematic coverage, which would be further enhanced through an upgrade at forward rapidities consisting of electromagnetic and a hadronic calorimetry as well as tracking. The program will bring to fruition the long-term campaign at RHIC on Cold QCD, with its recent achievements summarized in Section 1.1 and Ref. [5]. **It is especially stressed** that the final experimental accuracy achieved will enable quantitative tests of process dependence, factorization and universality by comparing lepton-proton with proton-proton collisions, providing critical checks of our understanding of QCD dynamics.

|  | Year | $\sqrt{s}$ (GeV) | Delivered Luminosity | Scientific Goals | Observable | Required Upgrade |
|---|---|---|---|---|---|---|
| **Scheduled RHIC running** | 2017 | p↑p @ 510 | 400 pb$^{-1}$ 12 weeks | Sensitive to Sivers effect non-universality through TMDs and Twist-3 $T_{q,F}(x,x)$ Sensitive to sea quark Sivers or ETQS function Evolution in TMD and Twist-3 formalism | $A_N$ for γ, W$^±$, Z$^0$, DY | $A_N^{DY}$: Postshower to FMS@STAR |
|  |  |  |  | Transversity, Collins FF, linearly pol. Gluons, Gluon Sivers in Twist-3 | $A_{UT}^{\sin(\phi_s-2\phi_h)}$ $A_{UT}^{\sin(\phi_s-\phi_h)}$ modulations of h$^±$ in jets, $A_{UT}^{\sin(\phi_s)}$ for jets | None |
|  |  |  |  | First look at GPD $E_g$ | $A_{UT}$ for J/Ψ in UPC | None |
|  | 2023 | p↑p @ 200 | 300 pb$^{-1}$ 8 weeks | subprocess driving the large $A_N$ at high $x_F$ and $\eta$ | $A_N$ for charged hadrons and flavor enhanced jets | Yes Forward instrum. |
|  |  |  |  | evolution of ETQS fct. properties and nature of the diffractive exchange in p+p collisions. | $A_N$ for γ $A_N$ for diffractive events | None None |
|  | 2023 | p↑Au @ 200 | 1.8 pb$^{-1}$ 8 weeks | What is the nature of the initial state and hadronization in nuclear collisions | $R_{pAu}$ direct photons and DY | $R_{pAu}$(DY):Yes Forward instrum. |
|  |  |  |  | Nuclear dependence of TMDs and nFF | $A_{UT}^{\sin(\phi_s-\phi_h)}$ modulations of h$^±$ in jets, nuclear FF | None |
|  |  |  |  | Clear signatures for Saturation | Dihadrons, γ-jet, h-jet, diffraction | Yes Forward instrum. |
|  | 2023 | p↑Al @ 200 | 12.6 pb$^{-1}$ 8 weeks | A-dependence of nPDF, | $R_{pAl}$: direct photons and DY | $R_{pAl}$(DY): Yes Forward instrum. |
|  |  |  |  | A-dependence of TMDs and nFF | $A_{UT}^{\sin(\phi_s-\phi_h)}$ modulations of h$^±$ in jets, nuclear FF | None |
|  |  |  |  | A-dependence for Saturation | Dihadrons, γ-jet, h-jet, diffraction | Yes Forward instrum. |
| **Potential future running** | 202X | p↑p @ 510 | 1.1 fb$^{-1}$ 10 weeks | TMDs at low and high $x$ quantitative comparisons of the validity and the limits of factorization and universality in lepton-proton and proton-proton collisions | $A_{UT}$ for Collins observables, i.e. hadron in jet modulations at $\eta > 1$ and mid-rapidity observables as in 2017 run | Yes Forward instrum. None |
|  | 202X | $\vec{p}\vec{p}$ @ 510 | 1.1 fb$^{-1}$ 10 weeks | $\Delta g(x)$ at small $x$ | $A_{LL}$ for jets, di-jets, h/γ-jets at $\eta > 1$ | Yes Forward instrum. |

Table 1-2: Summary of the Cold QCD physics program propsed in the years 2017 and 2023 and if an additional 500 GeV run would become possible.

## 1.1 RECENT ACHIEVEMENTS

A myriad of new techniques and technologies made it possible to inaugurate the Relativistic Heavy Ion Collider at Brookhaven National Laboratory as the world's first high-energy polarized proton collider in December 2001. This unique environment provides opportunities to study the polarized quark and gluon spin structure of the proton and QCD dynamics at a high energy scale and is therefore complementary to semi-inclusive deep inelastic scattering experiments. RHIC has completed very successful polarized p+p runs both at $\sqrt{s}$ = 200 GeV and 500(510) GeV. Table 1-3 summarizes the luminosities recorded by PHENIX and STAR and the average beam polarization (as measured by the Hydrogen-jet polarimeter) for runs since 2006. In addition to the p+p runs enumerated in Table 1-3, RHIC commenced polarized $p^\uparrow$+A running in 2015 with successful investigations of $p^\uparrow$+Au and $p^\uparrow$+Al collisions.

| Year | $\sqrt{s}$ (GeV) | Recorded Luminosity for longitudinally / transverse polarized p+p STAR | Recorded Luminosity for longitudinally / transverse polarized p+p PHENIX | <P> in % |
|---|---|---|---|---|
| 2006 | 62.4 | -- pb$^{-1}$ / 0.2 pb$^{-1}$ | 0.08 pb$^{-1}$ / 0.02 pb$^{-1}$ | 48 |
|  | 200 | 6.8 pb$^{-1}$ / 8.5 pb$^{-1}$ | 7.5 pb$^{-1}$ / 2.7 pb$^{-1}$ | 57 |
| 2008 | 200 | -- pb$^{-1}$ / 7.8 pb$^{-1}$ | -- pb$^{-1}$ / 5.2 pb$^{-1}$ | 45 |
| 2009 | 200 | 25 pb$^{-1}$ / -- pb$^{-1}$ | 16 pb$^{-1}$ / -- pb$^{-1}$ | 55 |
|  | 500 | 10 pb$^{-1}$ / -- pb$^{-1}$ | 14 pb$^{-1}$ / -- pb$^{-1}$ | 39 |
| 2011 | 500 | 12 pb$^{-1}$ / 25 pb$^{-1}$ | 18 pb$^{-1}$ / -- pb$^{-1}$ | 48 |
| 2012 | 200 | -- pb$^{-1}$ / 22 pb$^{-1}$ | -- pb$^{-1}$ / 9.7 pb$^{-1}$ | 61/56 |
|  | 510 | 82 pb$^{-1}$ / -- pb$^{-1}$ | 32 pb$^{-1}$ / -- pb$^{-1}$ | 50/53 |
| 2013 | 510 | 300 pb$^{-1}$ / -- pb$^{-1}$ | 155 pb$^{-1}$ / -- pb$^{-1}$ | 51/52 |
| 2015 | 200 | 52 pb$^{-1}$ / 52 pb$^{-1}$ | -- pb$^{-1}$ / 60 pb$^{-1}$ | 53/57 |

Table 1-3: Recorded luminosities for collisions of longitudinally and transverse polarized proton beams at the indicated center-of-mass energies for past RHIC runs since 2006. The PHENIX numbers are for |vtx| < 30cm. The average beam polarization as measured by the Hydrogen-jet polarimeter, two polarization numbers are given if the average polarization for the two beams was different

The polarized proton beam program at RHIC has and will continue to address several overarching questions, which have been discussed in detail in [5] and are summarized here.

- *What is the nature of the spin of the proton?*

RHIC has in the last years completed very successful polarized p+p runs both at $\sqrt{s}$ = 200 GeV and 500(510) GeV. The measurement of the gluon polarization in longitudinally polarized protons has been a major emphasis. Data from the RHIC run in 2009 have **for the first time shown that gluons inside a proton are polarized.** The integral of $\Delta g(x,Q^2=10$ GeV$^2)$ in the region $x > 0.05$ is $\mathbf{0.20^{+0.06}_{-0.07}}$ at 90% C.L.

Figure 1-5 shows clearly that the published data, recent preliminary data (see Figure 1-3 and Figure 1-4) and data currently under analysis (RHIC Run-2015) are expected to reduce the present uncertainties on the truncated integral even further by about a factor of 2 at $x_{min}$ = 10$^{-3}$. Measuring $A_{LL}$ versus the invariant mass $M_{inv}$ for di-jets does not only constrain the value of of $\Delta g(x,Q^2)$ but also its functional shape versus $x$ (see Section 3 and Figure 3-1).

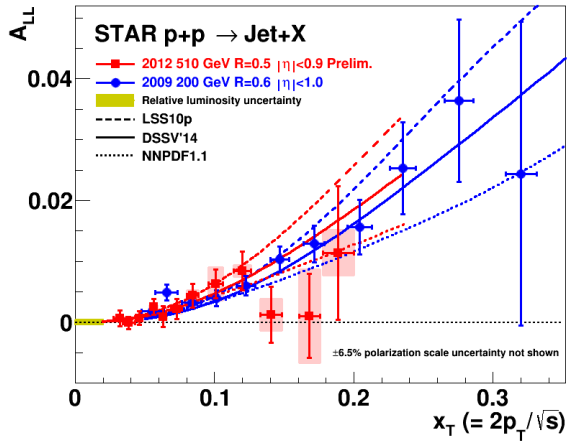

Figure 1-3: $A_{LL}$ vs. $x_T$ for inclusive jet production at mid-rapidity in 200 GeV (blue circles) [6] and 510 GeV (red squares) [7] p+p collisions, compared to NLO predictions [8,9] for three recent NLO global analyses [10,11,12] (blue curves for 200 GeV and red curves for 510 GeV).

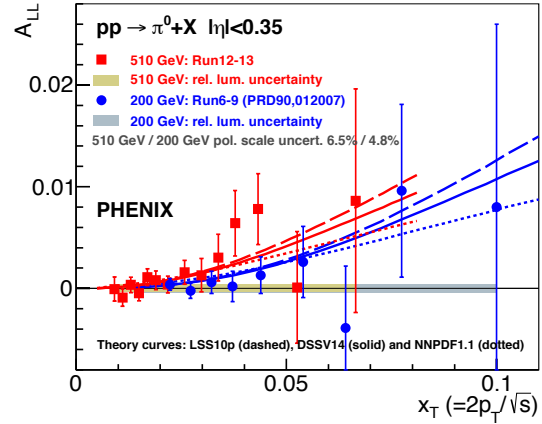

Figure 1-4: $A_{LL}$ vs. $x_T$ for $\pi^0$-meson production at mid rapidity with the point-to-point uncertainties in 200 GeV (blue circles) [13] and 510 GeV (red squares) [14] p+p collisions, compared to NLO predictions [15] for three recent NLO global analyses [10,11,12] (blue curves for 200 GeV and red curves for 510 GeV). The gray/gold bands give the correlated systematic uncertainties.

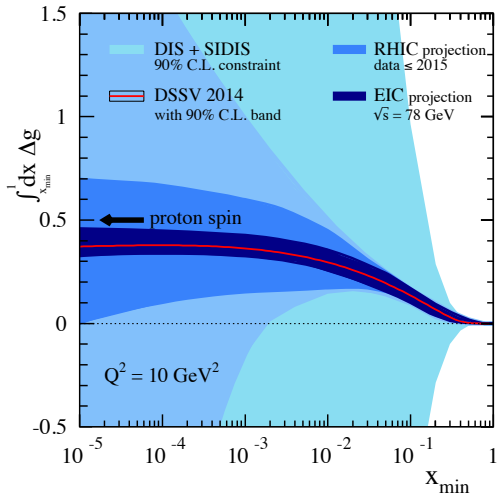

Figure 1-5: The running integral for $\Delta g$ as a function of $x_{min}$ at $Q^2 = 10$ GeV$^2$ as obtained in the DSSV global analysis framework. The different uncertainty bands at 90% C.L. are estimated from the world DIS and SIDIS data, with and without including the combined set of projected pseudo-data for preliminary and RHIC measurements up to Run-2015, respectively as well as including EIC DIS pseudo data (taken from Ref. [16]).

The production of $W^\pm$ bosons in longitudinally polarized proton-proton collisions serves as a powerful and elegant tool [17] to access valence and sea quark helicity distributions at a high scale, $Q \sim M_W$, and without the additional input of fragmentation functions as in semi-inclusive DIS. While the valence quark helicity densities are already well known at intermediate $x$ from DIS, the sea quark helicity PDFs are only poorly constrained. The latter are of special interest due to the differing predictions in various models of nucleon structure (see Ref. [18, 19]). The 2011 and the high statistics 2012 longitudinally polarized p+p data sets provided the first results for $W^\pm$ with substantial impact on our knowledge of the light sea (anti-) quark polarizations (see Figure 1-7 (left)). With the complete data from 2011 to 2013 analyzed by both the PHENIX (see Figure 1-6 (right)) and STAR experiments the final uncertainties will allow one to measure the integrals of the $\Delta\bar{u}$ and $\Delta\bar{d}$ helicity in the accessed $x$ range above 0.05. The uncertainty on the flavor asymmetry for the polarized light quark sea $\Delta\bar{u} - \Delta\bar{d}$ will also be further reduced and a measurement at the $2\sigma$ level will be possible (see Figure 1-7 (right)). These results demonstrate that the RHIC $W$ program will lead, once all the recorded data are fully analyzed, **to a substantial improvement in the understanding of the light sea quark and antiquark polarization in the nucleon.**

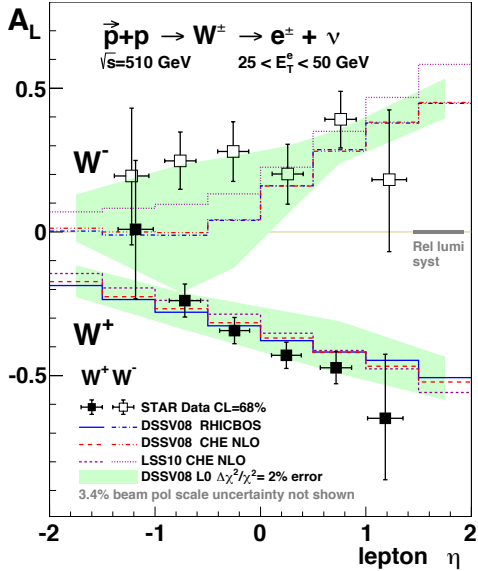
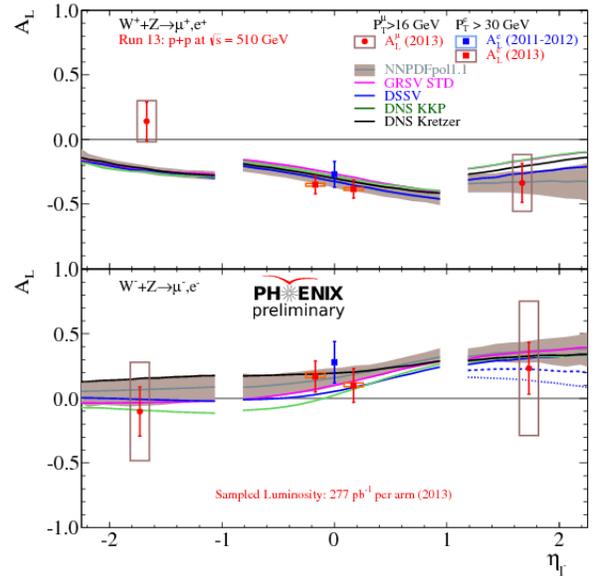

Figure 1-6: Longitudinal single-spin asymmetry $A_L$ for $W^\pm$ production as a function of lepton pseudorapidity $\eta_e$ measured by STAR [20] (left) and PHENIX [21] (right) in comparison to theory predictions [22,23] based on polarized PDFs extracted from only inclusive and semi-inclusive DIS data. The uncertainties for the PHENIX data correspond to the total delivered luminosity corresponding to 713 pb$^{-1}$ (2009 – 2013). The STAR statistical uncertainties will reduce by a factor 2 after the 2013 data set is completely analyzed.

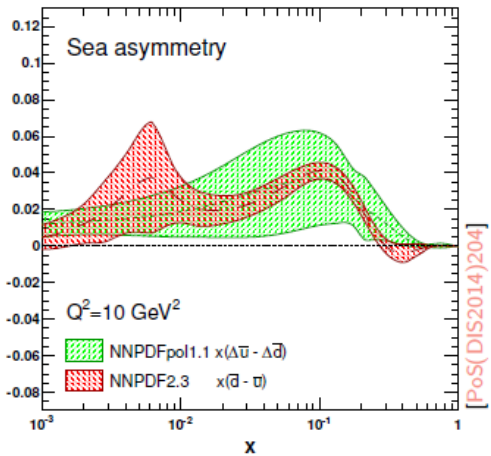
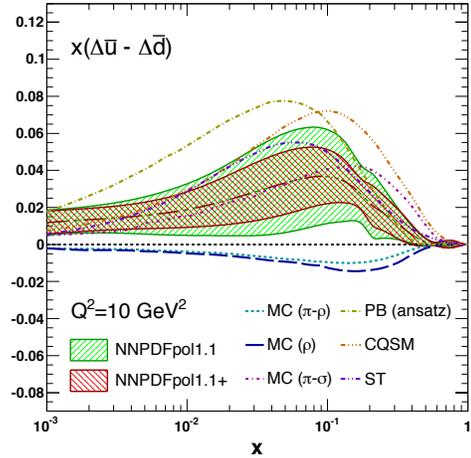

Figure 1-7: left: Light sea polarized (green) and unpolarized (red) differences between $\bar{u}$ and $\bar{d}$ quarks. The curves are extracted by NNPDF-2.3 for the unpolarized PDFs and by NNPDFpol1.1 for the polarized PDFs, which included the 2012 STAR $W$ single spin asymmetries in their fit [24]. right: The polarized light sea-quark asymmetry $x(\Delta\bar{u} - \Delta\bar{d})$ computed with NNPDFpol1.1 and NNPDFpol1.1+ PDFs, after including pseudo-data with projected uncertainties corresponding to the complete data sets from STAR and PHENIX at $Q^2 = 10$ GeV$^2$ compared to various models of nucleon structure (see Ref. [18] for a review).

- *How do quarks and gluons hadronize into final-state particles?*
  *How can we describe the multidimensional landscape of nucleons and nuclei?*

In recent years, transverse spin phenomena have gained substantial attention. They offer a host of opportunities to map out proton structure in three dimensions. Beyond this, they challenge and bring forward our understanding of the interplay between the structure of a hadron and the "color environment" in which this structure is probed. Results from BRAHMS, PHENIX and STAR have shown that large transverse single spin asymmetries (SSA) for inclusive hadron production that were seen in p+p collisions at fixed-target energies and modest $p_T$ extend to the highest RHIC energies and surprisingly large $p_T$. Recently the focus has shifted to observables that will help to separate the contributions from the initial and final state effects, and will give insight to the transverse spin structure of hadrons.

Latest results from transversely polarized data taken in 2006, 2011, and 2012, demonstrate for **the first time that transverse quark polarization is accessible in polarized proton collisions** at RHIC through observables involving the Collins fragmentation function (FF) convoluted with the quark transversity distribution and the interference fragmentation function convoluted with the quark transversity distribution accessed through SSA of the azimuthal distributions of hadrons inside a high energy jet and the azimuthal asymmetries of pairs of oppositely charged pions respectively (see Figure 1-8 and Figure 1-9) at $\sqrt{s}$ = 200 and 500 GeV.

Among the quantities of particular interest to give insight to the transverse spin structure of hadrons is the "Sivers function", which encapsulates the correlations between a parton's transverse momentum inside the proton and the proton spin. It was found that the Sivers function is not universal in hard-scattering processes, which has its physical origin in the rescattering of a parton in the color field of the remnant of the polarized proton (see Figure 2-1). Theory predicts that the Sivers distributions measured in Drell-Yan (DY) production, and in SIDIS are equal in magnitude but opposite in sign.

The experimental test of this prediction is an outstanding task in hadronic physics. It involves our very understanding of QCD factorization, which is among the most important concepts that convey predictive power to the theory. RHIC provides the unique opportunity for the ultimate test of the theoretical concepts of TMDs, factorization, evolution and non-universality, by measuring $A_N$ for $W^{\pm}$, $Z^0$ boson, $DY$ production, and direct photons (for details see Section 2.1).

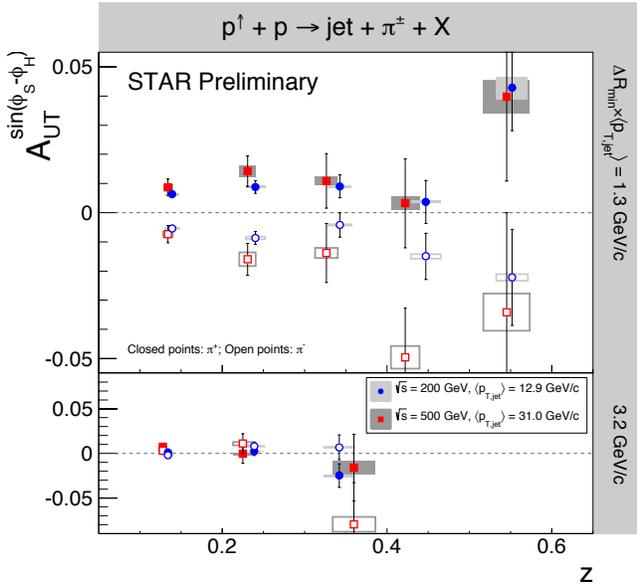

Figure 1-8: $A_{UT}^{\sin(\phi_s-\phi_h)}$ vs. $z$ for charged pions in jets at $0<\eta<1$ from p+p collisions at $\sqrt{s}$ = 200 GeV and 500 GeV by STAR. The $p_{T,jet}$ ranges have been chosen to sample the same parton $x$ values for both beam energies. The angular cuts, characterized by the minimum distance of the charged pion from the jet thrust axis, have been chosen to sample the same $j_T$–values ($j_T \sim z \times \Delta R \times p_{T,jet}$). These data show for the first time a nonzero asymmetry in p+p collisions sensitive to transversity x Collins FF.

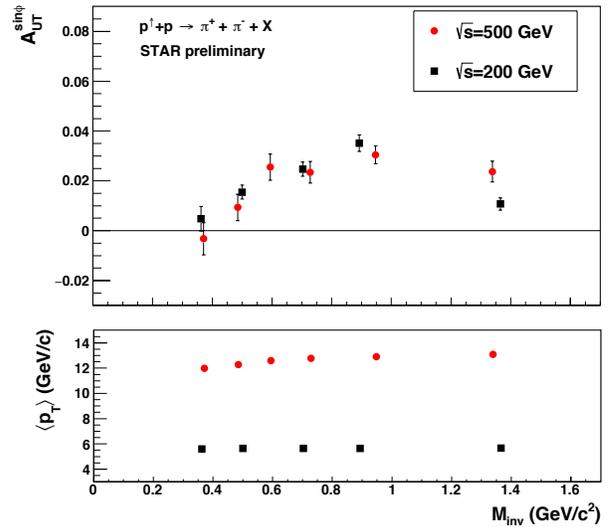

Figure 1-9: $A_{UT}^{\sin(\phi)}$ as a function of $M_{\pi^+\pi^-}$ (upper panel) and corresponding $p_{T(\pi^+\pi^-)}$ (lower panel). A clear enhancement of the signal around the ρ-mass region is observed both at $\sqrt{s}$ = 200 GeV and 500 GeV by STAR for $-1 < \eta < 1$. The $p_{T(\pi^+\pi^-)}$ was chosen to sample the same $x_T$ for $\sqrt{s}$ = 200 GeV and 500 GeV.

- ***What is the nature of the initial state in nuclear collisions?***

Using RHIC's unique capability of (un)polarized $p^{\uparrow}$+A collisions gives the unparalleled opportunity to make progress in our quest to understand QCD processes in Cold Nuclear Matter by studying the dynamics of partons at very small and very large momentum fractions $x$ in nuclei, and at high gluon-density to investigate the existence of nonlinear evolution effects.

First hints for the onset of saturation in d+Au collisions at RHIC have been observed by studying the rapidity dependence of the nuclear modification factor, $R_{dAu}$, as a function of $p_T$ for charged hadrons [25] and $\pi^0$-mesons [26], and more recently through forward-forward hadron-hadron correlations [27].

The nuclear modification factor $R_{pA}$ is equal to unity in the absence of collective nuclear effects. While the inclusive yields of hadrons ($\pi^0$-mesons) at $\sqrt{s} = 200$ GeV in p+p collisions generally agree with pQCD calculations based on DGLAP evolution and collinear factorization, in d+Au collisions, the yield per binary collision is suppressed with increasing $\eta$, decreasing to ~30% of the p+p yield at $<\eta>$ = 4, well below shadowing and multiple scattering expectations (see Figure 3.30 in Ref. [3]). The $p_T$ dependence of the hadron yield in d+Au collisions is found to be consistent with the gluon saturation picture of the Au nucleus (e.g., CGC model calculations [28]) although other interpretations cannot be ruled out based on this observable alone [29]. A more powerful technique than single inclusive measurements is the use of two particle azimuthal correlations, as discussed in Section 4.1.

Scattering a polarized probe on a saturated nuclear wave function provides a unique way of probing the gluon and quark transverse momentum distribution functionss. The single transverse spin asymmetry $A_N$ may provide access to an elusive nuclear gluon distribution function, which is fundamental to the CGC formalism. In particular the nuclear dependence of $A_N$ may shed light on the strong interaction dynamics in nuclear collisions. Theoretical approaches based on CGC physics predicted that hadronic $A_N$ should decrease with increasing size of the nuclear target [30,31,32], some approaches based on perturbative QCD factorization predicted that $A_N$ would stay approximately the same for all nuclear targets [33]. The asymmetry $A_N$ for prompt photons is equally important to measure. The contribution to the photon $A_N$ from the Sivers effect [34] is expected to be nonzero, while the contributions of the Collins effect [35] and of the CGC-specific multi-gluon-mediated contributions [36] to the photon $A_N$ are expected to be suppressed [31,37]. The measurement of $A_N$ for $\pi^0$-mesons was realized during the transversely polarized p+p and p+Au run in 2015. Figure 1-10 shows the results from STAR of $A_N$ for $\pi^0$-mesons measured in the rapidity range 2.5 < $\eta$ < 4.0 as function of $p_T$ and Feynman-x $(x_F = x_1 - x_2)$ for transversely polarized p+p and p+Au collisions [38]. No strong suppression effects have been observed for $A_N$ in p+Au collisions. In light of our latest understanding that a significant fraction of the large transverse single spin asymmetries in the forward direction are not of 2→2 parton scattering processes (see Section 2.3 and 2.1.2), this result supports the clear need for more data to understand the true physics origin for the large forward SSA and the missing nuclear dependence.

Another interesting measurement that stays aside but still may be connected to the discussions above is $A_N$ for neutrons in p+A collisions of a polar angle $\Theta$<2.2 mrad (or $\eta$>6.5), which revealed a surprisingly strong nucleus size dependence, see Figure 1-11. The SSA $A_N$ for inclusive neutrons increases with nucleus mass from being negative in p+p collisions to being large and positive in p+Au collisions. Placing an additional requirement to detect charged particles in the beam-beam counter acceptance (3.0<$|\eta|$<3.9) leads to a saturation-like effect for a heavy nucleus. This effect may be explained by accidental compensation between different mechanisms generating the forward neutron $A_N$. Among such mechanisms could be pion and other Reggeon exchange [39], photon-induced reactions in ultra peripheral collisions, or parton scattering with Delta resonance production. More theoretical developments to understand the sources of these asymmetries and its A-dependencies have just started.

In summary all these results show that spin is a key element in the exploration of fundamental physics. Spin-dependent observables have often revealed deficits in the assumed theoretical framework and have led to novel developments and concepts. The RHIC spin program has played and will continue to play a key role in this by using spin to study how a complex many-body system such as the proton emerge from the dynamics of QCD.

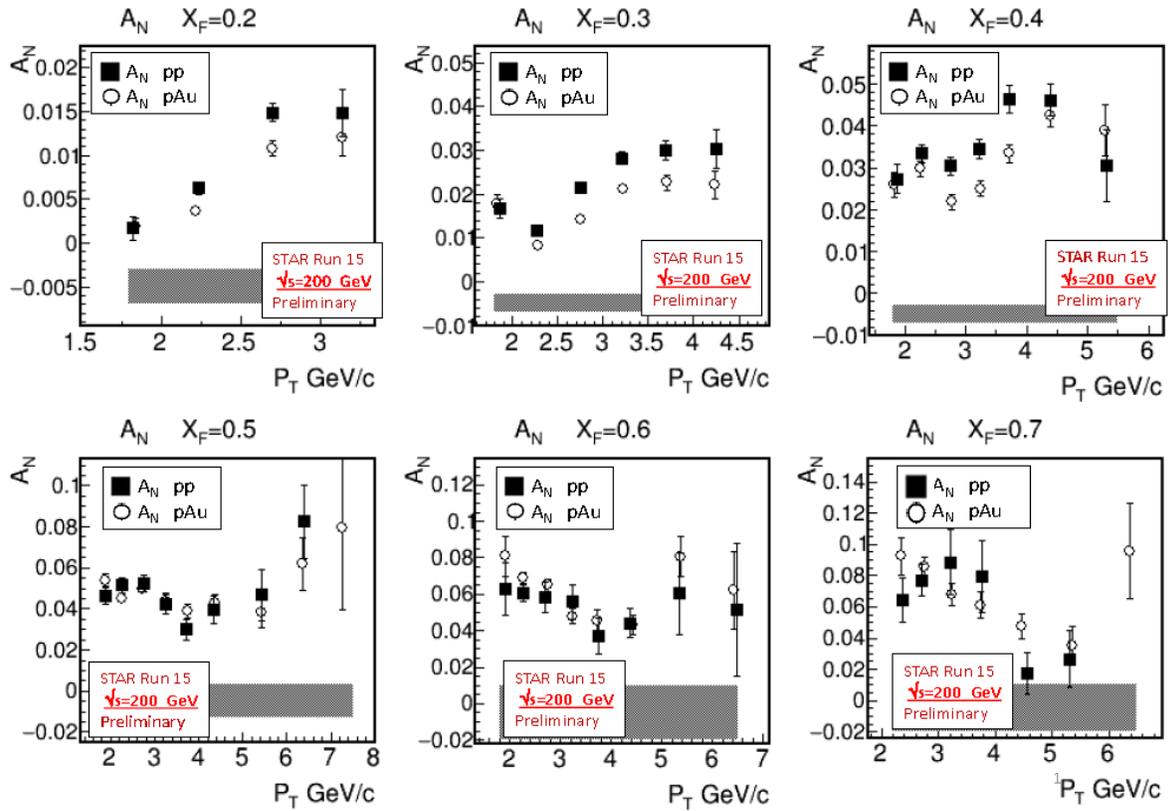

Figure 1-10: $A_N$ for $\pi^0$-mesons measured in the rapidity range $2.5 < \eta < 4.0$ as function of $p_T$ and Feynman-$x$ ($x_F = x_1 - x_2$) for transversely polarized p+p and p+Au collisions measured by STAR. Similar results are expected from the PHENIX MPC-EX [40]

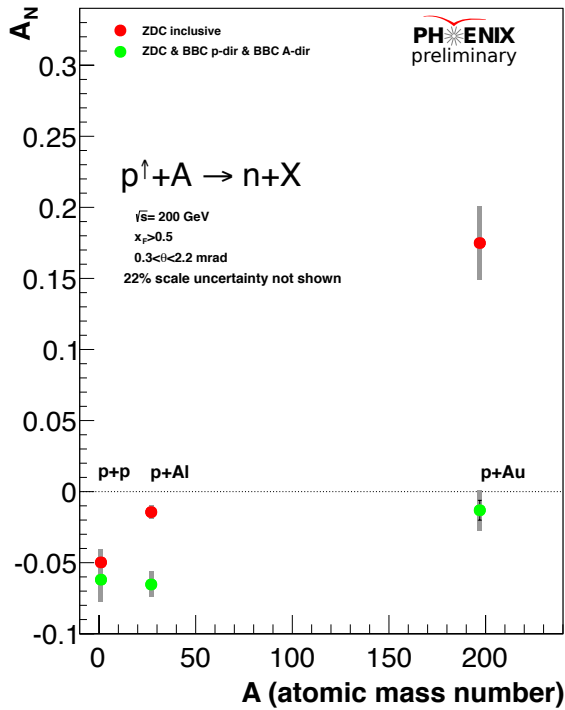

Figure 1-11: $A_N$ for forward neutron production in p+p, p+Al and p+Au collisions at $\sqrt{s}$=200 GeV with experimental cuts corresponding to neutron polar angle $0.3<\Theta<2.2$ mrad relative to polarized proton beam line and $x_F>0.5$, measured by PHENIX; red points - for inclusive neutrons, green points - with additional requirement of signals in both BBC detectors on either side from the collision point, covering pseudorapidity $3.0<|\eta|<3.9$.

# 2 PHYSICS OPPORTUNITIES WITH TRANSVERSLY POLARIZED PROTON - PROTON COLLISIONS

The investigation of nucleon structure will be revolutionized by imaging the proton in both momentum and impact parameter space. From TMD parton distributions we can obtain an "image" of the proton in transverse as well as in longitudinal momentum space (2+1 dimensions). In combination with transverse spin, the study of TMDs has challenged and greatly brought forward our understanding of the interplay between hadron structure and the process by which this structure manifests itself. This has attracted renewed interest, both experimentally and theoretically, in transverse single spin asymmetries in hadronic processes at high energies. The surprisingly large asymmetries seen are nearly independent of $\sqrt{s}$ over more than two orders of magnitude. To understand the observed SSAs one has to go beyond the conventional leading twist collinear parton picture in the hard processes. Two theoretical formalisms have been proposed to explain sizable SSAs in the QCD framework: These are transverse momentum dependent parton distributions and fragmentation functions, such as the Sivers and Collins functions discussed below, and transverse-momentum integrated (collinear) quark-gluon-quark correlations, which are twist-3 distributions in the initial state proton or in the fragmentation process. For many spin asymmetries, several of these functions can contribute and need to be disentangled to understand the experimental observations in detail, in particular the dependence on measured $p_T$. The functions express a spin dependence either in the initial state (such as the Sivers distribution or its Twist-3 analog, the Efremov-Teryaev-Qui-Sterman (ETQS) function [41]) or in the final state (via the fragmentation of polarized quarks, such as the Collins function).

The Sivers function, $f_{1T}^\perp$, describes the correlation of the parton transverse momentum with the transverse spin of the nucleon. A non-vanishing $f_{1T}^\perp$ means that the transverse parton momentum distribution is azimuthally asymmetric, with the nucleon spin providing a preferred transverse direction. The Sivers function, $f_{1T}^\perp$, is related with the ETQS functions, $T_{q,F}$, in two ways (see e.g. [42] for further discussion). On the one hand, there is an integral relation:

$$T_{q,F}(x,x) = -\int d^2k_\perp \frac{|k_\perp|^2}{M} f_{1T}^{\perp q}(x, k_\perp^2)|_{SIDIS}$$
[Eq. 2-1].

On the other hand, the large $k_T$ behavior of the Sivers function can be expressed in terms of $T_{q,F}(x_1, x_2)$ convolved with a known hard-scattering kernel. Given these relations, a measurement constraining the ETQS function also constrains the Sivers function. We will use this connection repeatedly in the following discussion.

The physical origin of these relations can be seen in Figure 2-1. The Sivers function includes the effect of the exchange of (any number of) gluons between the spectator partons in the polarized protons and the active partons taking part in the hard-scattering subprocess. At high transverse parton momentum $k_T$, the exchange of one such gluon becomes dominant, and one can use the ETQS function to describe the relevant long-distance physics.

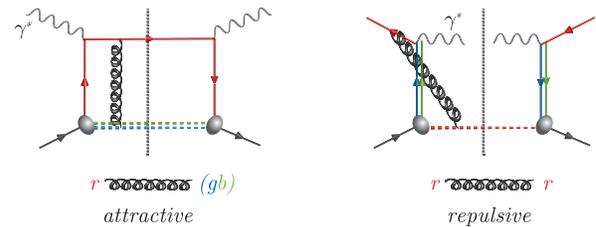

Figure 2-1: Graphs responsible for the Sivers asymmetry in deep-inelastic lepton nucleon scattering (left hand side) and the Drell Yan process (right hand side).

The Collins function, $H_1^\perp$, describes a correlation of the transverse spin of a fragmenting quark and the transverse momentum of a hadron singled out among its fragmentation products. A crucial difference between the Sivers and Collins functions is that the former has a specific process dependence (see Section 2.1) whereas the latter is process independent. This is a non-trivial theory result, first shown in [43] and extended to p+p collisions in [44], which applies to TMD distribution and fragmentation functions in general.

The universality of the Collins FF is of special importance to the p+p case where it is always coupled to the chirally odd quark transver-sity distribution $\delta q(x,Q^2)$, which describes the transverse spin preference of quarks in a transversely polarized proton.

In the last years observables have been identified that separate the contributions from the initial and final state effects, and will give much deeper insight to the transverse spin structure of hadrons.

To disentangle the different subprocesses it is important to identify less inclusive measurements. Table 2-1 identifies observables that allow separating the contributions from polarization effects in initial and final states, and will give insight to the transverse spin structure of hadrons. It is reemphasized that many observables in p+p collisions can only be related to the transverse spin structure of hadrons through the Twist-3 formalism, where only one hard scale is required. This is typically the $p_T$ of a produced particle or jet, which at RHIC is sufficiently large in much of the phase space. By contrast, the TMD framework requires two hard scales, $p_T$ and $Q$ with $p_T << Q$. Di-jets, azimuthal dependences of hadrons within a jet, $W$, $Z$, or Drell-Yan production are observables in p+p collisions providing two such scales. Moreover, TMD factorization in p+p collisions may be broken for processes with observed hadrons or jets in the final state, if there are more than two colored objects involved [45]. To measure the size of such factorization breaking effects stemming from "color entanglement" in these processes, is of interest in itself. It explores the limitations of the factorization concept and our understanding of the color flow in non-trivial QCD interactions in a quantitative way. Obviously, data from p+p collisions is indispensable for this.

| Initial State | Final State |
|---|---|
| $A_N$ as function of rapidity, $E_T$, $p_T$ and $x_F$ for inclusive jets, direct photons and charmed mesons | $A_{UT}$ as a function of the azimuthal dependence of the correlated hadron pair on the spin of the parent quark (transversity x interference fragmentation function) |
| $A_N$ as a function of rapidity, $p_T$ for $W^\pm$, $Z^0$ and DY | Azimuthal dependences of hadrons within a jet (transversity x Collins fragmentation function) |
| $A_N$ as function of rapidity, $p_T$ and $x_F$ for inclusive identified hadrons (transversity x Twist-3 fragmentation function) | |

Table 2-1: Observables to separate the contributions from initial and final states to the transverse single spin asymmetries. Two-scale observables are indicated in **green** and one-scale ones in **black**.

## 2.1 POLARIZATION EFFECTS IN THE PROTON: SIVERS AND TWIST-3

An important aspect of the Sivers effect that has emerged from theoretical study is its process dependence. In SIDIS, the quark Sivers function includes the physics of a final state effect from the exchange of (any number of) gluons between the struck quark and the remnants of the target nucleon. On the other hand, for the virtual photon production in the Drell-Yan process, the Sivers asymmetry appears as an initial state interaction effect (see Figure 2-1). As a consequence, the quark Sivers functions are of **equal size and of opposite sign** in these two processes. This non-universality is a fundamental prediction following from the symmetries of QCD and the space-time and color structure of the two processes. The experimental test of this sign change is one of the open questions in hadronic physics (NSAC performance measure HP13) and will deeply test our understanding of QCD factorization. The COMPASS experiment at CERN is pursuing this sign change through DY using a pion beam in the years 2015 and 2016 and new initiatives have been proposed e.g. at FNAL [46].

While the required luminosities and background suppressions for a precision measurement of a SSA in Drell-Yan production are challeng-ing, other channels can be exploited in p+p collisions, which provide the same sensitivity to the predicted sign change. These include in the TMD formalism the measurement of SSA of $W^{\pm}$ and $Z$ bosons and in the Twist-3 formalism the SSA for prompt photons and inclusive jets. These are already accessible with the existing STAR detector, but require continued polarized beam operations.

Figure 2-2 shows the predicted $A_N$ for DY (left) [52] and $W$ [54] (right) **before any TMD evolution is taken into account.** At this point, we must discuss a new theory development since the formulation of the Long Range Plan [1]. The equation describing the evolution of TMDs with the hard scale of the process is well known, but it involves an evolution kernel that is itself non-perturbative in the region relevant for small transverse parton momenta. The form of the kernel must hence be determined itself by experiment. Recent analyses of unpolarized data [47,48] have shown that the previously assumed form of the evolution kernel is most likely inadequate. This also puts into question the reliability of currently available theoretical predictions for the transverse single spin asymmetries for DY, $W^{\pm}$ and $Z^0$ bosons including TMD-evolution, for examples see [49,50,51] and references therein. In all cases the asymmetries have been significantly reduced. We are thus currently left with large uncertainties in the prediction for the DY, $W^{\pm}$ and $Z^0$ SSA, which can only be addressed by future measurements. Since the hard scale in typical DY events is very different from the one in $W$ and $Z$ production, their combination would provide an ideal setting for studying evolution effects. An indication of the magnitude of evolution effects in asymmetry measurements, where part of the effect might cancel in the ratio of the polarized over the unpolarized cross-section, can be taken from recent STAR results on the Collins effect (see Figure 1-8): Intriguingly virtually no reduction of the asymmetry is observed between measurements at $\sqrt{s}$=200 and $\sqrt{s}$=500 GeV and results are consistent with theory calculations using only collinear evolution effects.

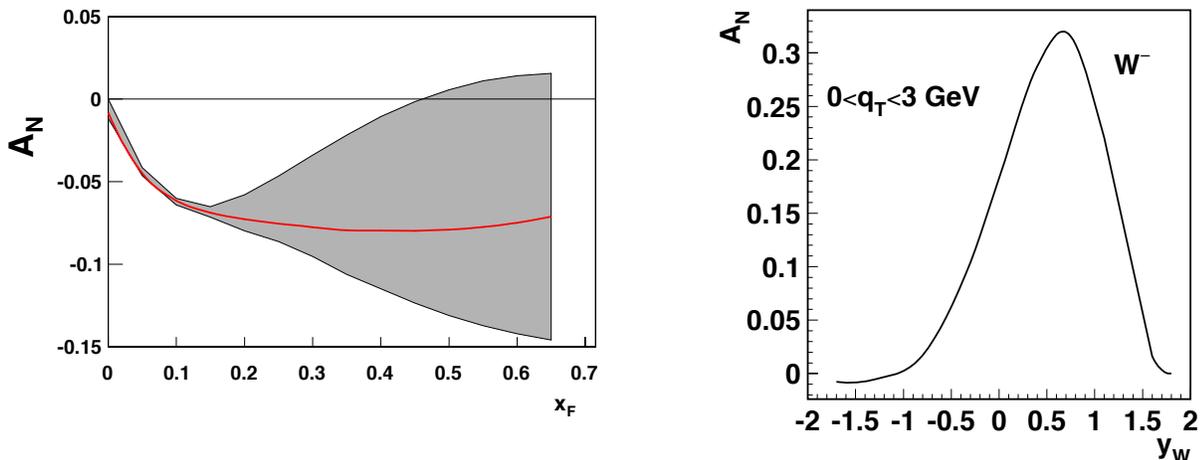

Figure 2-2: (left) Prediction for Sivers asymmetry $A_N$ for DY lepton pair production at $\sqrt{s}$=500 GeV, for the invariant mass 4<$Q$<8 GeV and transverse momenta $0 < q_T < 1$ GeV [52]. (right) $A_N$ as a function of $W^-$ boson rapidity at $\sqrt{s}$=500 GeV [54]. **Both predictions are before any TMD evolution is applied.**

### 2.1.1 Run-2017

*The Sivers Function*

The transversely polarized data set in Run-2011 at $\sqrt{s}$ = 500 GeV allowed STAR to make a first attempt to address the non-universality and the evolution of the Sivers function through a measurement of the transverse SSA $A_N$ for $W^{\pm}$ and $Z^0$ bosons [53]. It is noted that the measurement of the $A_N$ for $W^{\pm}$ bosons, contrary to the longitudinally polarized case, requires to completely reconstruct the $W$ bosons as the kinematic dependences of $A_N$ cannot easily be resolved through measurements of only the high $p_T$ decay lepton, for details see [54,55].

Due to the large STAR acceptance it is possible to reconstruct the $W$ boson kinematics from the recoil jet, a technique established at D0, CDF and the LHC experiments. Figure 2-3 shows the transverse single spin asymmetries $A_N$ for $W^{\pm}$ and $Z^0$ bosons, as functions of the boson rapidity $y$.

The asymmetries have also been reconstructed as functions of the $p_T$ of the $W$-bosons. For the $Z^0$-boson the asymmetry could only be reconstructed in one $y$-bin due to the currently limited statistics (25 pb$^{-1}$). Details for this analysis can be found in Ref. [53,56]. A combined fit to the $W^\pm$ asymmetries based on the theoretical predictions of the Kang-Qiu (KQ) model [54] favors a sign-change for the Sivers function relative to the Sivers function in SIDIS with $\chi^2$/ndf = 7.4/6 compared to $\chi^2$/ndf = 19.6/6 for no sign-change, if TMD evolution effects are small. The analysis represents an important proof of principle. Figure 2-4 shows the projected uncertainties for transverse single spin asymmetries for $W^\pm$ and $Z^0$ bosons as functions of rapidity for a delivered integrated luminosity of 400 pb$^{-1}$ and an average beam polarization of 55%, as expected in Run-2017. Such a measurement will provide the world wide first constraint on the light sea quark Sivers function. At the same time, this measurement is also able to access the sign change of the Sivers function, if the effect due to TMD evolution on the asymmetries is of the order of a factor five reduction. Figure 2-5 shows the unique $x$-$Q^2$ kinematics accessed by the $W$-bosons at RHIC compared to the $x$-$Q^2$ plane covered by a future EIC, JLab-12, and the current SIDIS world data. Combining the RHIC $W$-boson data with the future EIC SIDIS data accessing the Sivers function at the same $x$ but significant lower $Q^2$ will provide a unique opportunity for a stringent test of TMD evolution.

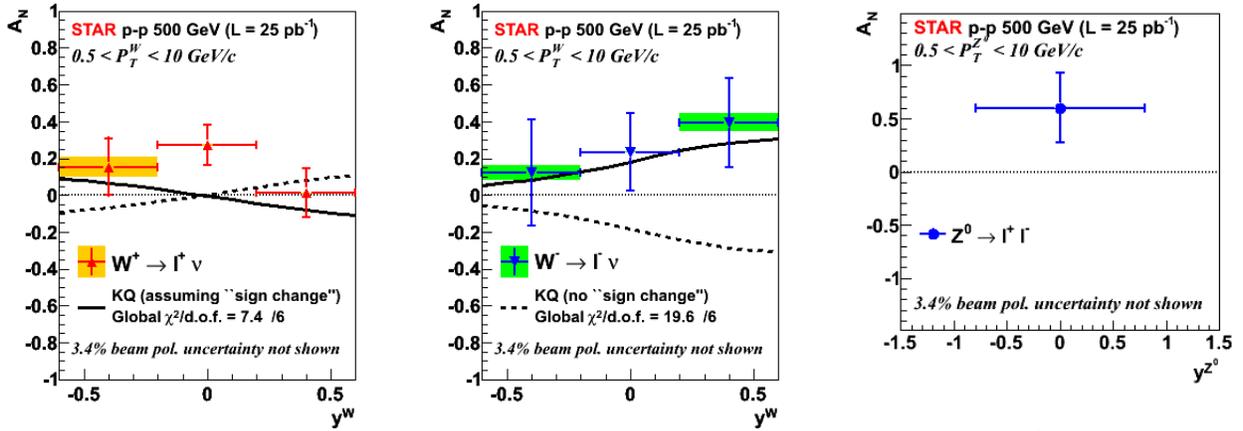

Figure 2-3: Transverse single-spin asymmetry amplitude for $W^+$ (left plot), $W^-$ (middle plot) and $Z^0$ boson versus $y_{W/Z}$ measured by STAR in proton-proton collisions at $\sqrt{s}$ = 500 GeV. The $W^\pm$ boson asymmetries are compared with the non TMD-evolved KQ [54] model, assuming (solid line) or excluding (dashed line) a sign change in the Sivers function.

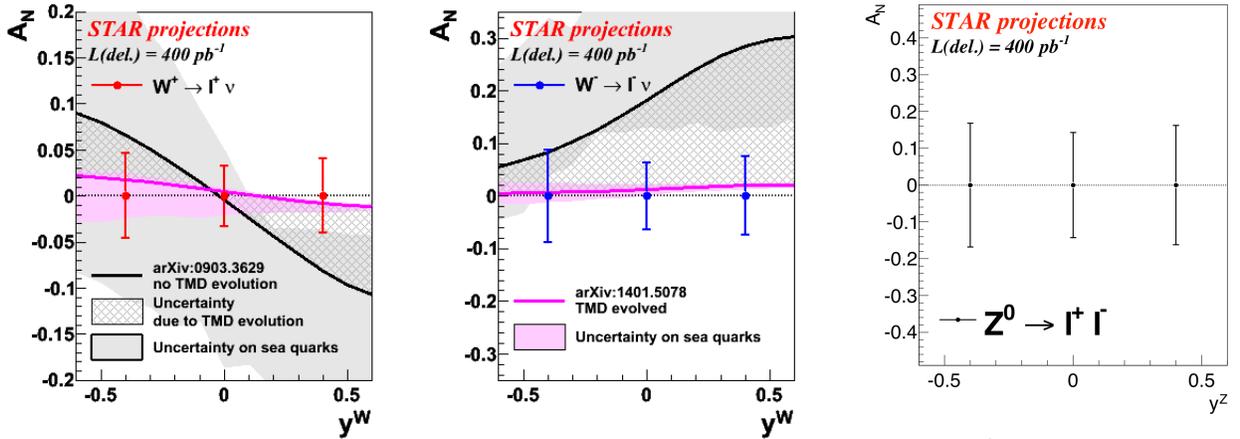

Figure 2-4: The projected uncertainties for transverse single-spin asymmetries of $W^\pm$ and $Z^0$ bosons as functions of rapidity for a delivered integrated luminosity of 400 pb$^{-1}$ and an average beam polarization of 55%. The solid light gray and pink bands represent the uncertainty on the KQ [54] and EIKV [51] model, respectively, due to the unknown sea quark Sivers function. The crosshatched dark grey region indicates the current uncertainty in the theoretical predictions due to TMD evolution.

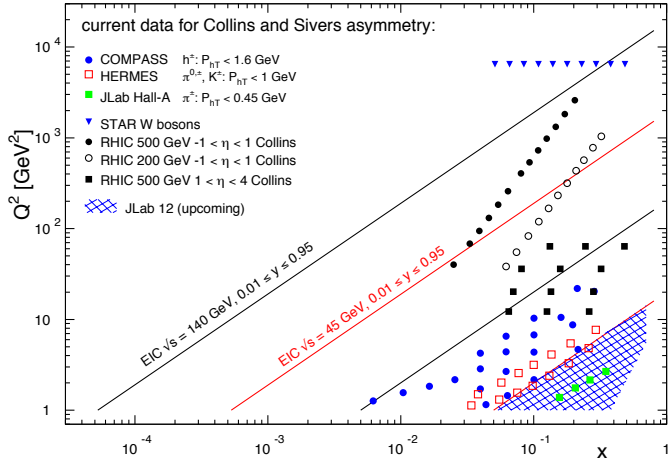

Figure 2-5: The $x$-$Q^2$ plane for data from the future EIC and Jlab-12 GeV as well as the current SIDIS data and the W-boson data from RHIC. All data are sensitive to the Sivers function and transversity times the Collins FF in the TMD formalism.

The ultimate test for the TMD evolution and the sign change of the Sivers function would be to measure $A_N$ for $W^{\pm}$, $Z^0$ boson and DY production simultaneously. To obtain a significant measurement of $A_N$ for DY production, the DY leptons need to be detected between rapidities 2 and 4 for a lepton pair mass of 4 GeV and bigger. This is a highly non-trivial measurement, as backgrounds mainly due to QCD 2→2 processes need to be suppressed by a factor of ~$10^6$. Figure 2-6 shows the achievable statistical precision measuring one point in the rapidity-range $2.5 < \eta < 4.0$ for the asymmetry for a delivered integrated luminosity of 400 pb$^{-1}$ in comparison to the theoretical predicted asymmetry with and without taking TMD evolution from a specific model into account.

The biggest challenge of DY measurements is the suppression of the overwhelming hadronic background which is of the order of $10^5 \sim 10^6$ larger than the total DY cross-section. The probability of misidentifying a hadron track as $e^+/e^-$ has to be suppressed down to the order of 0.01% while maintaining reasonable electron detection efficiencies. Due to the rarity of Drell-Yan events, the simulation of the both the Drell Yan process and the large QCD background are crucial to understanding how well we can distinguish the signal from the background. The combined electron/hadron discriminating power of the proposed calorimeter postshower and current calorimeter systems has been studied. We found that by applying multivariate analysis techniques to the features of EM/hadronic shower we can achieve hadron rejection powers of 800 to 14,000 for hadrons of 15 GeV to 60 GeV with 90% electron detection efficiency. The hadron rejection power has been parameterized as a function of hadron energy and has been used in a fast simulation to estimate DY signal-to-background ratios.

The current STAR detectors in this rapidity range $2.5 < \eta < 4.0$ are the Forward Meson Spectrometer (FMS), a Pb-glass electromagnetic detector with photomultiplier tubes, and the preshower, a simple hodoscope comprised of three layers of scintillator slats with silicon photomultipliers. The FMS is primarily sensitive to electrons and photons while hadrons will leave as minimum ionizing particles. The preshower provides photon and charged particle separation. The first two layers provide x and y positioning. A lead converter precedes the third scintillator layer causing photons and electrons to shower in lead and deposit significant energy in the third scintillator. To suppress photons, the signal should have energy deposition in each layer of the preshower. The three-detector setup (preshower, FMS and proposed postshower) distinguishes photons from minimum ionizing particles and provides electron/hadron discrimination.

These energy observables from the three detectors have been used as inputs to a Boosted Decision Trees (BDT) algorithm. The algorithm takes advantage of using not only the discriminating power of each single observable but also the correlations among them. The final background yields as a function of pair masses were then fit by an exponential function and rescaled to a total luminosity of 400 pb$^{-1}$, the results are shown in Figure 2-7.

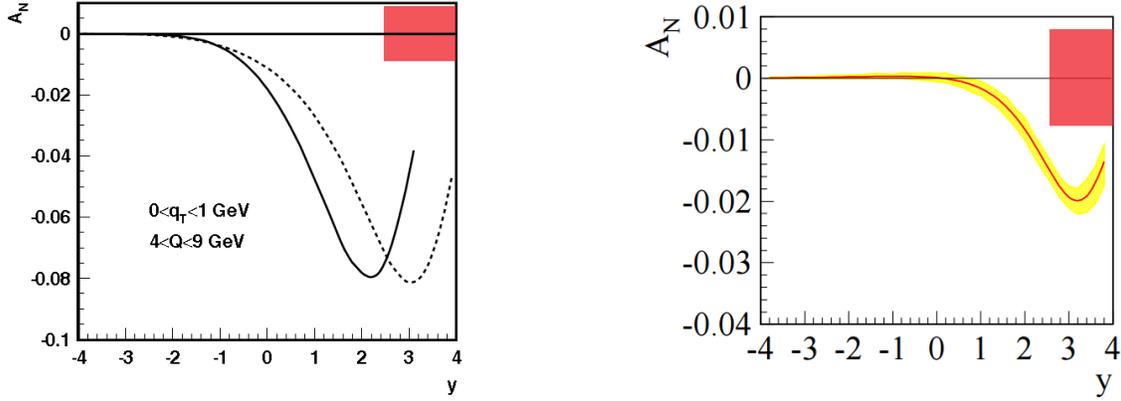

Figure 2-6: The orange square indicates the achievable statistical precision measuring the asymmetry integrated over the rapidity-range 2.5 < η < 4.0 (no $p_T$-cut) for a delivered integrated luminosity of 400 pb$^{-1}$ in comparison to the theoretical prediction for the Sivers asymmetry $A_N$ as a function of DY lepton-pair rapidity at √s=200 GeV (solid line) and √s=500 GeV (dotted line) [57] **before any TMD evolution is applied** (left). Theoretical predictions from reference [51] for DY for 0 GeV <$p_T$< 1 GeV and 4 GeV < Q < 9 GeV **after** TMD evolution is applied (right). The yellow band represents the expected uncertainty for the asymmetry.

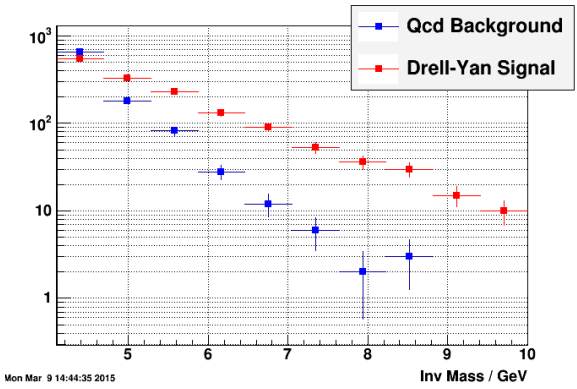

Figure 2-7: The background after BDT rejection (blue) along with a normalized Drell-Yan signal (red).

### *The Efremov-Teryaev-Qiu-Sterman Function*

Transverse single spin asymmetries in direct photon production provide a different path to access the sign change through the formalism utilizing the Twist-3 parton correlation functions. For the 2015 polarized p+p run both PHENIX and STAR installed a preshower in front of their forward electromagnetic calorimeters, the MPC [58] and the FMS [59]. These upgrades enabled a measurement of the SSA for direct photons up to $x_F \sim 0.7$ in Run-2015 at √s = 200 GeV, where the inclusive $\pi^0$ asymmetries are largest.

Figure 2-8 shows a theoretical prediction and the projected statistical and systematic uncertainties for a direct photon SSA at √s = 500 GeV. The theoretical predictions represent a calculation based on Twist-3 parton correlation functions, $T_{q,F}$, determined through Eq. (3-1) and thus constrained by the Sivers function obtained from a fit to the world SIDIS data [60]. At √s = 500 GeV the theoretical asymmetries are reduced by a factor two due to evolution effects compared to the one at √s = 200 GeV. In the quoted study, the evolution of the ETQS function was estimated in a simplified manner, using the DGLAP evolution equations for unpolarized PDFs. The full evolution of the Twist-3 functions is more difficult to implement, since it requires knowledge of $T_{q,F}(x_1, x_2)$ as a function of two independent momentum fractions. The comparison of the 200 GeV and 500 GeV data would provide a unique opportunity to obtain experimental constraints on Twist-3 functions and their evolution, a field that is only in its infancy at the current time. Due to the electromagnetic nature of the process the individual parton densities are weighted with the respective quark charge $e_q^2$, therefore the direct photon asymmetries are mainly sensitive to the $u$ quark Twist-3 correlation functions (in analogy

to Drell-Yan, which is mainly sensitive to the $u$-quark Sivers function).

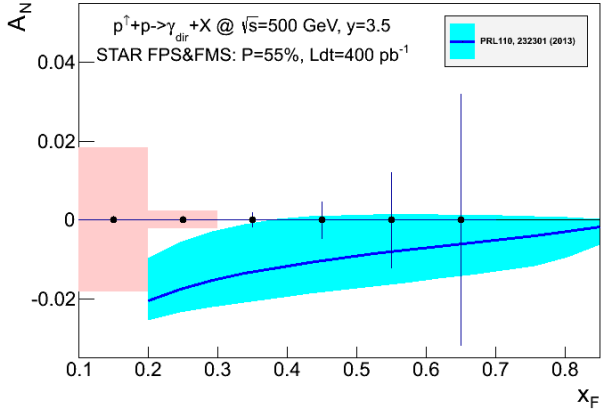

Figure 2-8: Statistical and systematic uncertainties for the direct photon $A_N$ after background subtraction compared to theoretical predictions from Ref. [52] for $\sqrt{s}$ = 500 GeV as projected for STAR. If the correlation, due to the different color interactions for initial and final state interaction between the Sivers function and the Twist-3 correlation function as described in [Eq.2-1] would be violated, the asymmetries would have the same magnitude but would be positive.

The ultimate test for the TMD factorization, evolution and the relation between the Sivers function and the Twist-3 correlation function (see Eq. 4.1) is to measure $A_N$ for $W^\pm$, $Z^0$ boson, DY production and direct photons. Table 2-2 summarizes the different observables and their sensitivity to the following main questions to be addressed with the transversely polarized p+p run in 2017:

- Can the sign change of the Sivers function between SIDIS and DY-production be experimentally verified?
- What are the effects on $A_N$ due to TMD evolution?
- Do sea quarks have significant Sivers and Twist-3 ETQS functions?
- Can the relation between the Sivers function and the Twist-3 ETQS distribution function be experimentally verified?
- Can the evolution of the Twist-3 ETQS distribution functions be experimentally constrained?

It is especially noted that answers to these questions are critical for the effective planning of the physics program of an electron-ion-collider.

|  | $A_N(W^{+/-},Z^0)$ | $A_N(DY)$ | $A_N(\gamma)$ |
|---|---|---|---|
| Sensitive to Sivers effect non-universality through TMDs | **Yes** | **Yes** | No |
| Sensitive to Sivers effect non-universality through Twist-3 $T_{q,F}(x,x)$ | No | No | **Yes** |
| Sensitive to TMD or Twist-3 evolution | **Yes** | **Yes** | **Yes** |
| Sensitive to sea quark Sivers or ETQS function | **Yes** | **Yes** ($x<10^{-3}$) | No |
| Detector upgrade needed | No | **Yes** FMS post-shower | No |
| Biggest experimental challenge | Integrated luminosity | Background suppression Integrated luminosity | ---- |

Table 2-2: Summary of all the processes accessible in STAR to access the sign change of the Sivers function.

## 2.1.2 Run-2023 and Opportunities with a Future Run at 500 GeV

First and foremost, a transversely polarized 500 GeV p+p run with anticipated delivered luminosity of 1 fb$^{-1}$ will reduce the statistical uncertainties of all observables discussed in Section 2.1.1 by a factor of two, including the flagship measurement of the Sivers effect in W and Z production. This experimental accuracy will significantly enhance the quantitative reach of testing the limits of factorization and universality in lepton-proton and proton-proton collisions.

Results from PHENIX and STAR have shown that large transverse single spin asymmetries for inclusive hadron production, $A_N$, that were first seen in p+p collisions at fixed-target energies and modest $p_T$ extend to the highest RHIC center-of-mass (c.m.) energies, $\sqrt{s}$ = 500 GeV and surprisingly large $p_T$. Figure 2-9 summarizes the world data as function of Feynman-x. Surprisingly the asymmetries are nearly independent of $\sqrt{s}$ over a very wide range ($\sqrt{s}$: 4.9 GeV to 500 GeV).

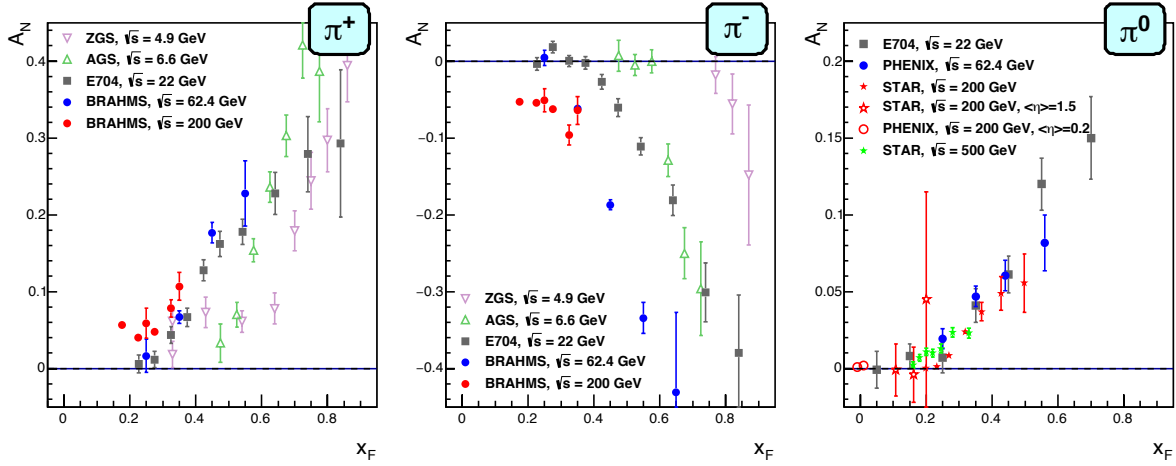

Figure 2-9: Transverse single spin asymmetry measurements for charged and neutral pions at different center-of-mass energies as a function of Feynman-x.

The latest attempt to explain $A_N$ for $\pi^0$ production at RHIC incorporated the fragmentation term within the collinear twist-3 approach [61]. In that work, the relevant (non-pole) 3-parton collinear fragmentation function $\hat{H}_{FU}^{\Im}(z, z_z)$ was fit to the RHIC data. The so-called soft-gluon pole term, involving the ETQS function $T_{q,F}(x_1,x_2)$, was also included by fixing $T_{q,F}$ through its well-known relation to the TMD Sivers function $f_{1T}^{\perp}$. The authors found a very good description of the data due to the inclusion of $\hat{H}_{FU}^{\Im}(z, z_z)$. Based on this work, one is able to make predictions for $\pi^+$ and $\pi^-$ production at forward rapidities covered by the forward upgrade. The results are shown in Figure 2-10 for two different center-of-mass energies (200 GeV and 500 GeV) and rapidity ranges (2 < $\eta$ < 3 and 3 < $\eta$ < 4).

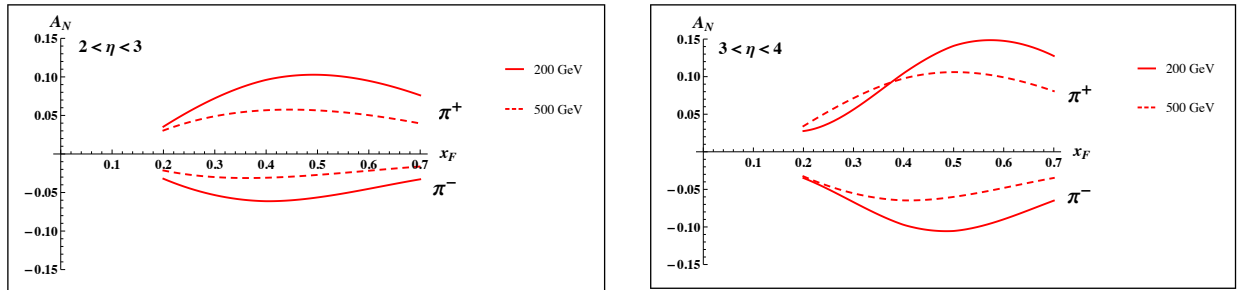

Figure 2-10: Predictions, based on the work in Ref. [61], for $A_N$ for $\pi^+$ and $\pi^-$ production for 2 < $\eta$ < 3 (left) and 3 < $\eta$ < 4 (right) at 200 GeV (solid lines) and 500 GeV (dashed lines).

The proposed forward upgrade, incorporating forward tracking (see Section 5), will enable us to access the previously measured charged hadron asymmetries [62] up to the highest center-of-mass energies at RHIC. It will be important to confirm that also the charge hadron asymmetries are basically independent of center-of-mass energy. The measurement of $A_N$ for charged hadrons together with the data from Run-2015 and 2017 on direct photons $A_N$ and $\pi^0$ should provide the best data set in the world to:

- Constrain the flavor dependence of the twist-3 ETQS distribution
- Constrain the evolution of the twist-3 ETQS distribution functions experimentally
- Determine if the 3-parton collinear fragmentation function $\widehat{H}_{FU}^{\Im}(z, z_z)$ is the main driver for the large forward $A_N$

Equally interesting is the possibility to test the relation of the ETQS correlation functions and the Sivers function by measuring $A_N$ for direct photon production and $A_N$ for forward jet production. Initial measurements from the $A_N$DY collaboration [93] indicated moderate asymmetries, which in [52] is argued to be consistent with the fact that the Twist-3 parton correlation functions for $u$ and $d$ valence quarks cancel, because their behavior follows the one obtained for the Sivers function from fits to the SIDIS data, which show the $u$ and $d$ quark Sivers function to have opposite sign but equal magnitude. To better quantitatively test the relation between the two regimes, jet asymmetries, which are biased towards up or down quark jets with the help of a high-$z$ charged hadron should be studied. In higher twist calculations of the jet asymmetries based on the Sivers function [48], sizeable asymmetries for the thus enhanced jets are predicted. This is experimentally accessible via forward jet reconstruction by tagging an additional charged hadron in the jet. Using realistic jet smearing in a forward calorimeter and tracking system and requiring a charged hadron with $z > 0.5$, the asymmetries can clearly be separated and compared to the predictions for the Sivers function based on the SIDIS data. The expected uncertainties, plotted at the predicted values can be seen in Figure 2-11. Dilutions by underlying event and beam remnants were taken into account. The simulations have assumed only an integrated luminosity of 100 pb$^{-1}$ at $\sqrt{s}$ = 200 GeV, which is significantly lower than what is currently expected for a 200 GeV polarized p-p run in 2023. The same measurement is possible at 500 GeV.

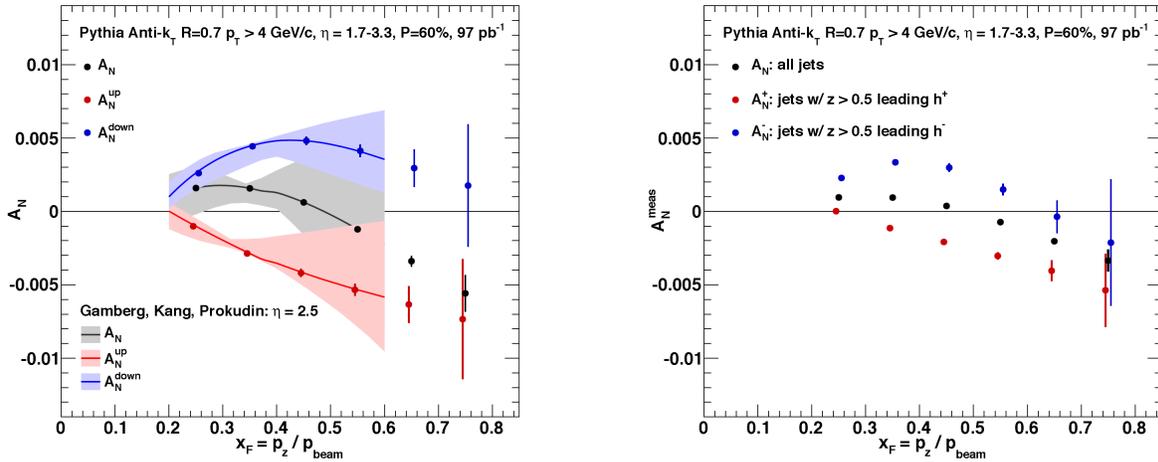

Figure 2-11: Left: up quark (red points), down quark (blue points) and all jet (black points) single spin asymmetries as a function of $x_f$ as calculated by the ETQS based on the SIDIS Sivers functions. Right: Expected experimental sensitivities for jet asymmetries tagging in addition a positive hadron with $z$ above 0.5 (red points), a negative hadron with $z$ above 0.5 (blue points) or all jets (black) as a function of $x_f$. Note: these figures are currently for 200 GeV center-of-mass energy proton collisions – the 500 GeV results are expected to be qualitatively similar but with reduced uncertainties due to the larger luminosities expected.

## 2.2 TRANSVERSITY, COLLINS FUNCTION AND INTERFERENCE FRAGMENTATION FUNCTION

As described above, for a complete picture of nucleon spin structure at leading twist one must consider not only unpolarized and helicity distributions, but also those involving transverse polarization, such as the transversity distribution, [63, 64, 65]. The transversity distribution can be interpreted as the net transverse polarization of quarks within a transversely polarized proton [64]. It is noted that the difference between the helicity distributions and the transversity distributions for quarks and antiquarks provides a direct, $x$-dependent, connection of nonzero orbital angular momentum components in the wave function of the proton [66]. Recently, the measurement of transversity has received renewed interest to access the so-called the tensor charge of the nucleon, defined as the integral over the valence quark transversity: $\int_0^1 (\delta q^a(x) - \delta \bar{q}^a(x)) dx = \delta q^a$ [64, 67]. Measuring the tensor charge is very important for two reasons: It is an essential quantity to our understanding of the spin structure of the nucleon. It can be calculated on the lattice with comparatively high precision, and due to the valence nature of transversity, it is one of the few quantities that allow us to compare experimental results on the spin structure of the nucleon to ab-initio QCD calculations. The second reason is that the tensor charge describes the sensitivity of observables in low energy hadronic reactions to beyond the standard model (BSM) physics processes with tensor couplings to hadrons. Examples are experiments with ultra-cold neutrons and nuclei.

Transversity is difficult to access due to its chiral-odd nature, requiring the coupling of this distribution to another chiral-odd distribution. SIDIS experiments have successfully probed transversity through two channels: asymmetric distributions of single pions, coupling transversity to the transverse-momentum-dependent Collins FF [68], and azimuthally asymmetric distributions of di-hadrons, coupling transversity to the so-called "interference fragmentation function" [69] in the framework of collinear factorization. Taking advantage of universality and robust proofs of TMD factorization for SIDIS, recent results [70,71,72,73] have been combined with $e^+e^-$ measurements [74,75] isolating the Collins and IFFs for the first global analyses to extract simultaneously the transversity distribution and polarized FF [76, 77]. In spite of this wealth of data, the kinematic reach of existing SIDIS experiments, where the range of Bjorken-$x$ values does not reach above $x \sim 0.3$, limits the current extractions of transversity.

Following the decomposition as described in [78,79,80] the Collins effect times the quark transversity distribution and the IFF times the quark transversity distribution may be accessed through single spin asymmetries of the azimuthal distributions of hadrons inside a high energy jet and the azimuthal asymmetries of pion pairs with different charges, respectively (for the current status see Section 1 and [5]). A comparison of the transversity signals extracted from the Collins effect and IFF measurements will explore questions about universality and factorization breaking, while comparisons of measurements at 200 and 500 GeV will provide experimental constraints on evolution effects. The first extraction of transversity from the STAR IFF data [81] has started (for details see Figure 14 in [82]).

By accessing the Collins asymmetry through the distribution of pions within a jet, one may also extract the $k_T$ dependence of transversity, giving insight into the multidimensional dependence of the distribution. Following the decomposition described in Ref. [79], that shows how to correlate different angular modulations to different TMDs, STAR has extracted several other angular modulations [83]. One example is the Collins-like asymmetry $A_{UT}^{\sin(\phi_s - 2\phi_h)}$. Currently all existing model predictions are unconstrained by measurements and suggest a maximum possible upper limit of ~2%. The present data fall well below this maximum with the best precision at lower values of $z$, where models suggest the largest effects may occur. Thus, these data should allow for the first phenomenological constraint on model predictions utilizing linearly polarized gluons beyond the positivity bounds.

While the measurements of transversity through the Collins FF need TMD factorization to hold in p+p scattering, di-hadron asymmetries utilize collinear factorization. Thus, not only can more precise measurements of these effects in p+p improve our knowledge of transversity, such measurements are invaluable to test the longstanding theoretical questions, such as the magnitude of any existing TMD factorization breaking. Extractions at RHIC kinematics also allow the possibility for understanding the TMD evolution of the Collins FF (e.g. Ref. [84]) by comparing to those extractions from SIDIS and $e^+e^-$ data. As noted earlier, extending measurements of di-hadron and Collins asymmetries in the forward direction will allow access to transversity in the region x >0.3, which is not probed by current experiments. This valence quark region is essential for the determination of the tensor charge, which receives 70% of its contributions from 0.1 < $x$ <1.0. In addition probing transversity in p+p collisions also provides better access to the d-quark transversity than is available in SIDIS, due to the fact that there is no charge weighting in the hard scattering QCD 2→2 process in p+p collisions. We want to note that this is a fundamental advantage of p+p collisions, as any SIDIS measurement of the d-quark transversity has to be on a bound system, i.e. He-3, which leads to nuclear corrections. The high scale we can reach in 500 GeV collisions at RHIC will also allow for the verification that previous SIDIS measurements at low scales in fact accessing the nucleon at leading twist. Figure 2-5 shows the $x$-$Q^2$ coverage spanned by the RHIC measurements compared to a future EIC, JLab-12, and the current SIDIS world data.

Another fundamental advantage of p+p collisions is the ability to access gluons directly. While gluons cannot carry any transverse spin, there is a strong analogy between quark transversity and the linear polarization of gluons. Similarly, there exists an equivalent of the Collins fragmentation function for the fragmentation of linearly polarized gluons into unpolarized hadrons [85]. The linear polarization of gluons is a largely unexplored phenomenon, but it has been a focus of recent theoretical work, in particular due to the relevance of linearly polarized gluons in unpolarized hadrons for the $p_T$ spectrum of the Higgs boson measured at the LHC. Polarized proton collisions with √s = 500 GeV at RHIC, in particular for asymmetric parton scattering if jets are detected in the backward direction are an ideal place to study the linearly polarized gluon distribution in **polarized** protons (Note: that the distributions of linearly polarized gluons inside an unpolarized and a polarized proton provide independent information). A first measurement of the "Collins-like" effect for linearly polarized gluons has been done by STAR with data from Run-2011, providing constraints on this function for the first time.

### 2.2.1 Run-2017

STAR has three times as much data at 200 GeV than shown in Figure 1-8 after the 2015 RHIC run, and has proposed to record over an order of magnitude more data at 500 GeV in 2017. This will enable far more detailed, multidimensional examination of the different asymmetries probing different combinations of several transverse momentum dependent PDFs and FFs, i.e., Transversity x Collins and linearly polarized gluons.

As discussed in Section 1, significant asymmetries have been measured in the Interference Fragmentation and Collins Function channel in the Run-2011 √s = 500 GeV data. Asymmetries sensitive to the gluon Sivers function and gluon linear polarization (Collins-like) have also been measured for the first time in hadronic collisions. The transverse data in Run-2011 (25 pb$^{-1}$) was initially recorded to estimate the systematic uncertainties in the helicity asymmetries from residual transverse components of the polarization and $A_{TT}$. A high luminosity run at √s = 500 GeV will provide the opportunity to increase the precision of the measurements in all of these channels and to make a high precision measurement of the Twist-3 ETQS function for gluons through $A_N$ for inclusive jets. Given that the same relation between the Twist-3 ETQS function and the Sivers function for quarks (see Eq. 2-1) exists also for gluons, a measurement constraining the ETQS function for gluons also constrains the gluon Sivers function (for more details see [86]). The uncertainties for all these measurements will shrink by a factor 4 with the proposed Run-2017 (see Figure 2-12 and Figure 2-13). It is noted that for some of these observables Run-2015 will provide the first statistical significant enough data set at √s = 200 GeV.

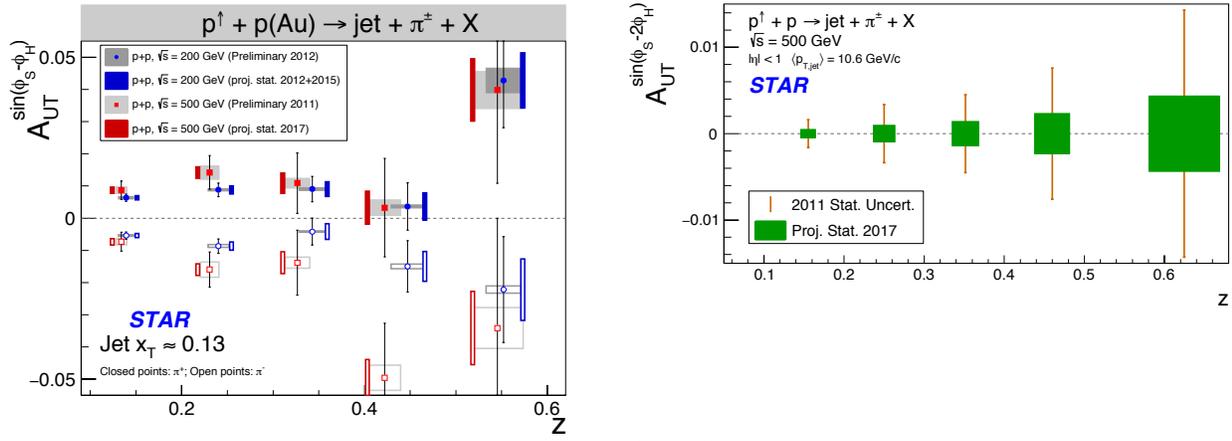

Figure 2-12: The improved statistical uncertainties for $A_{UT}^{\sin(\phi_s-\phi_h)}$ sensitive to the Collins effect and quark transversity (left) and $A_{UT}^{\sin(\phi_s-2\phi_h)}$ sensitive to gluon linear polarization (right), as function of $z$ for charged pions in jets at $0 < \eta < 1$ measured in STAR for transversely polarized p+p collisions at $\sqrt{s}$ = 200 GeV (Run-2012 to Run-2015) and 500 GeV (Run-2011 to Run-2017), respectively.

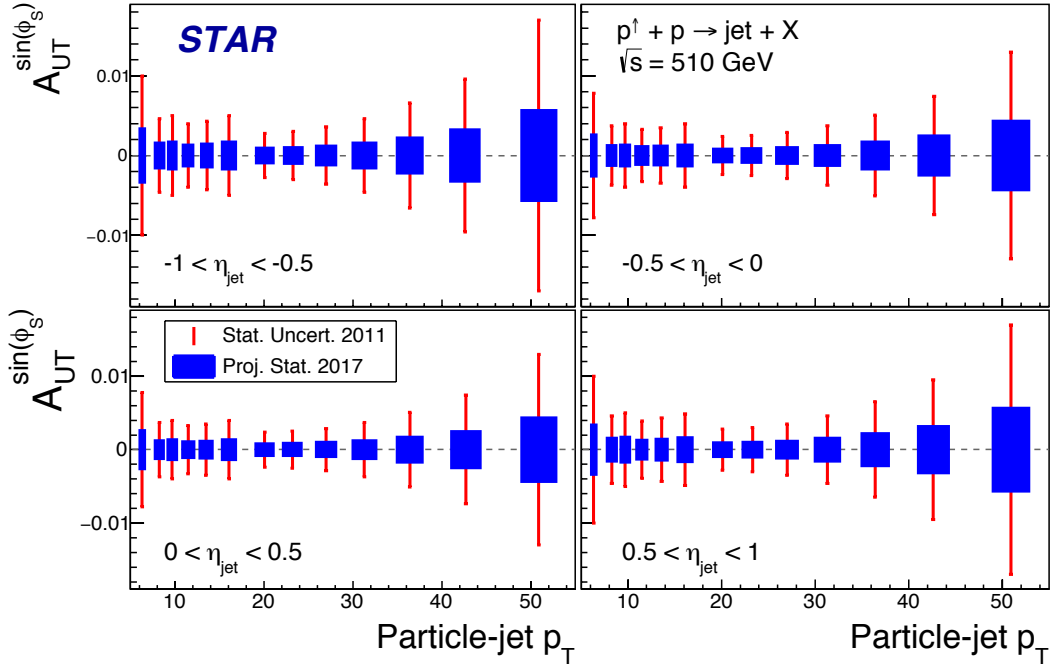

Figure 2-13: The improved statistical uncertainties for $A_{UT}^{\sin(\phi_s)}$ sensitive to the Twist-3 ETQS function for gluons, as function of particle-jet $p_T$ for 4 bins in rapidity measured in STAR for transversely polarized p+p collisions at 500 GeV (Run-2011 to Run-2017)

## 2.2.2 Opportunities with a Future Run at 500 GeV

First and foremost, a transversely polarized 500 GeV p+p run with anticipated delivered luminosity of 1 fb$^{-1}$ will reduce the statistical uncertainties of all observables discussed in Section 2.2.1 by a factor 2, This experimental accuracy will significantly enhance the quantitative reach of testing the limits of factorization and universality in lepton-proton and proton-proton collisions.

In order to further advance our understanding of transverse momentum dependent effects it is critical to enhance the current kinematical reach to lower or higher $x$. This can only be realized by either going to substantially higher jet transverse momenta or by measuring jets at forward rapidities where more asymmetric collisions allow larger $x$ and larger quark contributions in the hard process (see Figure 5-7) or to go to lower $x$ and tag on gluon contributions in the hard scattering. Assuming rapidity coverage between 1 and 4 it will be possible to extend the currently accessed coverage in $x$ substantially above 0.3 for reasonably high scales as well as quantitatively test universality in the $x$ range below which is overlapping the range accessed in SIDIS experiments. On the other end of the partonic momentum spectrum, which is important for the study of linearly polarized gluons, $x$ values below $2 \times 10^{-3}$ can be reached.

A realistic momentum smearing of final state hadrons as well as jets in this rapidity range was assumed and dilutions due to beam remnants (which become substantial at high rapidities) and underlying event contributions have been taken into account. As currently no dedicated particle identification at forward rapidities is feasible for these measurements only charged hadrons were taken into account that mostly reduces the expected asymmetries due to dilution by protons (10-14%) and a moderate amount of kaons (12-13%). As antiprotons are suppressed compared to protons in the beam remnants, especially the negative hadrons can be considered a good proxy for negative pions (~78% purity accd. to PYTHIA6). Given their sensitivity to the down quark transversity via favored fragmentation they are in particular important since SIDIS measurements due to their electromagnetic interaction, are naturally dominated by up-quarks.

We have estimated our statistical uncertainties based on an accumulated luminosity of 268 pb$^{-1}$, which leaves nearly invisible uncertainties after smearing. The uncertainties were evaluated in a very fine binning in jet transverse momentum, jet rapidity and the fractional energy z of the hadrons relative to the jet-$p_T$. These expected uncertainties are compared in Figure 2-14 to the asymmetries obtained from the transversity extractions based on SIDIS and Belle data [76] as well as from using the Soffer positivity bound for the transversity PDF [87]. More recent global fits [88] have slightly different central up and down quark transversity distributions, but due to the lack of any data for $x>0.3$ the upper uncertainties are compatible with the Soffer bounds. As can be seen from the average partonic $x$ probed in the hard two-to-two process, $x$ is increasing with increasing jet transverse momentum as well as rapidity. As discussed earlier (see Section 2.2) it is this high-$x$ coverage that allows giving important insights into the tensor charge essential to understand the nucleon structure at leading twist. It is important to emphasize, that even though the studies presented here are for the Collins asymmetries, the resulting statistical uncertainties will be similar for other measurements using azimuthal correlations of hadrons in jets. One important example is the measurement of "Collins-like" asymmetries to access the distribution of linearly polarized gluons. As described earlier, the best kinematic region to access this distribution is at backward angles with respect to the polarized proton and at small jet $p_T$. With the instrumentation assumed for the forward Collins asymmetry studies, therefore a high precision measurement of the distribution of linearly polarized gluons can be performed as well.

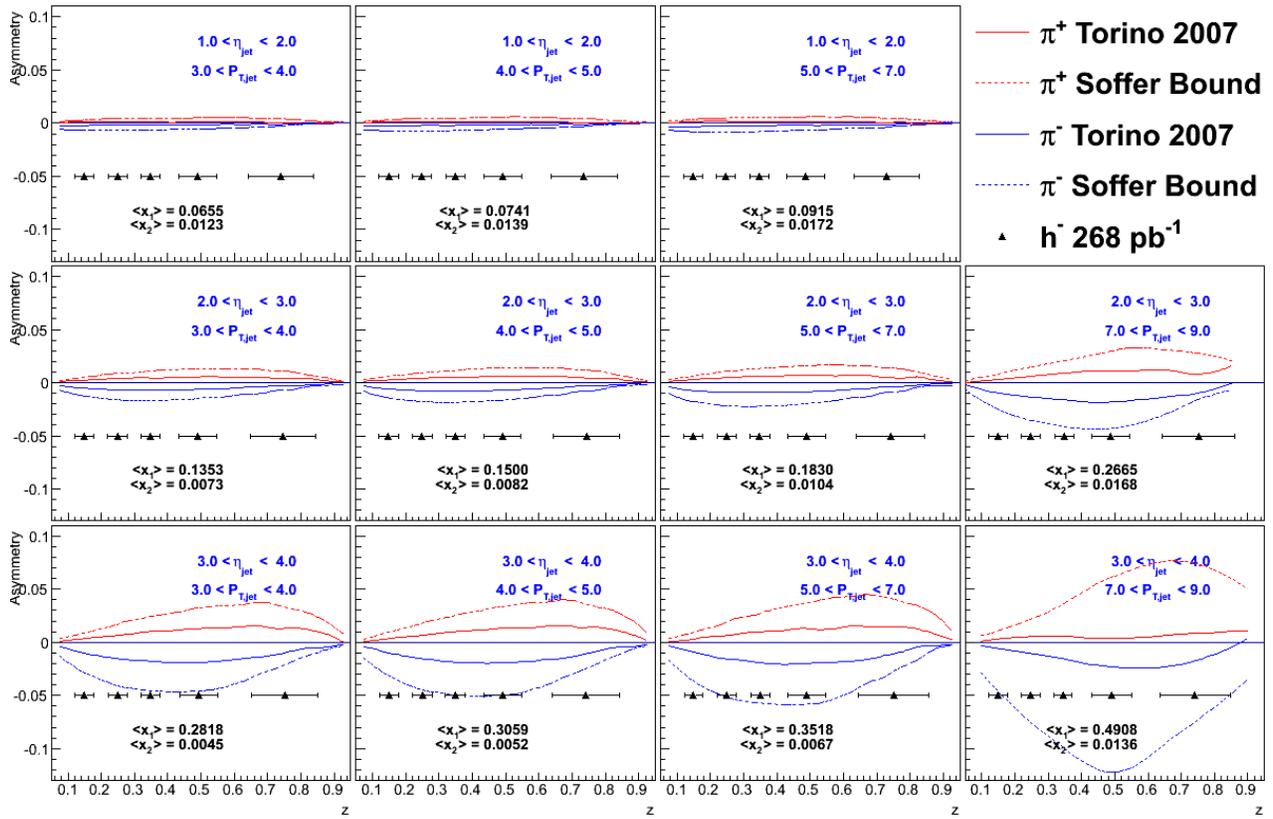

Figure 2-14: Expected h⁻ Collins asymmetry uncertainties (black points) compared to positive (red) and negative (blue) pion asymmetries based on the Torino extraction [45] (full lines) and the Soffer bound [83] (dashed lines) as a function of fractional energy *z* for various bins in jet rapidity and transverse momentum.

## 2.3 DIFFRACTION

Diffractive processes at RHIC are an essential tool that can elucidate the origin of single-spin asymmetries in polarized p+p collisions and access the orbital motion of partons inside the proton. Also at an EIC diffractive processes have been identified as the golden tool to study several key physics programs

- the spatial structure of nucleons and nuclei
- to access the orbital motion of small-$x$ partons inside the proton
- saturation in nuclei.

The essential characteristics of diffraction in QCD are summarized by two facts:

- The event is still called diffractive if there is a rapidity gap. Due to the presence of a rapidity gap, the diffractive cross-section can be thought of as arising from an exchange of several partons with zero net color between the target and the projectile. In high-energy scattering, which is dominated by gluons, this color neutral exchange (at the lowest order) consists of at least two exchanged gluons. This color singlet exchange has historically been called the pomeron, which had a specific interpretation in Regge theory. A crucial question in diffraction is the nature of the color neutral exchange between the protons. This interaction probes, in a novel fashion, the nature of confining interactions within hadrons.
- The proton/nuclear target is not always an opaque "black disk" obstacle of geometric optics. A projectile, which interacts more weakly due to color-screening and asymptotic freedom, is likely to produce a different diffractive pattern from a larger, more strongly interacting, projectile.

HERA discovered that 15% of the total ep cross-section is given by diffractive events (for details see [89] and references therein), basically independent of kinematics. At RHIC center-of-mass energies diffractive scattering events constitute ~25% of the total inelastic p+p cross-section [90]. As described above diffraction is defined as an interaction that is mediated by the exchange of the quantum numbers of the vacuum, as shown in Figure 2-15. Experimentally these events can be characterized by the detection of a very forward scattered proton and jet (singly diffractive) or two jets (doubly diffractive) separated by a large rapidity gap. Central diffraction, where two protons, separated by a rapidity gap, are reconstructed along with a jet at mid-rapidity, are also present, but suppressed compared to singly and doubly diffractive events. To date, there have been no data in p+p collisions studying spin effects in diffractive events at high √s apart from measuring single spin asymmetries in elastic p+p scattering [91].

The discovery of large transverse single spin asymmetries in diffractive processes would open a new avenue to study the properties and understand the nature of the diffractive exchange in p+p collisions.

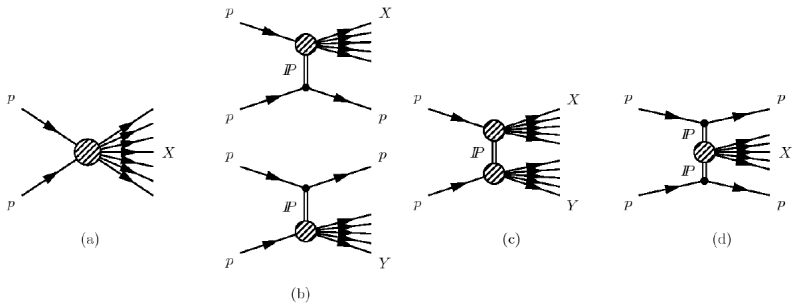

Figure 2-15: Schematic diagrams of (a) nondiffractive, pp→X, (b) singly diffractive, pp→Xp or pp→pY, (c) doubly diffractive, pp→XY, and (d) centrally diffracted, pp→pXp, events.

## 2.3.1 Run-2017, Run-2023 and Opportunities with a Future Run at 500 GeV

The primary observable of PHENIX and STAR to access transverse spin phenomena has been forward neutral pion production in transversely polarized p+p collisions. This effort has been extended to include the first measurements at √s = 500 GeV. The STAR Run-2011 data taken with transverse polarization at √s = 500 GeV have revealed several surprising results.

Figure 2-16 shows the transverse single spin asymmetry $A_N$ for "electromagnetic jets" (i.e. jets with their energy only measured in an electromagnetic calorimeter) detected in the STAR FMS at 2.5 < $\eta$ < 4.0 as a function of the jet $p_T$ for different photon multiplicities and ranges in jet energy [92]. It can be clearly seen that with increasing number of photons in the "electromagnetic jet" the asymmetry decreases. Jets with an isolated $\pi^0$ exhibit the largest asymmetry, consistent with the asymmetry in inclusive $\pi^0$-events. For all jet energies and photon multiplicities in the jet, the asymmetries are basically flat as a function of jet $p_T$, a feature also already seen for the inclusive $\pi^0$ asymmetries. This behavior is very different from what would be naively expected for an asymmetry driven by QCD subprocesses, which would follow a dependence of $1/p_T$. This STAR result is in agreement with preliminary observations from the $A_N$DY collaboration at RHIC, which measured $A_N$ for inclusive jets at √s = 500 GeV in the order of ~5x10$^{-3}$ [93].

In Ref. [60] it is argued that the behavior of $A_N$ for inclusive jets is consistent with the fact that the parton asymmetries for $u$ and $d$ valence quarks cancel, because the $u$ and $d$ quark asymmetries have opposite sign but equal magnitude. **If 2→2 parton-level subprocesses drive the inclusive $\pi^0$ production at forward rapidities then the same cancelation should occur!**

In addition the transverse single spin asymmetry $A_N$ of these electromagnetic jets in correlation with an away side jet in the rapidity range −1 < $\eta$ < 2 is reduced, the same behavior is seen correlating an away-side jet with the isolated forward $\pi^0$ mesons.

All these observations might indicate that the underlying subprocess causing a significant fraction of the large transverse single spin asymmetries in the forward direction are not of 2→2 parton scattering processes but of diffractive nature.

Figure 2-17 shows the inclusive $\pi^0$ cross-section for p+p collisions at √s=200 GeV versus the leading $\pi^0$ energy for three fixed forward pseudorapidities [97]. PYTHIA-8 [94] was used to evaluate the fraction of hard diffractive events [95] contributing to the inclusive $\pi^0$ cross-section at forward rapidities. Figure 2-18 shows the hard diffractive cross-section for $\pi^0$ production at √s=200 GeV and 500 GeV for a rapidity range of 2.8 < $\eta$ < 3.8 with and without applying several experimental cuts, i.e. the proton in the STAR Roman Pot acceptance. The prediction from this PYTHIA-8 simulation is that ~20% of the total inclusive cross-section at forward rapidities is of diffractive nature. This result is in agreement with measurements done over a wide range of √s (see Figure 12 in Ref. [90]).

In 2015 STAR collected data that will permit the measurement of correlations between forward scattered $\pi^0$ and tagged protons in its new forward Roman Pots [59, 96], this will constitute the first exploratory measurement of $A_N$ $\pi^0$ for single and double diffractive events by tagging one or both protons in the Roman Pots and a measurement of the hard diffractive cross-section at √s=200 GeV. In the 2017 transversely polarized p+p run at √s = 500 GeV it will be crucial to establish the observations for $A_N$ $\pi^0$ for single and double diffractive events made at √s = 200 GeV in the 2015 survive at higher center-of-mass energy.

The proposed forward upgrades will be a game changer for diffractive measurements at RHIC. It will allow the reconstruction of full jets both at √s=200 GeV and 500 GeV (see Section 5.3). As at HERA we will be able to reconstruct jets produced with the scattered proton tagged in Roman Pots and/or requiring rapidity gaps. Measuring spin asymmetries for diffractive events as function of √s might reveal surprises, which will inspire new physics opportunities for EIC, i.e SSA in polarized eA collisions.

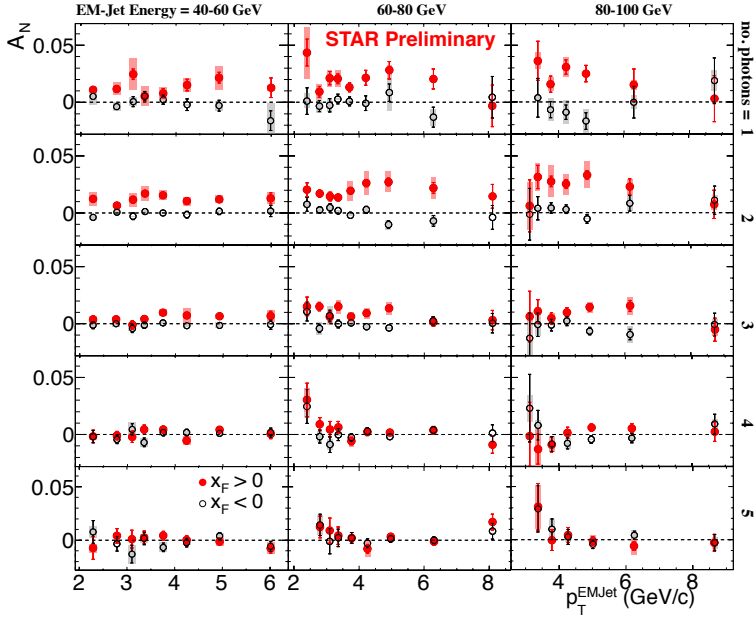

Figure 2-16: The transverse single spin asymmetry $A_N$ for "electromagnetic" jets detected in the FMS ($2.5 < \eta < 4.0$) as function of the jet $p_T$ and the photon multiplicity in the jet in bins of the jet energy. This behavior raises serious questions regarding how much of the large forward $\pi^0$ asymmetries are due to the same underlying dynamics as jet production namely 2→2 parton scattering processes.

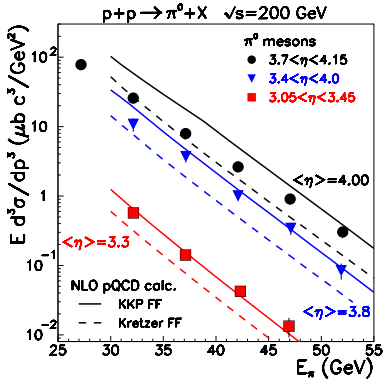

Figure 2-17: Inclusive $\pi^0$ cross-section for p+p collisions at $\sqrt{s}$=200 GeV versus the leading $\pi^0$ energy averaged over 5 GeV bins at fixed pseudorapidity [97]. The error bars combine statistical and point-to-point systematic errors. The curves are NLO pQCD calculations using two sets of fragmentation functions.

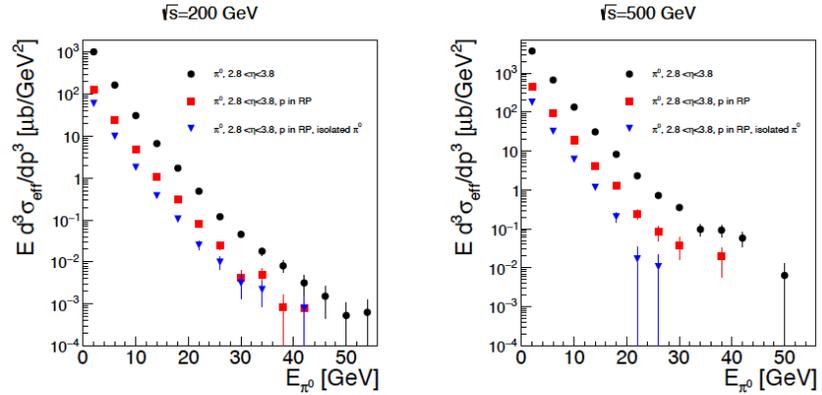

Figure 2-18: Estimate of the cross-section for hard diffractive processes at $\sqrt{s}$=200 GeV and 500 GeV using Pythia 8. The different points reflect different analysis cuts applied: $\pi^0$ in rapidity $2.8 < \eta < 3.8$ (black), one proton is required to be detected in the STAR Roman Pot acceptance (see Figure 2-19 (left)) (red) and an isolation cut of 35 mrad around the $\pi^0$ (blue).

## *Ultra Peripheral Collisions to access the Generalized Parton Distribution $E_g$:*

Two key questions, which need to be answered to understand overall nucleon properties like the spin structure of the proton, can be summarized as:

- How are the quarks and gluons, and their spins distributed in space and momentum inside the nucleon?
- What is the role of orbital motion of sea quarks and gluons in building the nucleon spin?

The formalism of generalized parton distributions provides a theoretical framework, which allows some answers to the above questions [98]. Exclusive reactions in DIS, i.e., deeply virtual Compton scattering, have been mainly used to constrain GPDs. RHIC, with its capability to collide transversely polarized protons at $\sqrt{s}$=500 GeV, has the unique opportunity to measure $A_N$ for exclusive $J/\psi$ in ultra-peripheral $p^\uparrow$+p collisions (UPC) [99]. The measurement is at a fixed $Q^2$ of 9 GeV$^2$ and $10^{-4} < x < 10^{-1}$. A nonzero asymmetry would be the first signature of a nonzero GPD *E* for gluons, which is sensitive to spin-orbit correlations and is intimately connected with the orbital angular momentum carried by partons in the nucleon and thus with the proton spin puzzle. Detecting one of the scattered polarized protons in "Roman Pots" (RP) ensures an elastic process. The event generator SARTRE [100], which also describes well the STAR results for $\rho^0$ production in UPC in Au+Au collisions, has been used to simulate exclusive $J/\psi$ -production in p$^\uparrow$+p UPC. The acceptance of the STAR RP PHASE-II* system in *t*, the momentum transfer between the incoming and outgoing proton, matches well the *t* spectrum in UPC collisions (see Figure 2-19). To select the $J/\psi$ in UPC, at least one of the two protons are required in the STAR RPs. The $J/\psi$ is reconstructed from its decay electrons, in the STAR EMCals between $-1 < \eta < 2$. Accounting for all trigger and reconstruction efficiencies the total number of $J/\psi$'s for a delivered luminosity of 400 pb$^{-1}$ is ~11k in Run-2017.

This measurement can be further improved with a high statistics transversely polarized p$^\uparrow$+Au run in 2023. For the process where the Au emits the virtual photon scattering off the p$^\uparrow$, this will provide an advantage in rate enhanced by $Z^2$ compared to ultra-peripheral p$^\uparrow$+p collisions at the same $\sqrt{s}$=200 GeV. The process where the p$^\uparrow$ emits the virtual photon can be suppressed by requiring a hit in the RP in the proton direction. The rapidity distribution of the $J/\psi$'s for these two processes is shown in Figure 2-20.

The total number of $J/\psi$'s for a delivered luminosity of 1.75 pb$^{-1}$ is ~13k for the Au as photon source with a background of ~5k for the p$^\uparrow$ as photon source. This measurement will provide important input for a future EIC, where the same process can be studied in photoproduction. Knowing the size of the asymmetry will help planning for the experimental needs, i.e. designing the detector to control systematics at an appropriate level, and planning for the luminosity needed to obtain data with adequate precision.

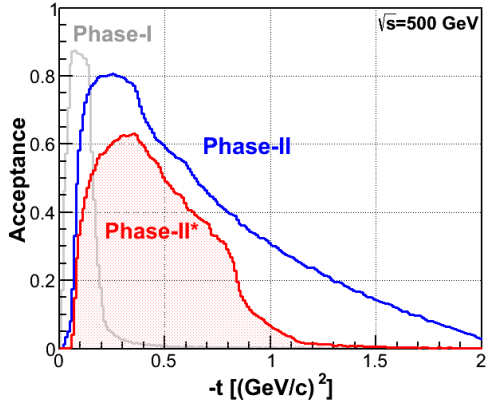 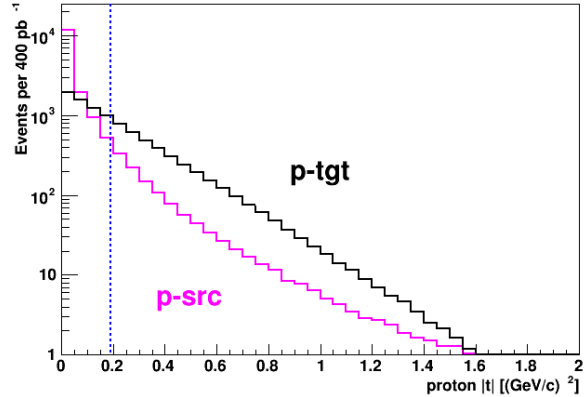

Figure 2-19: (left) Acceptance of protons in exclusive p+p scattering at $\sqrt{s}$ = 500 GeV as function of *t* for a possible future upgrade (blue) and the STAR set up since 2015 (PHASE-II*) (red) configuration. The acceptance for the original STAR Phase-I setup is also shown (grey). (right) The *t* spectrum of the proton emitting the photon (pink) as well as the one from the scattered proton (black).

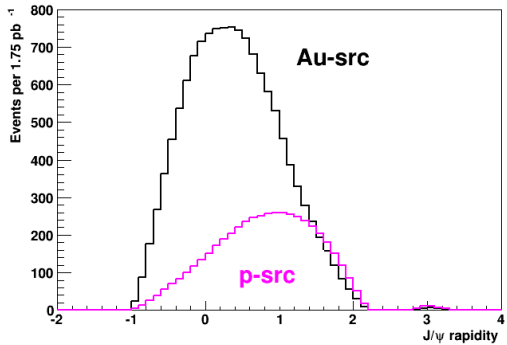

Figure 2-20: The rapidity distribution of the $J/\psi$'s for the process where the Au emits the virtual photon (black) and where the p$^\uparrow$ emits the virtual photon (pink).

# 3 PHYSICS OPPORTUNITIES WITH LONGITUDINALLY POLARIZED PROTON - PROTON COLLISIONS

## 3.1 RUN 2023 AND OPPORTUNITIES WITH A FUTURE RUN AT 500 GEV

If the beams are longitudinally polarized at either STAR or sPHENIX during the proposed $\sqrt{s}$ = 200 GeV p+p running in 2023, it would be possible to increase the data sample for the two main channels of the RHIC $\Delta G$ program, inclusive mid-rapidity jets and neutral pions, by a factor of 3. With the projected integrated luminosity of 300 pb$^{-1}$ (see Table 1-2) the other channels such as direct photons and charged pions are expected to show sensitivity to a non-zero $\Delta G$ for moderate $x$ ($x>0.05$), though with significantly smaller statistical power compared to jets and neutral pions.

The current RHIC plan does not include collisions above $\sqrt{s}$ = 200 GeV in the years after 2020. **If the timeline should change, making additional running feasible**, proton-proton collisions at $\sqrt{s}$= 500 GeV would allow RHIC to explore the low $x$ region of the gluon helicity distribution $\Delta g(x)$. A future 500 GeV longitudinal polarized p+p run (8 weeks with a delivered integrated luminosity of 1.1 fb$^{-1}$) would further reduce the statistical uncertainties of the two workhorses of the RHIC $\Delta g$ program, inclusive mid-rapidity jets and neutral pions, by a factor of 1.25 compared to the existing data sets (Figure 1-3 and Figure 1-4).

The existing mid-rapidity analyses are sensitive to gluons in the range of 0.01 < $x$ < 1. While these measurements clearly point to a positive $\Delta g(x)$ for moderate $x$ values, they do little to constrain the functional form of the distribution at lower $x$. This lack of data translates directly into a large uncertainty on the total gluon contribution to the spin of the proton $\Delta G = \int_0^1 \Delta g(x, Q^2) dx$, as shown in Figure 1-5.

Di-jet measurements provide a more direct connection to the probed values of momentum fractions $x$, and if extended to forward region, allow us to access $x$ down to $10^{-3}$. Figure 3-2 shows the projected precision for the asymmetries $A_{LL}$ as a function of the scaled invariant di-jet mass $M_{inv}/\sqrt{s}$ for four topological di-jet configurations involving a generic forward calorimeter system (FCS) in combination with either -1.0 < $\eta$ < 0.0, 0.0 < $\eta$ < 1.0, 1.0 < $\eta$ < 2.0, and the FCS (2.5 < $\eta$ < 4.0). In particular the 1.0 < $\eta$ < 2.0 / FCS and FCS / FCS configurations would allow one to probe $x$ values as low as a few times $10^{-3}$, as shown in Figure 3-3. The systematic uncertainty, which is assumed to be driven by the relative luminosity uncertainty of $\delta R = 5 \cdot 10^{-4}$, is clearly dominating over the statistical uncertainties. Any future measurements in these topological configurations including very forward measurements would clearly benefit from an improved relative luminosity measurement.

Di-jet measurements would provide theoretically well-controlled insight into the nature of the proton spin compared to the current forward rapidity (2.8 < $\eta$ < 4.0) inclusive $\pi^0$ $A_{LL}$. Jet reconstruction in the region will require electromagnetic and hadronic calorimetry, as well as some nominal tracking to associate charged particles with a single vertex (see Section 5).

The STAR collaboration has already established di-jet double spin asymmetry $A_{LL}$ measurements in the pseudorapidity range -1 < $\eta$ < 2. Figure 3-1 shows preliminary results for p+p collisions at $\sqrt{s}$ = 200 GeV (blue) and $\sqrt{s}$ = 510 GeV (red), based on data that were recorded in 2009 and 2012, respectively. The impact of the measurements from 2009+2015 ($\sqrt{s}$ = 200 GeV) and 2012 + 2013 ($\sqrt{s}$ = 500 GeV) on the helicity gluon distribution is currently being assessed by the DSSV collaboration in the context of a global QCD analysis at next-to-leading order accuracy, which matches the experimental cuts and jet parameters.

An EIC is anticipated to resolve the individual contributions to the spin of the nucleon with unprecedented precision in the x range down to a few times $10^{-5}$ [3,5]. Hence, RHIC mid- and forward-rapidity $A_{LL}$ measurements would continue providing unique and compelling sensitivity to the gluon helicity distribution of the proton **only** if an EIC were not realized or be significantly delayed due to external constraints. By themselves, these measurements do not provide a sufficiently compelling reason for another longitudinally polarized p+p run at $\sqrt{s}$ = 500 GeV past 2020 and certainly do not justify delay in the planning or construction of an EIC in the 2025 timeframe.

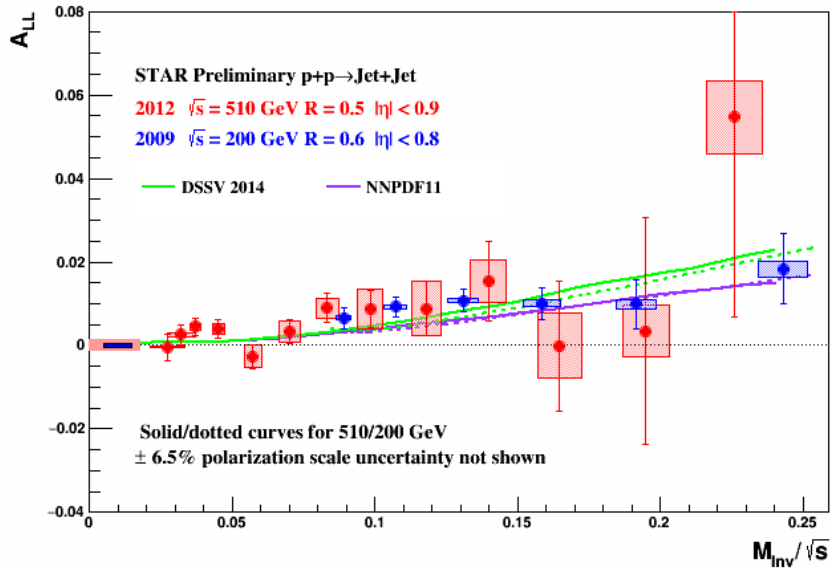

Figure 3-1: STAR preliminary measurements of the di-jet double spin asymmetry $A_{LL}$ versus $M_{inv}/\sqrt{s}$ of the pair for mid-rapidity p+p collisions at $\sqrt{s}$ = 200 (blue) GeV and $\sqrt{s}$ = 510 (red) GeV, compared to model predictions based on DSSV14 and NNPDFpol1.1. The uncertainties will be reduced by a factor of approximately 1.7 after additional data recorded during 2013 (510 GeV) and 2015 (200 GeV) are included.

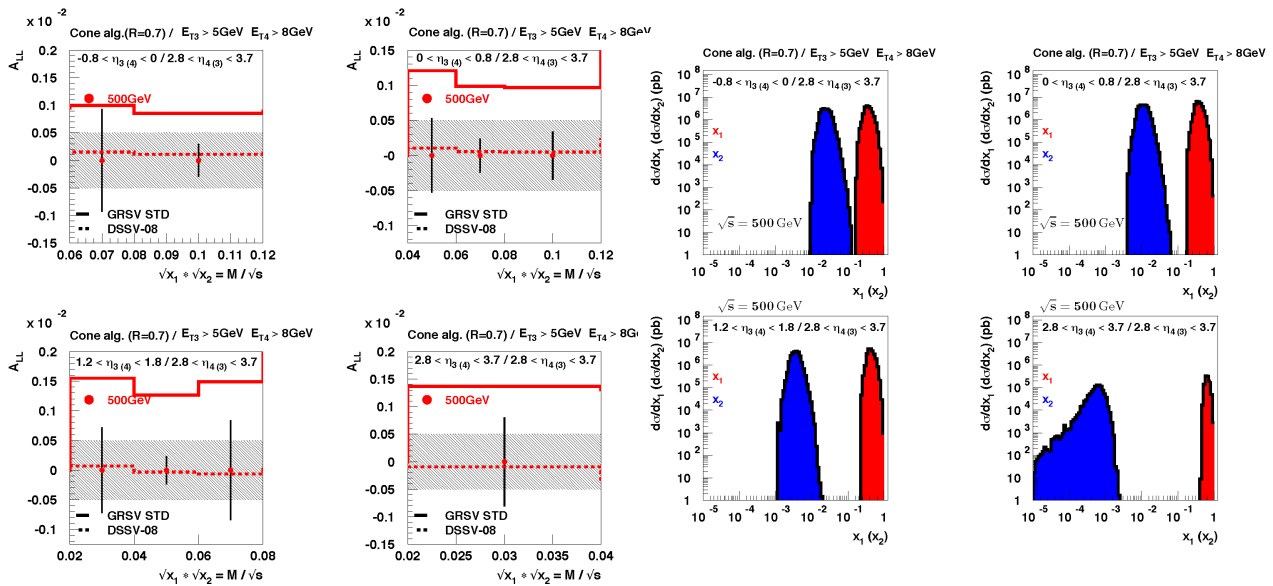

Figure 3-2: $A_{LL}$ NLO calculations as a function of $M_{inv}/\sqrt{s}$ for 2.8 < $\eta$ < 3.7 together with projected statistical and systematic uncertainties. An uncertainty $5 \cdot 10^{-4}$ has been assumed for the systematic uncertainty due to relative luminosity. A beam polarization of 60% and a total delivered luminosity of 1 fb$^{-1}$ have been assumed with a ratio of 2/3 for the ratio of recorded to delivered luminosity.

Figure 3-3: $x_1 / x_2$ range for the forward acceptance region of 2.8 < $\eta$ < 3.7.

# 4 PHYSICS OPPORTUNITIES WITH (UN)POLARIZED PROTON-NUCLEUS COLLISIONS

Our quest to understand QCD processes in Cold Nuclear Matter (CNM) centers on the following fundamental questions:

- Can we experimentally find evidence of a novel universal regime of non-linear QCD dynamics in nuclei?
- What is the role of saturated strong gluon fields, and what are the degrees of freedom in this high gluon density regime?
- What is the fundamental quark-gluon structure of light and heavy nuclei?
- Can a nucleus, serving as a color filter, provide novel insight into the propagation, attenuation and hadronization of colored quarks and gluons?

Various aspects of these questions have been addressed by numerous experiments and facilities around the world, most of them at significantly lower center-of-mass energies and kinematic reach then RHIC. Deep inelastic scattering on nuclei addresses some of these questions with results from, for instance, HERMES at DESY [101], CLAS at JLab [102], and in the future at the JLab 12 GeV. This program is complemented by hadron-nucleus reactions in fixed target p+A at Fermilab (E772, E886, and E906) [*103*] and at the CERN-SPS.

In the following we propose a measurement program unique to RHIC to separate initial and final state effects in strong interactions in the nuclear environment. We also highlight the complementarity to the LHC p+Pb program and stress why RHIC data are essential and unique in the quest to further our understanding of nuclei.

## 4.1 THE INITIAL STATE OF NUCLEAR COLLISIONS

### 4.1.1 Run-2023

*Nuclear Parton Distribution Functions*

A main emphasis of the 2015 and later p+A runs is to determine the initial conditions of the heavy ion nucleus before the collision to support the theoretical understanding of the A+A program both at RHIC and the LHC. In the following, the current status of nPDFs will be discussed, including where the unique contribution of RHIC in comparison to the LHC and the future EIC lies

Our current understanding of nuclear parton distribution functions (nPDFs) is still very limited, in particular, when compared with the rather precise knowledge of PDFs for free protons collected over the past 30 years. Figure 4-1 shows a summary of the most recent extractions of nPDFs from available data, along with estimates of uncertainties. All results are shown in terms of the nuclear modification ratios, i.e., scaled by the respective PDF of the free proton. The yellow bands indicate regions in $x$ where the fits are not constrained by data [104] and merely reflect the freedom in the functional form *assumed* in the different fits. Clearly, high precision data at small $x$ and for various different values of $Q^2$ are urgently needed to better constraint the magnitude of suppression in the $x$ region where non-linear effects in the scale evolution are expected. In addition, such data are needed for several different nuclei, as the A-dependence of nPDFs cannot be predicted from first principles in pQCD and, again, currently relies on assumptions. Note that the difference between DSSZ [105] and EPS09 for the gluon modification arise from the different treatment of the PHENIX midrapidity $\pi^0$ $R_{dAu}$ data [106], which in the EPS09 [107] fit are included with an extra weight of 20. The $\pi^0$ $R_{dAu}$ data are the only data, which can probe the gluon in the nucleus directly, but these data also suffer from unknown nuclear effects in the final state (see Section 4.2). Therefore, it is absolutely critical to have high precision data only sensitive to nuclear modification in the initial state over a wide range in $x$ and intermediate values of $Q^2$ (away from the saturation regime) to establish the nuclear modification of gluons in this kinematic range.

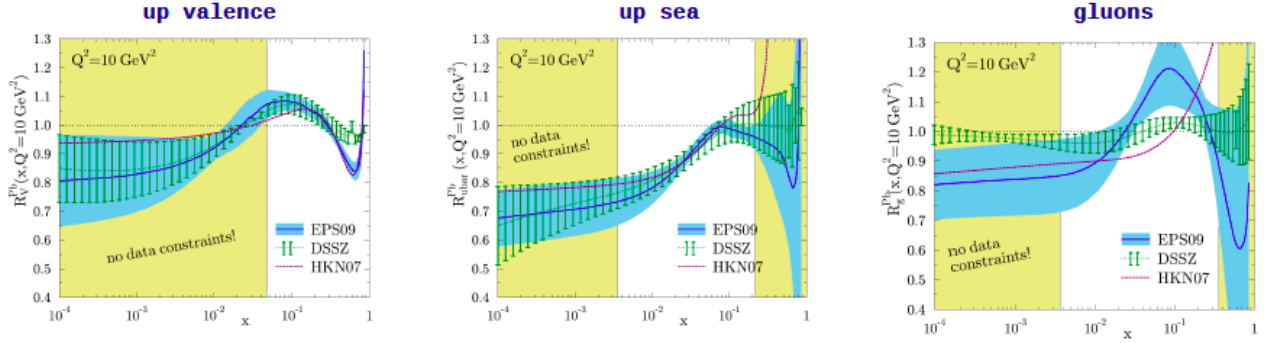

Figure 4-1: Summary of the most recent sets of nPDFs. The central values and their uncertainty estimates are given for the up valence quark, up sea quark, and the gluon. The yellow bands indicate regions in $x$ where the fits are not constrained by any data (taken from Ref. [104]).

Of course, it is important to realize that the measurements from RHIC are compelling and essential even when compared to what can be achieved in p+Pb collisions at the LHC. Due to the higher center-of-mass system energy most of the LHC data have very high $Q^2$, where the nuclear effects are already reduced significantly by evolution and are therefore very difficult to constrain. A recent article [108] assessed the impact of the available LHC Run-I p+Pb data on determinations of nPDFs. The rather moderate impact of these data is illustrated in Figure 4-2.

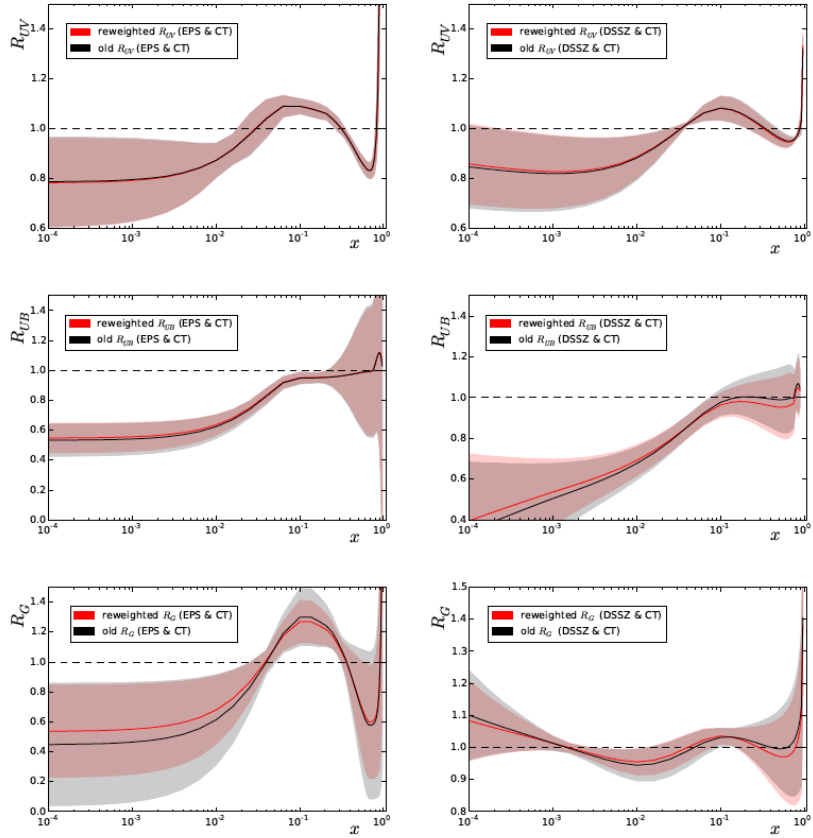

Figure 4-2: Impact of the LHC Run-I data on the nPDFs of EPS09 (left panels) and DSSZ (right panels) before (black curves/ gray bands) and after the reweighting (red/light red), for the valence up quark (upper row), up sea quark (middle row), and the gluon (lower row) distributions at $Q^2$=1.69 GeV$^2$, except for the DSSZ gluons which are plotted at $Q^2$=2 GeV$^2$ (taken from [108])

RHIC has the *unique* opportunity to provide data in a kinematic regime (moderate $Q^2$ and medium-to-low $x$) where the nuclear modification of the sea quark and the gluon is expected to be sizable and currently completely unconstrained. In addition, and unlike the LHC, RHIC can vary the nucleus in p+A collisions and as such also constrain the $A$-dependence of nPDFs.

The two golden channels to achieve these goals at RHIC are a measurement of $R_{pA}$ for Drell-Yan production at forward pseudo-rapidities with respect to the proton direction ($2.5 < \eta_p < 4.5$) to constrain the nuclear modifications of sea-quarks and of $R_{pA}$ for direct photon production in the same kinematic regime to constrain the nuclear gluon distribution. The first measurement of $R_{pA}$ for direct photon production has been done already during the p+Au and p+Al runs in 2015, with a recorded luminosity of $L_{pAu}$ = 0.45 pb$^{-1}$ (STAR and PHENIX) and $L_{pAl}$ = 1 pb$^{-1}$ (STAR), respectively. The anticipated statistical precision for pA runs in 2015 and projections for a run in 2023 are shown in Figure 4-3.

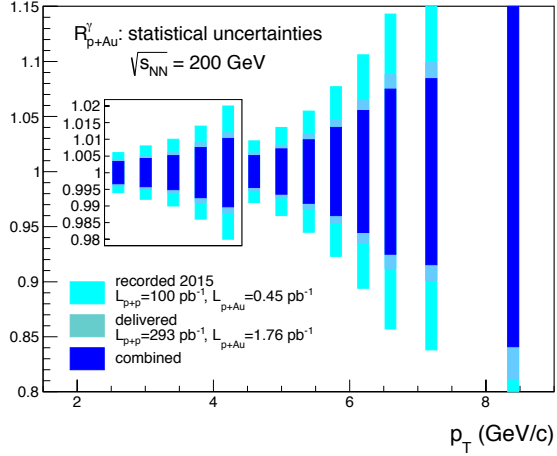

Figure 4-3: Projected statistical uncertainties for $R_{pAu}$ for direct photons in Run-2015 (light blue) and a run in 2023 (blue) and the sum of both (dark blue). The recorded luminosity for Run-2015 was $L_{pAu}$ = 450 nb$^{-1}$ and $L_{pp}$ = 100 pb$^{-1}$. The delivered luminosity for Run-2023 is assumed to be $L_{pAu}$ = 1.8 pb$^{-1}$ and $L_{pp}$ = 300 pb$^{-1}$. A p+Al run of 8 weeks in 2023 would have matched parton luminosity resulting in an equal statistical precision.

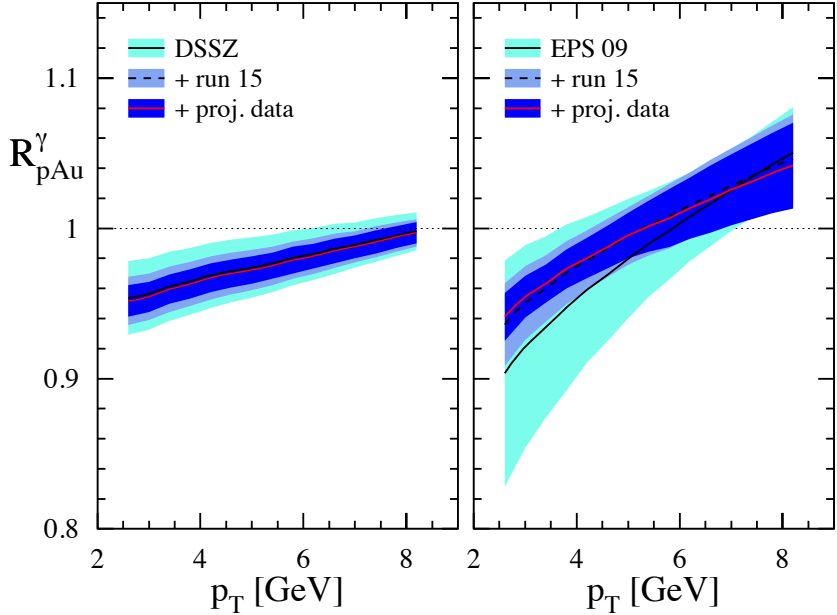

Figure 4-4: The impact of the direct photon $R_{pA}$ data measured in Run-2015 (blue band) and for the anticipated statistics for a future p+Au run in 2023 (dark blue band) compared with the current uncertainties (cyan band) from DSSZ (left) and EPS-09 (right).

Figure 4-4 shows the significant impact of the Run-2015 $R_{pA}$ for direct photon production and a future run in the 2023 on the corresponding theoretical expectations and their uncertainties obtained with both the EPS09 and DSSZ sets of nPDFs. The uncertainty bands are obtained through a reweighting procedure [109] by using the projected data shown in Figure 4-3 and randomizing them according to their expected statistical uncertainties around the central values obtained with the current set of DSSZ nPDFs. These measurements will help significantly in further constraining the nuclear gluon distribution in a broad range of $x$ that is roughly correlated with accessible transverse momenta of the photon, i.e., few times $10^{-3} < x <$ few times $10^{-2}$. The relevant scale $Q^2$ is set be $\sim p_T^2$ and ranges from 6 GeV$^2$ to about 40 GeV$^2$. Like all other inclusive probes in p+p and pA collisions, e.g., jets, no access to the exact parton kinematics can be provided event-by-event but global QCD analyses easily account for that. After the p+Au run in 2023, the statistical precision of the prompt photon data will be sufficient to contribute to a stringent test of the universality of nuclear PDFs when combined with the expected data from an EIC (see Figure 2.22 and 2.23 in Ref [110]).

Figure 4-5 shows the kinematic coverage in $x$–$Q^2$ of past, present, and future experiments capable of constraining nuclear parton distribution functions. The experiments shown provide measurements that access the initial state parton kinematics on an event-by event basis (in a leading order approximation) while remaining insensitive to any nuclear effects in final state. Some of the LHC experiments cover the same $x$-range as DY at forward pseudo-rapidities at RHIC but at a much higher scale $Q^2$, where nuclear modifications are already significantly reduced [108,111]. At intermediate $Q^2$, DY at RHIC will extend the low-$x$ reach by nearly one decade compared to EIC.

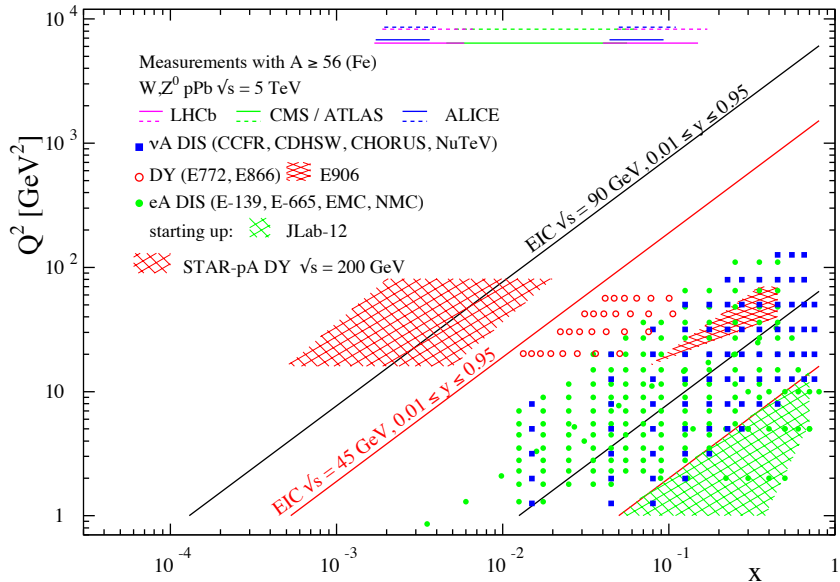

Figure 4-5: The kinematic coverage in $x$–$Q^2$ of past, present and future experiments constraining nPDFs with access to the exact parton kinematics event-by-event and no fragmentation in the final state.

The biggest challenge of a DY measurement is to suppress the overwhelming hadronic background: the total DY cross-section is about $10^{-5}$ to $10^{-6}$ smaller than the corresponding hadron production cross-sections. Therefore, the probability of misidentifying a hadron track as a lepton has to be suppressed to the order of 0.1% while maintaining reasonable electron detection efficiencies. To that end, we have studied the combined electron/hadron discriminating power of the proposed forward tracking and calorimeter systems. It was found that by applying multivariate analysis techniques to the features of EM/hadronic shower development and momentum measurements we can achieve hadron rejection powers of 200 to 2000 for hadrons of 15 GeV to 50 GeV with 80% electron detection efficiency.

The left panel in Figure 4-6 shows the normalized background yields along with the expected DY production and their uncertainties for a delivered luminosity of 2.3 pb$^{-1}$ and assuming the performance of the upgraded forward instrumentation as described in detail in Section 5. The green band represents the statistical uncertainties of the background yield and its shape. The right panel shows the DY signal to QCD background ratio as a function of the lepton pair mass.

The same procedure as for the direct photon $R_{pA}$ was used to study the potential impact of the DY $R_{pA}$ data. For the DSSZ and EPS-09 sets of nPDFs both the predicted nuclear modifications and the current uncertainties are very similar. This is because both groups use the same DIS and DY data without any special weight factors in constraining sea-quarks. As can be inferred from Figure 4-7 we expect again a significant impact on the uncertainties of $R_{pA}$ DY upon including the projected and properly randomized data. Clearly, the DY data from RHIC will be instrumental in reducing present uncertainties in nuclear modifications of sea quarks. Again, these data will prove to be essential in testing the fundamental universality proper-ty of nPDFs in the future when EIC data become available.

STAR's unique detector capabilities, i.e., the FMS+FPS and the Roman Pot detectors, provided the first data on J/ψ-production in ultra-peripheral collisions for the 2015 polarized p+Au run. Like direct photon measurements, the J/ψ is detected through its leptonic decay channel to study solely the effects of strong interactions in the initial state [112]. This measurement provides access to the spatial gluon distribution by measuring the *t*-dependence of $d\sigma/dt$. As follows from the optical analogy, the Fourier-transform of the square root of this distribution yields the source distribution of the object probed. To study the gluon distribution in the gold nucleus, events need to be tagged where the photon is emitted from the proton. For both observables a measurement with different nuclei is required to pin down the A-dependence of nPDFs. The J/ψ-production in ultra-peripheral collisions requires significantly more statistics than accumulated to date.

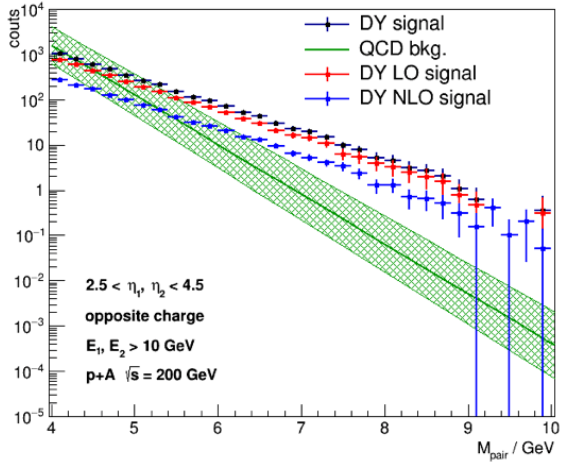
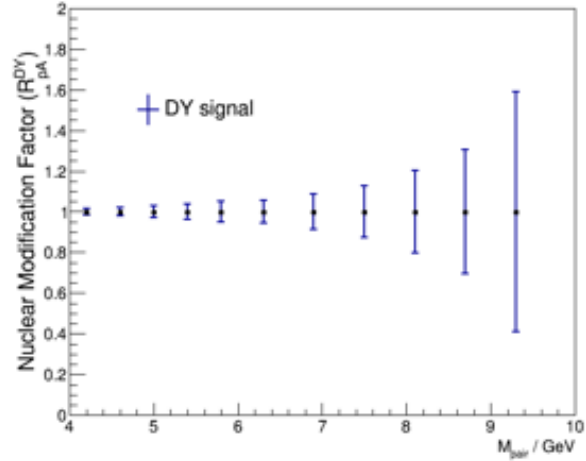

Figure 4-6: (left) DY signal and background yield from 2.3 pb$^{-1}$ p+Au 200 GeV collisions. (right) The expected $R_{pA}$ based on the 2.3 pb$^{-1}$ p+Au and 383 pb$^{-1}$ p+p reference data.

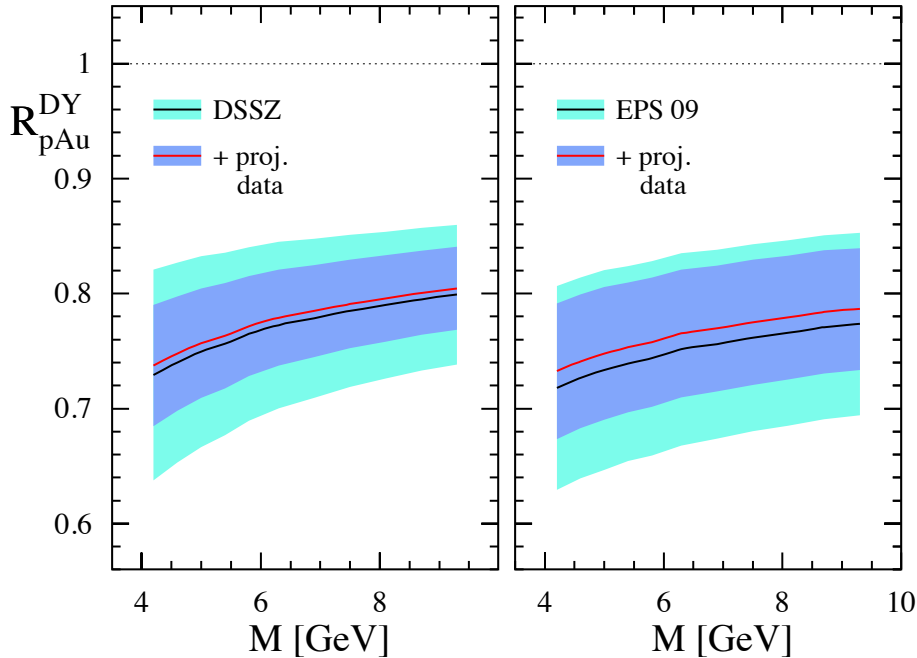

Figure 4-7: The impact of the DY $R_{pA}$ data for the anticipated statistics for a p+Au run in 2023 (dark blue band) compared to the current uncertainties (cyan band) from DSSZ and EPS-09.

## Gluon Saturation

Our understanding of the proton structure and of the nuclear interactions at high energy would be advanced significantly with the definitive discovery of the saturation regime [113]. Saturation physics would provide an infrared cutoff for perturbative calculations, the saturation scale $Q_s$, which grows with the atomic number of the nucleus $A$ and with decreasing value of $x$. If $Q_s$ is large it makes the strong coupling constant small, $\alpha_s(Q_s^2) \ll 1$ allowing for perturbative QCD calculations to be under theoretical control.

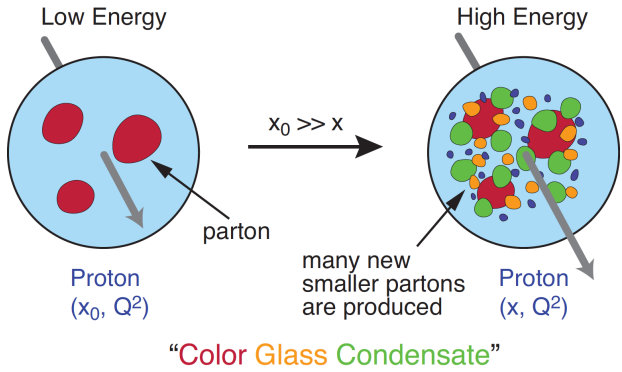

Figure 4-8: Proton wave function evolution towards small-x.

It is well known that PDFs grow at small-$x$ (see Figure 1-2). If one imagines how such a high number of small-$x$ partons would fit in the (almost) unchanged proton radius, one arrives at the picture presented in Figure 4-8: the gluons and quarks are packed very tightly in the transverse plane. The typical distance between the partons decreases as the number of partons increases, and can get small at low-$x$ (or for a large nucleus instead of the proton). One can define the saturation scale as the inverse of this typical transverse interparton distance. Hence $Q_s$ indeed grows with $A$ and decreasing $x$.

The actual calculations in saturation physics start with the classical gluon fields (as gluons dominate quarks at small-x) [114], which are then evolved using the nonlinear small-x BK/JIMWLK evolution equations [115]. The saturation region is depicted in Figure 4-9 in the $(x, Q^2)$ plane and can be well-approximated by the following formula: $Q_s^2 \sim (A/x)^{1/3}$. Note again that at small enough x the saturation scale provides an IR cutoff, justifying the use of perturbative calculations. This is important beyond saturation physics, and may help us better understand small-x evolution of the TMDs.

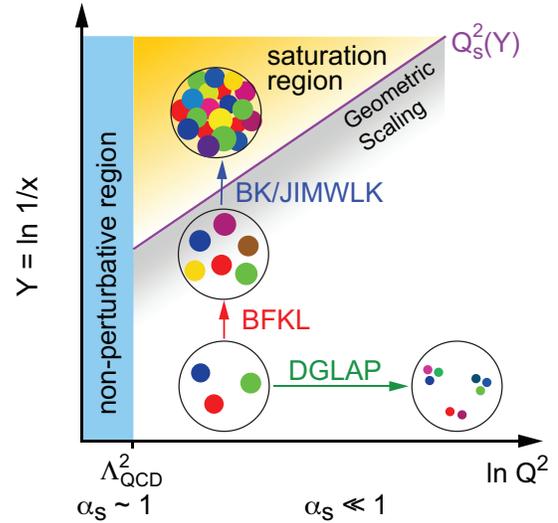

Figure 4-9: Saturation region in the (x,Q2) plane.

While the evidence in favor of saturation physics has been gleaned from the data collected at HERA, RHIC and the LHC, the case for saturation is not sealed and alternative explanations of these data exist. The EIC is slated to provide more definitive evidence for saturation physics [3]. To help the EIC complete the case for saturation, it is mandatory to generate higher-precision measurements in p+A collisions at RHIC. These higher-precision measurements would significantly enhance the discovery potential of the EIC as they would enable a stringent test of universality of the CGC. We stress again that a lot of theoretical predictions and results in the earlier Sections of this document would greatly benefit from saturation physics: the small-$x$ evolution of TMDs in a longitudinally or transversely polarized proton, or in an unpolarized proton, can all be derived in the saturation framework [116] in a theoretically better-controlled way due to the presence of $Q_s$. Hence saturation physics may help us understand both the quark and gluon helicity PDFs as well as the Sivers and Boer-Mulders functions.

The saturation momentum is predicted to grow approximately like a power of energy, $Q_s^2 \sim E^{\lambda/2}$ with $\lambda \sim 0.2$-$0.3$, as phase space for small-$x$ (quantum) evolution opens up. The saturation scale is also expected to grow in proportion to the valence charge density at the onset of small-$x$ quantum evolution. Hence, the saturation scale of a large nucleus should exceed that of a nucleon by a fac-tor of $A^{1/3}$ ~5 (on average over impact parameters). RHIC is capable of running p+A collisions for different nuclei to check this dependence on the mass number. This avoids potential issues with dividing say p+Pb collisions in $N_{part}$ classes [117]. Figure 4-10 shows the kinematic coverage in the $x$-$Q^2$ plane for p+A collisions at RHIC, along with previous e+A measurements and the kinematic reach of an EIC. The saturation scale for a Au nucleus and the proton is also shown. To access at RHIC a kinematic regime sensitive to saturation with $Q^2 > 1$ GeV$^2$ requires measurements at forward rapidities. For these kinematics the saturation scale is moderate, on the order of a few GeV$^2$, so measurements sensitive to the saturation scale are by necessity limited to semi-hard processes.

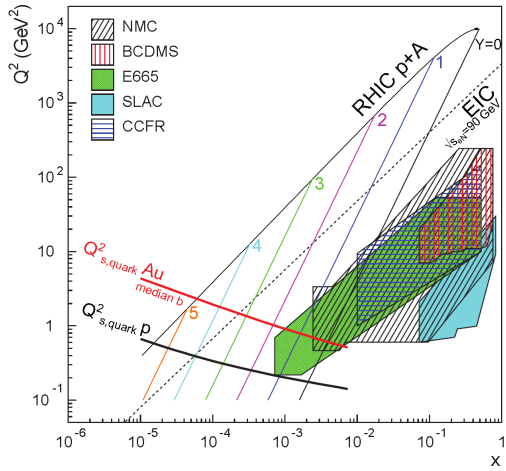

Figure 4-10: Kinematic coverage in the $x$-$Q^2$ plane for p+A collisions at RHIC, along with previous e+A measurements, the kinematic reach of an electron-ion collider, and estimates for the saturation scale $Q_s$ in Au nuclei and protons. Lines are illustrative of the range in $x$ and $Q^2$ covered with hadrons at various rapidities.

Until today the golden channel at RHIC to observe strong hints of saturation has been the angular dependence of two-particle correlations, because it is an essential tool for testing the underlying QCD dynamics [118]. In forward-forward correlations facing the p(d) beam direction one selects a large-$x$ parton in the p(d) interacting with a low-$x$ parton in the nucleus. For $x < 0.01$ the low-$x$ parton will be back-scattered in the direction of the large-$x$ parton. Due to the abundance of gluons at small $x$, the backwards-scattered partons are dominantly gluons, while the large-$x$ partons from the p(d) are dominantly quarks. The measurements of di-hadron correlations by STAR and PHENIX [119,120] have been compared with theoretical expectations using the CGC framework based on a fixed saturation scale $Q_s$ and con-sidering valence quarks in the deuteron scattering off low-$x$ gluons in the nucleus with impact parameter $b = 0$ [121,122]. Alternative calculations [123] based on both initial and final state multiple scattering, which determine the strength of this transverse momentum imbalance, in which the suppression of the cross-section in d+Au collisions arises from cold nuclear matter energy loss and coherent power corrections have also been very successful to describe the data.

The 2015 p+Au run at RHIC has provided unique opportunities to study this channel in more detail both at STAR and PHENIX. The high delivered integrated luminosities allow one to vary the trigger and associated particle $p_T$ from low to high values and thus crossing the saturation boundary as shown in Figure 4-10 and reinstate the correlations for central p+A collisions for forward-forward $\pi^0$'s.

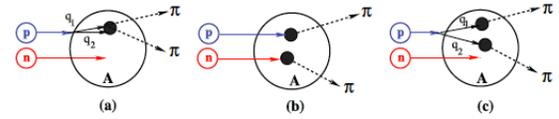

Figure 4-11: Contributions to two-pion production in d+A collisions through the double-interaction mechanism [124].

Studying di-hadron correlations in p+A collisions instead of d+A collisions has a further advantage. In reference [124], the authors point out that the contributions from double-parton interactions to the cross-sections for d+A → $\pi^0\pi^0$X are not negligible. This mechanism is illustrated in Figure 4-11. They find that such contributions become important at large forward rapidities, and especially in the case of d+A scattering. Whether or not this mechanism provides an alternative explanation of the suppression of the away-side peak in $\pi^0$-$\pi^0$ can be settled with the 2015 p+A data.

It is very important to note that for the measurements to date in p(d)+A collisions both initial and final states interact strongly, leading to severe complications in the theoretical treatment (see [125, 126] and references therein). As described in detail in the Section above in p+A collisions, these complications can be ameliorated by removing the strong interaction from the final state, by using photons and Drell-Yan electrons. The Run-2015 p+A run will for the first time (see Figure 4-3) provide data on $R_{pA}$ for direct photons and therefore allow one to test CGC based predictions on this observable as depicted in Figure 4-12 (taken from Ref. [127]). The higher delivered integrated luminosity for the upcoming p+Au and p+Al run in 2023 together with the proposed for-ward upgrade will enable one to study more luminosity hungry processes and/or complementary probes to the di-hadron correlations, i.e. photon-jet, photon-hadron and di-jet correlations.

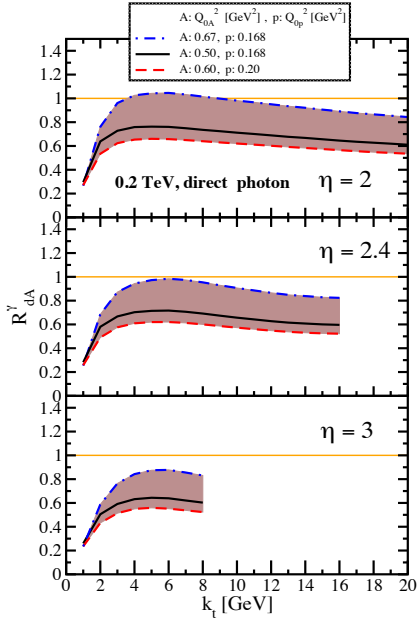

Figure 4-12: Nuclear modification factor for direct photon production in p(d)A collisions at various rapidities at RHIC √s = 0.2 TeV. The curves are the results obtained from Eq. (12) in Ref. [127] and the solution to rcBK equation using different initial saturation scales for a proton $Q_{op}$ and a nucleus $Q_{oA}$. The band shows our theoretical uncertainties arising from allowing a variation of the initial saturation scale of the nucleus in a range consistent with previous studies of DIS structure functions as well as particle production in minimum-bias p+p, p+A and A+A collisions in the CGC formalism, see Ref. [127] for details.

We use direct photon plus jet (direct γ+jet) events as an example channel to indicate what can be done in 2023. These events are dominantly produced through the gluon Compton scattering process, g+q→γ+q, and are sensitive to the gluon densities of the nucleon and nuclei in p+p and p+A collisions. Through measurements of the azimuthal correlations in p+A collisions for direct γ+jet production, one can study gluon saturation phenomena at small-x. Unlike di-jet production that is governed by both the Weizsäcker-Williams and dipole gluon densities, direct γ+jet production only accesses the dipole gluon density, which is better understood theoretically [127,128]. On the other hand, direct γ+jet production is experimentally more challenging due to its small cross-section and large background contribution from di-jet events in which photons from fragmentation or hadron decay could be misidentified as direct photons. The feasibility to perform direct γ+jet measurements with the proposed forward upgrade in unpolarized p+p and p+Au collisions at √$s_{NN}$=200 GeV has been studied. PYTHIA-8.189 [129] was used to produce direct γ+jet and di-jet events. In order to suppress the di-jet background, the leading photon and jet are required to be balanced in transverse momentum, $|\phi^\gamma - \phi^{jet}| > 2\pi/3$ and $0.5 < p_T^\gamma/p_T^{jet} < 2$. Both the photon and jet have to be in the forward acceptance $1.3 < \eta < 4.0$ with $p_T > 3.2$ GeV/c in 200 GeV p+p collisions. The photon needs to be isolated from other particle activities by requiring the fraction of electromagnetic energy deposition in the cone of ΔR=0.1 around the photon is more than 95% of that in the cone of ΔR=0.5. Jets are reconstructed by an anti-$k_T$ algorithm with ΔR=0.5. After applying these selection cuts, the signal-to-background ratio is around 3:1 [130]. The expected number of selected direct γ+jet events is around 1.0M/0.9M at √$s_{NN}$=200 GeV in p+Au/p+Al collisions for the proposed run in 2023. We conclude that a measurement of direct photon-hadron correlation from p+A collisions is feasible, which is sensitive to the gluon density in 0.001<x<0.005 in the Au nucleus (see Figure 4-13) where parton saturation is expected.

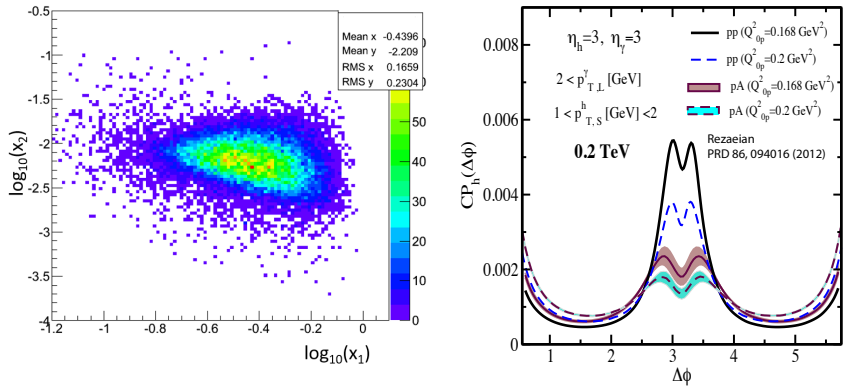

Figure 4-13: Left: Bjorken-x distributions of hard scattering partons in direct γ+jet production after event selections described in the text in p+p collisions at √s=200. Right: γ-hadron azimuthal correlation in minimum bias p+p and p+Au collisions at √$s_{NN}$=200 GeV. The curves are obtained with two different initial saturation scale of proton $Q^2_{0p}$=0.168 and 0.2 GeV$^2$ and the corresponding initial saturation scale in the nucleus within $Q^2_{0A}$~3-4$Q^2_{0p}$ (c.f. [127,128]).

# 4.2 THE FINAL STATE: NUCLEAR FRAGMENTATION FUNCTIONS

## 4.2.1 Run-2023

In spite of the remarkable phenomenological successes of QCD, a quantitative understanding of the hadronization process is still one of the great challenges for the theory. Hadronization describes the transition of a quark or gluon into a final state hadron. It is a poorly understood process even in elementary collisions. RHIC's unique versatility will make it possible to study hadronization in vacuum and in the nuclear medium, and additionally with polarized beams.

It has long been recognized that the hadron distributions within jets produced in p+p collisions are closely related to the fragmentation functions that have typically been measured in $e^+e^-$ collisions and SIDIS. The key feature of this type of observable is the possibility to determine the relevant momentum fraction $z$ experimentally as the ratio of the hadron to the jet transverse momentum. But only within the past year [131] has the quantitative relationship been derived in a form that enables measurements of identified hadrons in jets in p+p collisions to be included in fragmentation function fits on an equal footing with $e^+e^-$ and SIDIS data. Furthermore, hadrons in p+p jets provide unique access to the gluon fragmentation function, which is poorly determined in current fits [132], in part due to some tension found in the inclusive high $p_T$ pion yields measured by the PHENIX and ALICE collaborations. Here, the proposed measurements can provide valuable new insight into the nature of this discrepancy.

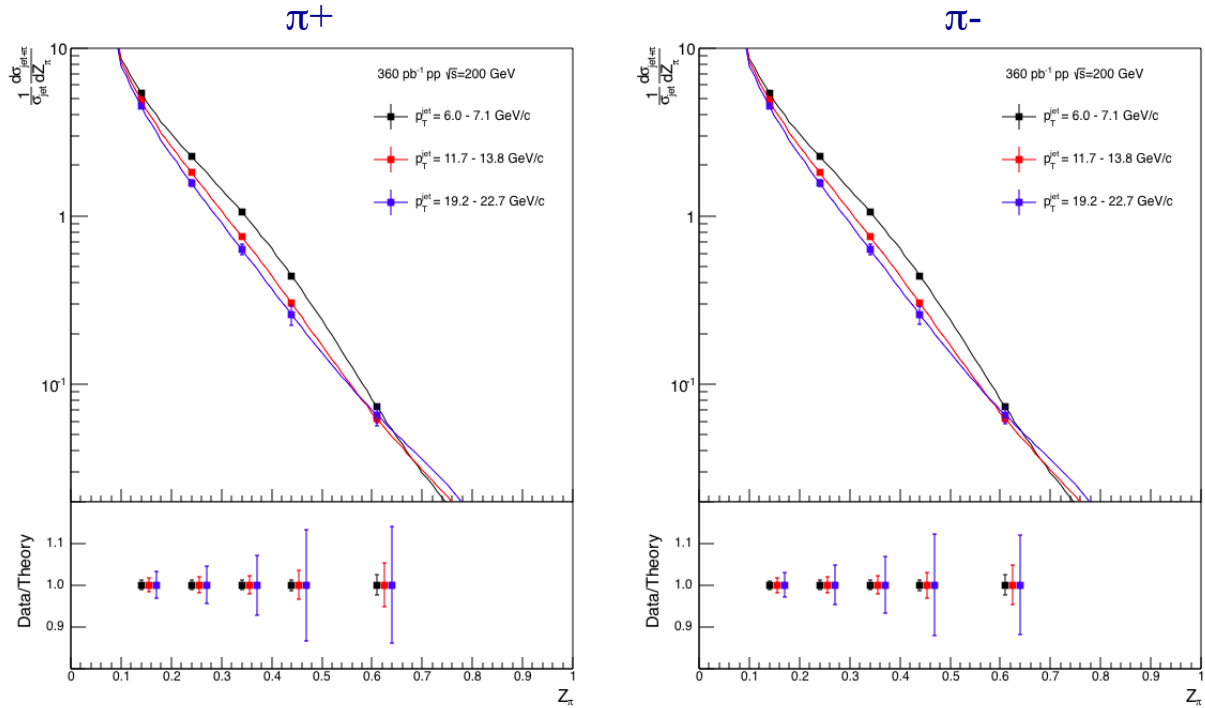

Figure 4-14: Anticipated precision for identified pions within jets at $|\eta| < 0.4$ in 200 GeV p+p collisions for three representative jet $p_T$ bins. The data points are plotted on theoretical predictions based on the DSS14 pion fragmentation functions [131,132]. Kaons and (anti)protons will also be measured, over the range from $z < 0.5$ at low jet $p_T$ to $z < 0.2$ at high jet $p_T$, with uncertainties a factor of ~3 larger than those for pions.

This development motivated STAR to initiate a program of identified particle fragmentation function measurements using p+p jet data at 200 and 500 GeV from 2011, 2012, and 2015. Figure 4-14 shows the precision that is anticipated for identified $\pi^+$ and $\pi^-$ in 200 GeV p+p collisions for three representative jet $p_T$ bins after the existing data from 2012 and 2015 are combined with future 200 GeV p+p data from 2023. Identified kaon and (anti)proton yields will also be obtained, with somewhat less precision, over a more limited range of hadron $z$. Following Run-2017, the uncertainties for 500 GeV p+p collisions will be compara-ble to that shown in Figure 4-14 at high jet $p_T$, and a factor of ~2 larger than shown in Figure 4-14 at low jet $p_T$. Identified hadron yields will also be measured multi-dimensionally vs. $j_T$, $z$, and jet $p_T$, which will provide important input for unpolarized TMD fits.

Data from the HERMES experiment [101] have shown that production rates of identified hadrons in semi-inclusive deep inelastic e+A scattering differ from those in e+p scattering. These differences cannot be explained by nuclear PDFs, as nuclear effects of strong interactions in the initial state should cancel in this observable. Only the inclusion of nuclear effects in the hadronization process allows theory to reproduce all of the dependencies ($z$, $x$, and $Q^2$) of $R_{eA}$ seen in SIDIS, as shown in Figure 4-15.

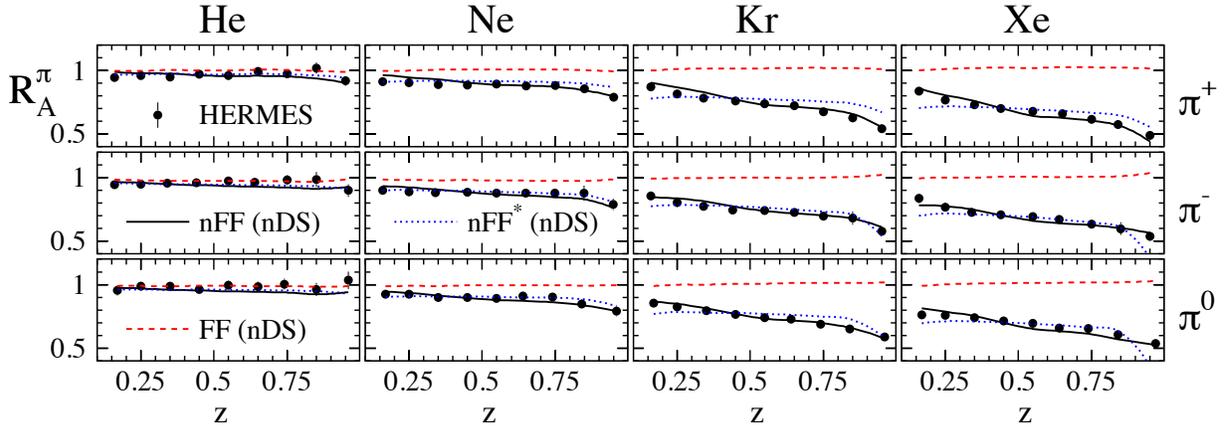

Figure 4-15: $R_{eA}$ in SIDIS for different nuclei in bins of z as measured by HERMES [101]. The solid lines correspond to the results using effective nuclear FF [133] and the nDS medium modified parton densities [134]. The red dashed lines are estimates assuming the nDS medium modified PDFs but standard DSS vacuum FFs [135] and indicate that nPDFs are insufficient to explain the data

It is critical to see if these hadronization effects in cold nuclear matter persist at the higher $\sqrt{s}$ and $Q^2$ accessed at RHIC and EIC – both to probe the underlying mechanism, which is not understood currently, and to explore its possible universality. The combination of p+p jet data from RHIC and future SIDIS data from EIC will also provide a much clearer picture of modified gluon hadronization than will be possible with EIC data alone. Using the 200 GeV p+Au data collected in 2015, STAR will be able to make a first opportunistic measurement of these hadron-jet fragmentation functions in nuclei, but the precision will be limited. Additional data will be needed in 2023 in order to provide a sensitive test for universality, as shown in Figure 4-16. Unfortunately, almost no suitable p+Al data were recorded during 2015. Thus, it will also be critical to collect data with a lighter nuclear target in 2023, such as Al, to establish the nuclear dependence of possible medium modifications in the final state, which is not predicted by current models.

STAR has provided the first ever observation of the Collins effect in p+p collisions, as shown in Figure 1-8. RHIC has the unique opportunity to extend the Collins effect measurements to nuclei, thereby exploring the spin-dependence of the hadronization process in cold nuclear matter. This will shed additional light on the mechanism that underlies modified nuclear hadronization. STAR collected a proof-of-principle set of transversely polarized p+Au data during Run-2015. While these data should provide a first estimate of the size of medium-induced effects, a high statistics polarized p+Au dataset and a scan in A is essential to precisely determine the mass dependence of these effects. Figure 4-17 shows the anticipated precision for p+Au and p+Al during the 2023 RHIC run.

It's important to note that all of the measurements discussed in this subsection involve jet detection at mid-rapidity. As such, they don't require forward upgrades to either STAR or sPHENIX. However, they do require good particle identification over quite a wide momentum range, such as that achieved by combining *dE/dx* and TOF measurements in STAR. If higher precision particle identification can be achieved, as is anticipated with the addition of the iTPC to STAR, the fragmentation function uncertainties will be further reduced, especially for kaons.

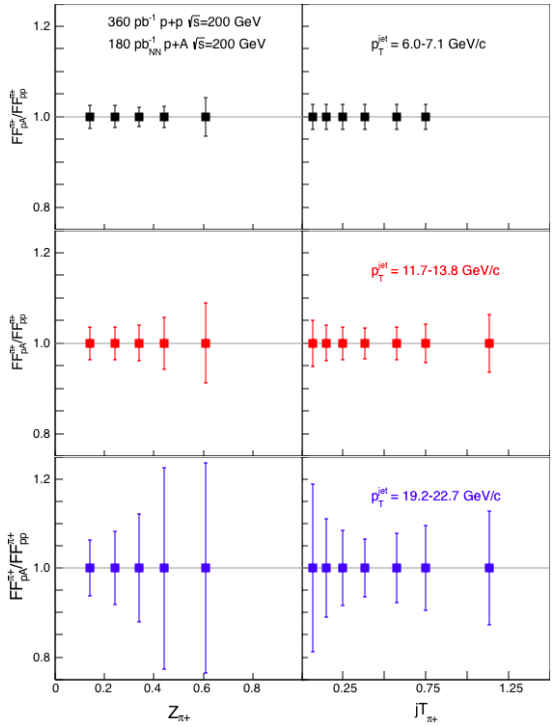 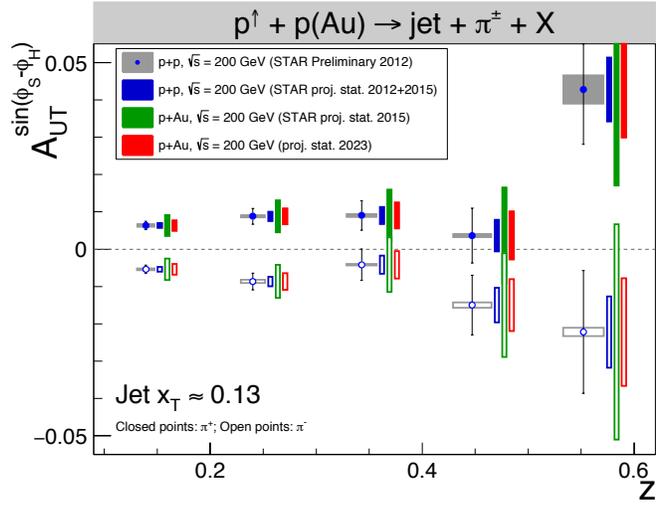

Figure 4-16: Anticipated precision for measurements of $\pi^+$ fragmentation functions in p+A/p+p at $|\eta| < 0.4$ vs. $z$ and $j_T$ in 2023 for three representative jet $p_T$ bins. Uncertainties for $\pi^-$ will be similar to those shown here for $\pi^+$, while those for kaons and (anti)protons will be a factor of ~3 larger. Comparable precision is expected for p+Au and p+Al collision systems.

Figure 4-17: Anticipated uncertainties for Collins effect measurements in p+p and p+A at $\sqrt{s_{NN}} = 200$ GeV for $0 < \eta < 1$. All points are plotted at the preliminary values found by STAR for data recorded during 2012. Similar precision will be obtained for p+Au and p+Al during the 2023 run.

# 5 TECHNICAL REALISATIONS FOR FORWARD UPGRADES

In response to a charge from the BNL Associate Lab Director Berndt Mueller, the STAR and PHENIX Collaborations documented their plans for future p+p and p+A running at RHIC in 2021 and beyond [136,96]. The time period covered by the charge coincides with scheduled p+p and p+A running at $\sqrt{s} = 200$ GeV as part of the proposed sPHENIX [137] run plan and assumes modest increases in RHIC performance [138]. A summary of the detailed white papers from PHENIX and STAR is given in the following Sections, including changes since the white papers were originally submitted. While the white papers were consistent with the current plan for RHIC running in the next decade, both realizations also consider compelling opportunities that could be exploited with additional polarized proton running at 510 GeV.

## 5.1 THE fsPHENIX FORWARD DETECTOR

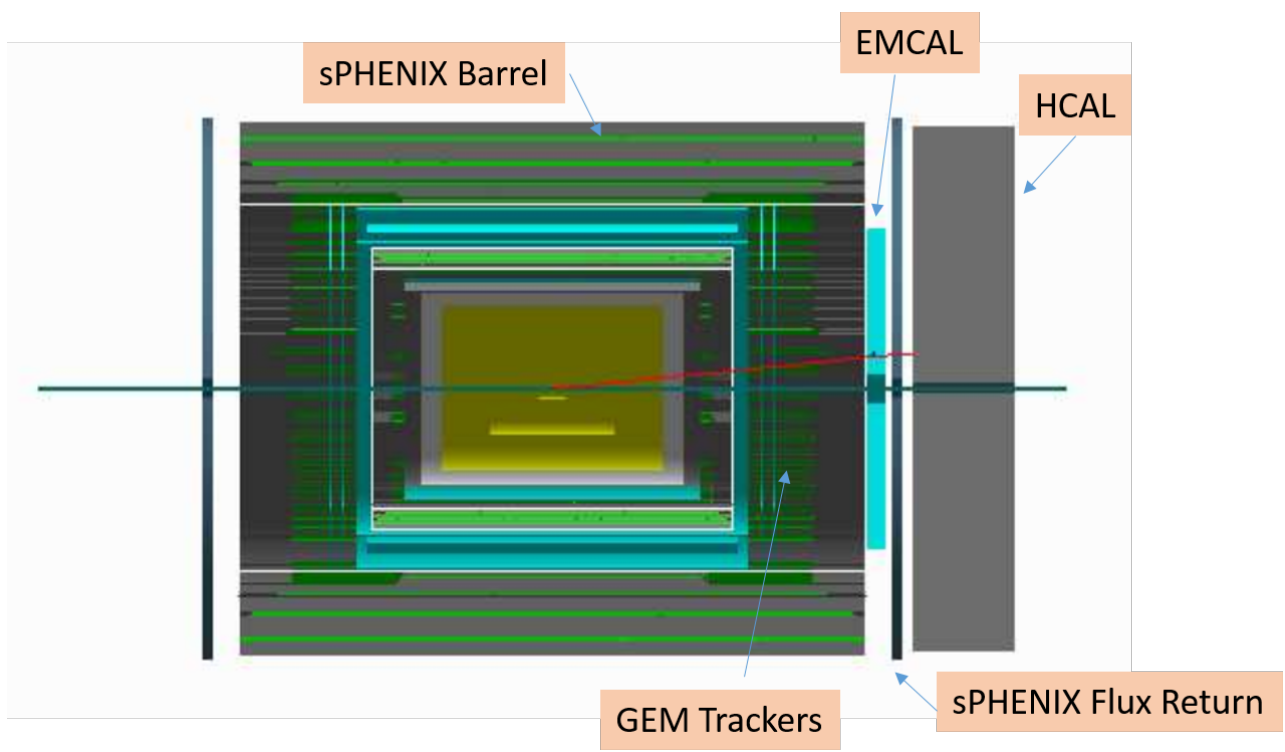

Figure 5-1: GEANT4 rendering of the sPHENIX (central barrel) and fsPHENIX (forward spectrometer) apparatus. See text for details of the fsPHENIX spectrometer arm.

The PHENIX collaboration has previously presented plans for a next-generation heavy-ion detector known as sPHENIX [137], and potentially to an EIC detector [139]. The time period covered by the charge from the BNL ALD is intermediate to these existing proposals and would coincide with scheduled p+p and p+A running at $\sqrt{s} = 200$ GeV as part of the proposed sPHENIX run plan. One possible realization of the physics goals described in this document would be to upgrade the sPHENIX detector with additional forward instrumentation.

The fsPHENIX ("forward sPHENIX") physics program focuses on physics observables at forward angles with additional detectors augmenting the sPHENIX detector. With the possibility that in the transition to the EIC, if it is realized at RHIC, the RHIC collider may not maintain the ability to provide hadron collisions, opportunities for significant new discoveries would be lost if the existing investment in RHIC is not fully exploited with measurements in the forward region. The fsPHE-NIX physics program centers on the comprehensive set of measurements in transversely spin-polarized p+p and p+A collisions described in this document, exploiting the unique capability of the RHIC collider to provide beams of protons with high polarization in addition to a variety of unpolarized nuclear beams.

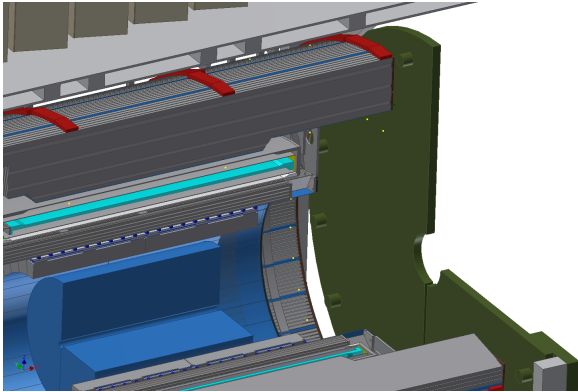

Figure 5-2: 3D engineering model of the sPHENIX forward region showing the available space for the fsPHENIX detectors between the central tracking volume (blue) and the forward flux return (green).

The sPHENIX central barrel also includes a tungsten/scintillating fiber electromagnetic calorimeter and a central tracker, completing the barrel acceptance for jets and electromagnetic probes in $|\eta|<1.1$. Within the region from $1.1 < \eta < 4.0$ is sufficient space to implement a suite of forward calorimetry ($1.4 < \eta < 4.0$ for electromagnetic calorimetry) and tracking detectors. It is currently anticipated that tracking and electromagnetic calorimetry would be inside the magnetic field volume, while a hadronic calorimeter would be positioned outside the forward magnetic flux return. Magnetic field simulations indicate that the forward flux return could be as thin as 4-5" of steel, which should not substantially degrade the resolution for hadronic showers.

In order to optimize the use of available resources, the conceptual design of the fsPHENIX detector (as shown in Figure 5-1), has been developed around the proposed sPHENIX central detector and the re-use of existing PHENIX detector systems (such as the muon identifier and the existing PHENIX EMCal), as well as elements of a possible future EIC detector forward hadron arm. The conceptual design of the fsPHENIX apparatus is as follows. A magnetic field for particle tracking and charge identification is provided by shaping the sPHENIX superconducting solenoid field with a high permeability piston located around the beam line in the forward region. Three new GEM stations at z = 150, 200 and 300cm with a position resolution of $rd\phi$ = 50-100μm would provide tracking and excellent momentum determination for charged particles ($\delta p/p < 0.3\%*p$ with $rd\phi$ = 50μm) over the full pseudorapidity range. The shaping of the magnetic field provided by the high permeability piston improves the momentum resolution at high pseudorapidity by more than a factor of two (see Figure 5-3). An electromagnetic calorimeter will provide measurements of photons and electrons, while a hadronic calorimeter measures total jet energy, position and size. We envision the electromagnetic calorimeter could be based on refurbishing the existing PHENIX PbSc EMCal towers with SiPM readout to allow operation in the magnetic field. The PHENIX EMCal has achieved an energy resolution of $\delta E/E \sim 8\%/\sqrt{E}$. It would also be possible to re-use the PHENIX MPC/MPC-EX detectors to provide high resolution prompt photon identification at the highest pseudorapidities ($3.0 < \eta < 4.0$). Because of the location of the support ring for the sPHENIX inner hadronic calorimeter, the combined fsPHENIX electromagnetic calorimeter would cover $1.4 < \eta < 4.0$. The hadronic calorimeter, with a jet energy resolution of $\delta E/E \sim 100\%/\sqrt{E}$, follows the electromagnetic calorimeter and flux return and cover $1.1 < \eta < 4.0$.

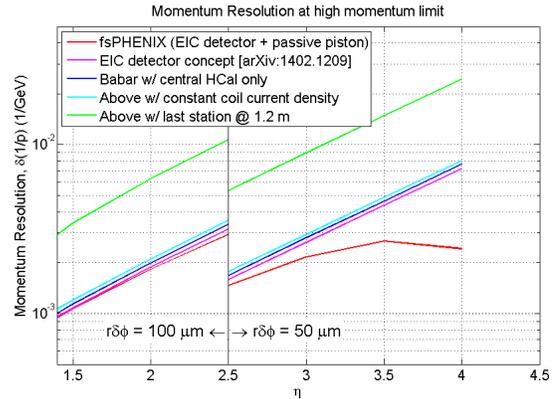

Figure 5-3: Momentum resolution as a function of pseudorapidity in the fsPHENIX forward arm for different configurations in the high momentum limit. The fsPHENIX tracking resolution is highlighted in red, showing the improvement gained with the addition of the high permeability magnetic piston.

The existing PHENIX North Muon Identifier (MuID) system, covering $1.2<\eta<2.4$, will be used for muon identification for Drell-Yan measurements. We propose to extend the pseudorapidity region for muon identification from $\eta$=2.4 to $\eta$=4, by building a "miniMuID" with a design similar to the existing PHENIX MuID. Finally, a set of Roman Pots installed in the IP8 beam line, similar to those used by the STAR experiment, enable the diffractive physics program described in this document.

The majority of the cost of the fsPHENIX detector, previously estimated to be $12M including overhead and contingency, but not including labor, the Roman Pots, nor the cost to refurbish the PHENIX EMCal [136], can be viewed as a down payment on potential EIC detector (assuming the EIC is realized at BNL). A major fraction of the cost of the fsPHENIX detector could be shared with a potential EIC detector. The new detector subsystems to be built as part of fsPHENIX that would be applicable to a potential EIC detector include:

- A forward hadronic calorimeter
- Three stations of GEM tracking chambers
- Roman Pots in the IP8 beamline

The hadronic calorimeter and GEM tracking stations have been designed jointly with the EIC detector LOI and fulfill the EIC detector requirements [139]. The "miniMuID" detector would be specific to the fsPHENIX detector and would not be applicable to a future EIC detector. As the EMCal would be about thiry years old by the time the EIC is realized it is likely that a new detector would be constructed to take advantage of advances in technology. In addition to providing an important set of new physics measurements on its own, an investment in fsPHENIX would be a step towards day-1 readiness for a potential EIC detector.

The baseline fsPHENIX design could be upgraded with a RICH detector (the one described in [139] perfectly fits the outlined physics goals) to provide charged kaon and pion identification. Such an addition would allow exciting new physics measurements, such as the Collins asymmetry for identified particles in jets.

There is strong interest within the newly formed sPHENIX collaboration to pursue the physics program enabled by fsPHENIX. At the present time the fsPHENIX detector geometry is being implemented within the sPHENIX simulations GEANT4 framework in order to provide detailed physics performance studies and further evolve the fsPHENIX design. Design work will advance in consultation with all groups interested in the physics program enabled by the detector for hadronic collisions, and with groups focused on the physics of a future EIC.

## 5.2 STAR FORWARD DETECTOR UPGRADE

### *Forward Calorimeter System*

The STAR forward upgrade is mainly driven by the desire to explore QCD physics in the very high or low region of Bjorken *x*. Previous STAR efforts using the FPD and FMS detectors, in particular the refurbished FMS with pre-shower detector upgrade in Run-2015 and a postshower for Run-2017, have demonstrated that there are unique and highly compelling QCD physics opportunities in the forward region as outlined in the previous Sections. In order to go much beyond what STAR would achieve with the improved FMS detector, STAR proposes a forward detector upgrade with superior detection capability for neutral pions, photons, electrons, jets and leading hadrons covering a pseudorapidity region of 2.5-4.5 in the years beyond 2020.

The design of the FCS' is a follow up development of the original proposed FCS system and is driven by detector performance, integration into STAR and cost optimization. The big reduction in the cost for the FCS' compared to the FCS is achieved by replacing the originally proposed W/ScFi SPACAL ECal with the refurbished PHENIX sampling ECal. In addition, the FSC' will utilize the existing Forward Preshower Detector (2.5 < $\eta$ < 4) that has been operated successfully in STAR since 2015. The proposed FCS' system will have very good (~ 8%$\sqrt{E}$) electromagnetic and (~ 70%/$\sqrt{E}$) hadronic energy resolutions. The proposed FCS' consists of 2000 of the 15552 existing PHENIX EMCal towers and 480 HCal towers covering an area of approximately 3 m × 2 m. The hadronic calorimeter is a sandwich lead scintillator plate sampling type, based on the extensive STAR Forward Upgrade and EIC Calorimeter Consortium R&Ds. Both calorimeters will share the same cost effective readout electronics and APDs as photo-sensors. It can operate without shielding in a magnetic field and in a high radiation environment. By design the system is scalable and easily re-configurable. Integration into STAR will require minimal modification of existing infrastructure.

In the past three years we carried out an extensive R&D program to develop sampling calorimeters for the STAR forward upgrade and the EIC barrel and forward/backward calorimeters including successful test beam runs of full-scale prototypes at FNAL. To have an easy re-configurable calorimeter system was one of the main design goals for the system. The HCal is made of Lead and Scintillator tiles with a tower size of 10×10×81 cm$^3$ corresponding to 4-interaction lengths. Figure 5-4 shows the location of the proposed FCS at the West side of the STAR detector system.

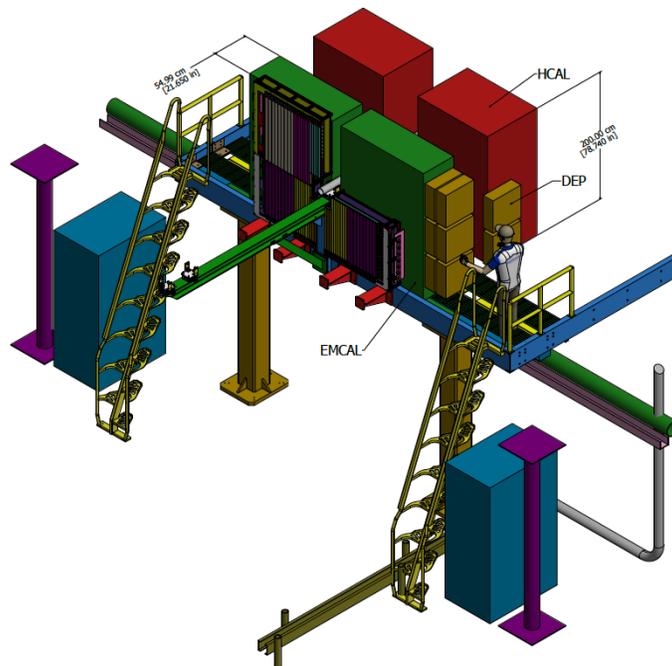

Figure 5-4: Location of the FCS at the West side of the STAR Detector system

Wavelength shifting slats are used to collect light from the HCAL scintillating plates to be detected by photon sensors at the end of the HCal. Multiple Silicon PMTs will be used to read out each SPACal and HCal module, 4 for SPACal and 8 for HCal, respectively.

A novel construction technique has been developed for the HCal by stacking Lead and Scintillator plate in-situ. Students and post-docs just before the test run constructed an array of 4×4 prototype HCal modules at the FNAL test beam site. We envision that a full HCal detector can be assembled at the STAR experimental hall within a few months during the summer shut-down period.

Figure 5-5 shows a newly constructed array of 4×4 HCal modules at the FNAL test beam facility. The right panel shows the energy resolutions for the FCS SPACal and HCal detectors as a function of the beam energy. SiPMTs were used for the read-out of both calorimeter detectors. The anticipated hadron energy resolution for these calorimeters, being combined with electromagnetic calorimeter response, is expected to be of an order of ~40-45%/$\sqrt{E}$ (see Figure 5-5 (right)), which was confirmed in the recent test run at FNAL.

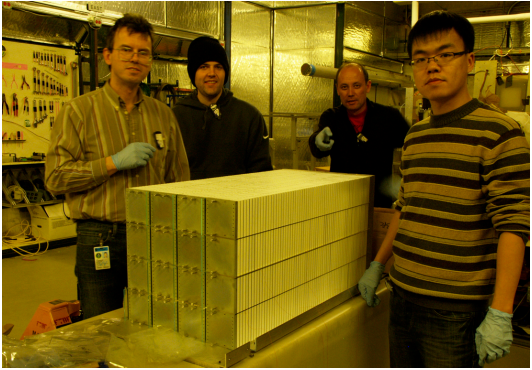
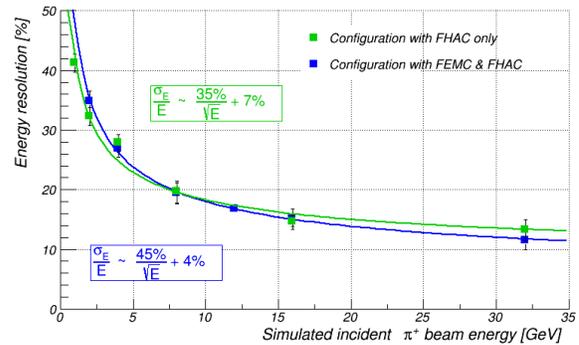

Figure 5-5: Prototype of HCAL calorimeter being assembled at FNAL for the test run. Simulated energy resolution of the forward calorimeter system for pions. Shown is the response from the lead-plastic hadronic sandwich calorimeter alone, as well as when the response from the electromagnetic tungsten powder scintillating fiber calorimeter installed in front of it, is added with a proper weight. The numbers are consistent with the results of T1018 test beam at FNAL in February 2014, as well as with [140].

It has been proposed to design and build a generic digitizer system ("Detector Electronics Platform", DEP) as part of the development of a readout platform for the future EIC. This system would be cost-effective, fast & modular. It would be available and could be used for many different applications within STAR for the 2022-2023 running periods. The basic board would consist of 32 12bit ADCs running in sampling mode at 8x the RHIC clock. The ADC would be followed by a fast FPGA capable of running various digital filters and other typical trigger algorithms such as: pedestal & zero subtraction, charge integration, moderate timing information (to <1ns), highest-tower, tower sums etc. The system will be capable of connecting up to 5 such boards (for a total of 160 channels) into a compact & cost-effective chassis. The data will be sent to a DAQ PC over a fast optical link and will have enough bandwidth to work in full streaming mode for typical occupancies, if so desired. It would also house the STAR TCD interface for the RHIC clock and Trigger command, which would also act as a Slow Controls Interface if needed. An interface to current or future STAR DSM boards will also be provided. Readout of FCS' will be based on DEP with a backup option based on extending the existing QT readout system currently used in the FMS and FPS. Both options of the FCS readout schemes are cost wise the same.

The calorimeter will use 1000 FEE boards each providing readout for one (HCAL) or four (EMCAL) towers using S8664-55 Hamamatsu APD's. The elements of the FEE board will include: Low noise preamplifiers based on BF862 JFET's, pulse shaping circuits, cable drivers designed to bridge either the DEP or existing QT boards, and temperature-compensated bias voltage regulators to provide a stable (<<1%) gain of the APD's. The bias voltage regulator and slow controls interface will be based on the successful FEE design for the STAR FPS; the preamplifier will be based on an Indiana University design for the aCORN experiment and on other BF862-based folded cascode designs. A multi-drop power & control interface cable will connect ~20 or more FEE boards to an output of the control interface box (TUFF-II), based on the TUFF box of FPS.

This development is closely tied with an ongoing EIC generic detector R&D. The current preliminary cost estimate for the FCS' is $2.06 M, including overhead and contingencies.

## Forward Tracking System

In addition to the FCS, a Forward Tracking System (FTS) is also under consideration for the STAR forward upgrade project. Such an FTS needs to be designed for the small field-integral from the STAR 0.5 T Solenoid magnet field in the forward region to discriminate charge sign for transverse asymmetry studies and those of electrons and positrons for Drell-Yan measurements. It needs to find primary vertices for tracks and point them towards the calorimeters in order to suppress pile-up events in the anticipated high luminosity collisions, or to select particles from Lambda decays. It should also help with electron and photon identification by providing momentum and track veto information. In order to keep multiple scattering and photon conversion background under control, the material budget of the FTS has to be small. These requirements present a major challenge for detector design in terms of position resolution, fast readout, high efficiency and low material budget.

STAR has considered two possible detector technology choices: the Silicon detector technology and Gas Electron Multiplier (GEM) technology. STAR has gained considerable experience in both technologies from the FGT (Forward GEM Tracker) construction and the Intermediate Silicon Tracker (IST) construction in recent years. Silicon detector technology is the currently preferred choice for the realization of the FTS. Several groups are pursuing further GEM-based detector R&D under the auspices of generic EIC R&D program.

Silicon detectors have been widely used in high-energy experiments for tracking in the forward direction. For example, Silicon strip detectors have been successfully used at many experiments: the D0 experiment at the Tevatron, CMS and LHCb at the LHC, and PHENIX at the RHIC. More recent designs incorporate hybrid Silicon pixel detectors, which resulted in the improvement of position resolutions and removal of ghost hits, but unfortunately they also significantly increased the cost and material budget. According to preliminary Monte Carlo simulations, charge sign discrimination power and momentum resolution for the FTS in the STAR Solenoid magnet depends mostly on phi resolution, and is insensitive to the r-position resolution. Therefore a Silicon mini-strip detector design would be more appropriate than a pixel design. STAR is evaluating designs that consist of four or more disks at z locations along the beam direction between 70 and 140 cm from the nominal interaction point. Each disk has wedges covering the full $2\pi$ range in $\phi$ and 2.5-4 in $\eta$. The wedge will use Silicon mini-strip sensors read out from the larger radius of the sensors. Compared to the configuration of reading out from the edges along the radial direction, the material budget in the detector acceptance will be smaller since the frontend readout chips, power and signal buses and cooling lines can be placed outside of the detector acceptance.

The current preliminary cost estimate for a 4-disk FTS is $3.8 M including overheads and contingencies.

## Roman Pot Upgrade

The diffractive physics opportunities in Section 2.3 rely crucially on detection capabilities for the scattered beam proton. Since 2015, STAR has successfully operated a Roman Pot subsystem that enables operation with the central detector and with default beam-optics in RHIC. Funding and scheduling constraints have thus far limited the acceptance of its Silicon tracker packages. It is proposed to upgrade these packages prior to the running periods in 2022-2023 and realize the full "phase-II" acceptance as illustrated in Figure 2-19.

The current preliminary cost estimate for this upgrade is $1M including overheads and contingencies.

# 5.3 KINEMATICS OF INCLUSIVE FORWARD JETS IN p+p WITH THE PROPOSED FORWARD UPGRADE

Both the measurement of the helicity structure and the transverse spin structure of the nucleon use reconstructed jets and di-jets to narrow the phase space of partonic kinematics. Since jets serve as proxies for the scattered partons, reconstructed jets allow the selection of events with a specific weighting of the fractional momenta of the parent protons carried by the scattering partons, assuming a 2-to-2 process. Here we call the fractional momentum carried by the parton coming from the beam along the $z$-axis (towards the proposed forward upgrade instrumentation) $x_1$, and the fractional momentum carried by the other parton $x_2$.

For measurements of $A_{LL}$ it is important to select events where one $x$ is determined with good accuracy within the kinematic region in which one wants to measure $\Delta g(x)$ and the other $x$ is in a region where the helicity distribution is well known, i.e. in the region of medium to high values of $x$. The transverse spin structure of the nucleon on the other hand is usually accessed using transverse single spin asymmetries and semi-inclusive measurements. This means one measures azimuthal asymmetries of the final state where the distribtion function of interest couples to a spin dependent FF that serves as a polarimeter. Consequently we studied how in single- and di-jet events, the jet pseudorapidity $\eta$ and $p_T$ are related to the underlying partonic variables $x_1$ and $x_2$. We also studied the matching between reconstructed jets and scattered partons and the resolutions with which the parton axis can be reconstructed from the reconstructed detector jets. The latter is important to evaluate how well azimuthal asymmetries around the outgoing parton axis will be reconstructed by looking at asymmetries of reconstructed particles around the reconstructed jet axis.

The simulations are based on PYTHIA Tune A at $\sqrt{s}$=500 GeV and a minimum partonic $p_T$ of 3 GeV. Fast detector simulations have been used to account for the resolutions of the STAR barrel and the forward upgrade detectors (for details see [96]). Jets are reconstructed with an anti-$k_T$ algorithm with a radius of 0.7. An association between reconstructed jets and scattered partons is defined to be a distance in $\eta$-$\phi$ space of less than 0.5. In the following, reconstructed jets are referred to as "detector jets" and jets found using stable, final state particles "particle jets."

Figure 5-7 (left) shows the regions of $x$ that can be accessed by jets in the forward region. A minimum jet $p_T$ of 3 GeV/$c$ was chosen to ensure that the momentum transfer is sufficiently high for pQCD calculations to be valid. At high $x$, values of $x$~0.6 should be reachable. This compares well with the current limit of SIDIS measurements, $x$~0.3, and encompasses the region in $x$ that dominates the tensor charge. To investigate the possibility of selecting specific $x$ regions, in particular high $x$, the dependence of $x$ on the jet $p_T$ and pseudorapidity was studied. Figure 5-7 (right) shows $x_1$ as a function of jet $p_T$.

For measurements of azimuthal asymmetries of jets or hadrons within a jet to probe the transverse spin structure of the nucleon it is important to reconstruct reliably the outgoing parton direction. Therefore, the matching of reconstructed jets to scattered partons was studied (Figure 5-6). In general, matching and parton axis smearing improves with $p_T$, which may be connected to the jet multiplicity that rises with transverse momentum.

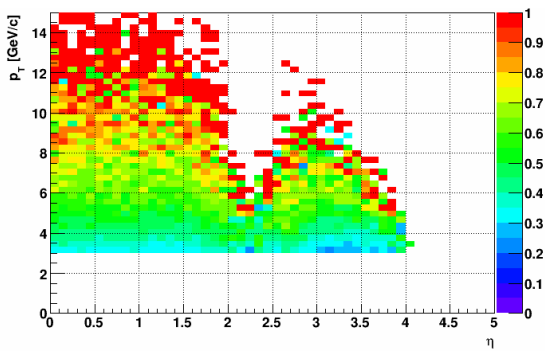

Figure 5-6: Matching Fraction between detector jets and partons. The matching fraction at low $p_T$ is only around 50%, but grows to over 90% for high $p_T$.

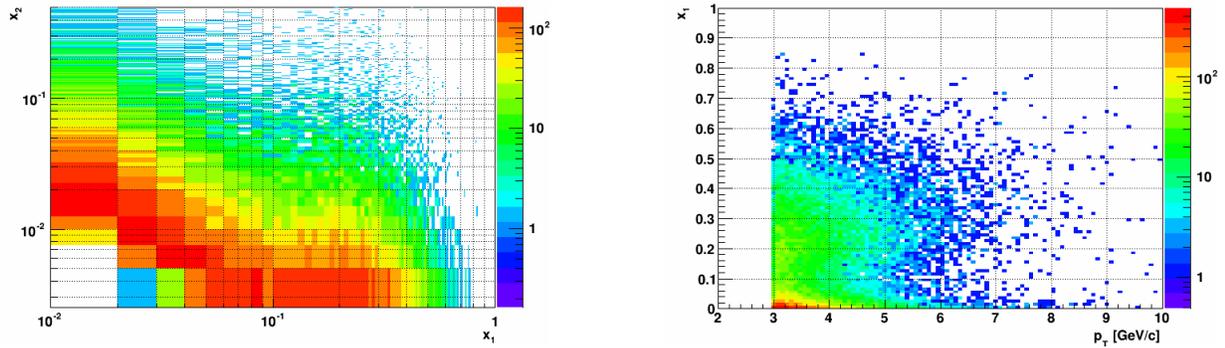

Figure 5-7: (left) Distribution of the partonic variables $x_1$ and $x_2$ for events with a jet with $p_T > 3$ GeV/$c$ and $2.8 < \eta < 3.5$. $x_1$ values of around 0.6 can be reached whereas $x_2$ goes as low as $7\times10^{-3}$. (right) $x_1$ versus jet $p_T$. As expected, there is a correlation between the x accessed and the $p_T$ of the jet. However, there is an underlying band of low $x_1$ values. This can be improved by further restricting the $\eta$ range of the jet. Here $2.8 < \eta < 3.8$.

# 6 APPENDIX

## 6.1 KINEMATIC VARIABLES

| Variable | Description |
|---|---|
| $x$ | longitudinal momentum fraction |
| $x_B$ | Bjorken scaling variable |
| $x_T$ | $x_T = 2p_T/\sqrt{s}$ |
| $x_F$ | Feynman-$x$ ($x_F \sim x_1 - x_2$) |
| $Q$ | virtuality of the exchanged photon in DIS |
| $s$ | The squared collision energy $s = 4 E_{p1} E_{p2}$ |
| $\sqrt{s}$ | Center-of-mass energy |
| $p_T$ | Transverse momentum of final state particles, i.e. jet, hadrons |
| $k_T$ | Transverse momentum of partons |
| $j_T$ | Momentum of a hadron transverse to the jet thrust axis |
| $b_T$ | Transverse position of parton inside the proton |
| $\xi$ | Parton skewness: $x_B/(2 - x_B)$ |
| $\eta$ | Pseudorapidity |
| $y$ | Rapidity |
| $z$ | the fractional energy of a hadron relative to the jet energy |
| $\cos\theta$ | $\theta$: polar angle of the decay electron in the partonic c.m.s., with $\theta > 0$ in the forward direction of the polarized parton |
| $\phi_s, \phi_h$ | Azimuthal angles of the final state hadron and the transverse polarization vector of the nucleon with respect to the proton beam |
| $\phi$ | Azimuthal angles of the final state hadron with respect to the proton beam |

Table 6-1: Definition of all kinematic variables used in the document

## 6.2 RHIC SPIN PUBLICATIONS

1. The RHIC Spin Program: Achievements and Future Opportunities
   E.C. Aschenauer et al., arXiv:1501.01220
2. The RHIC Spin Program: Achievements and Future Opportunities
   E.C. Aschenauer et al., arXiv:1304.0079
3. Plans for the RHIC Spin Physics Program
   G. Bunce et al., http://www.bnl.gov/npp/docs/RHICst08_notes/spinplan08_really_final_060908.pdf
4. Single transverse-spin asymmetry in very forward and very backward neutral particle production for polarized proton collisions at √s=200 GeV
   Y. Fukao et al, Phys. Lett. B650 (2007) 325

### ANDY:
1. Cross Sections and Transverse Single-Spin Asymmetries in Forward Jet Production from Proton Collisions at √s=500 GeV.
   L. Bland et al, Phys.Lett. B750 (2015) 660

### BRAHMS:
1. Cross-Sections and single spin asymmetries of identified hadrons in p+p at √s=200 GeV.
   J.H. Lee and F. Videbaek  arXiv:0908.4551.
2. Single transverse spin asymmetries of identified charged hadrons in polarized p+p collisions at √s=62.4

GeV.
I. Arsene et al. Phys. Rev. Lett. 101 (2008) 042001.
**Citations: 128**

## *pp2pp:*
1. Double Spin Asymmetries $A_{NN}$ and $A_{SS}$ at $\sqrt{s}=200$ GeV in Polarized Proton-Proton Elastic Scattering at RHIC.
   pp2pp Collaboration, Phys. Lett. B647 (2007) 98
2. First Measurement of $A_N$ at $\sqrt{s}=200$ GeV in Polarized Proton-Proton Elastic Scattering at RHIC.
   pp2pp Collaboration, Phys. Lett. B632 (2006) 167
3. First Measurement of Proton-Proton Elastic Scattering at RHIC.
   pp2pp Collaboration, Phys. Lett. B579 (2004) 245

## *PHENIX:*
1. Inclusive cross sections, charge ratio and double-helicity asymmetries for $\pi^0$ production in p+p collisions at $\sqrt{s}=510$ GeV.
   PHENIX Collaboration, Phys. Rev. D93, (2016) 011501.
2. Measurement of parity-violating spin asymmetries in $W^{+/-}$ production at midrapidity in longitudinally polarized *p+p* collisions
   PHENIX Collaboration, submitted to PRD; arXiv:1504.07451
3. Inclusive cross sections, charge ratio and double-helicity asymmetries for $\pi^+$ and $\pi^-$ production in p+p collisions at $\sqrt{s}=200$ GeV.
   PHENIX Collaboration, Phys. Rev. D91 (2015) 032001
4. Cross Section and Transverse Single-Spin Asymmetry of η-Mesons in $p^{\uparrow}$+p Collisions at $\sqrt{s}=200$ GeV at Forward Rapidity.
   PHENIX Collaboration, Phys.Rev. D90 (2014) 072008
5. Inclusive double-helicity asymmetries in neutral pion and eta meson production in collisions at $\sqrt{s}=200$ GeV.
   PHENIX Collaboration, Phys.Rev. D90 (2014) 012007.
6. Measurement of transverse-single-spin asymmetries for midrapidity and forward-rapidity production of hadrons in polarized p+p collisions at $\sqrt{s}=200$ and 62.4 GeV.
   PHENIX Collaboration, Phys. Rev. D90 (2014) 012006
7. Inclusive cross section and single transverse spin asymmetry for very forward neutron production in polarized p+p collisions at $\sqrt{s}=200$ GeV.
   PHENIX Collaboration, Phys.Rev. D88 (2013) 3, 032006.
8. Double Spin Asymmetry of Electrons from Heavy Flavor Decays in p+p Collisions at $\sqrt{s}=200$ GeV.
   PHENIX Collaboration, Phys.Rev. D87 (2013) 012011.
9. Cross sections and double-helicity asymmetries of midrapidity inclusive charged hadrons in p+p collisions at $\sqrt{s}=62.4$ GeV.
   PHENIX Collaboration, Phys.Rev. D86 (2012) 092006.
10. Cross section and double helicity asymmetry for $\eta$ mesons and their comparison to neutral pion production in p+p collisions at $\sqrt{s}=200$ GeV.
    PHENIX Collaboration, Phys.Rev. D83 (2011) 032001.
11. Measurement of Transverse Single-Spin Asymmetries for J/Ψ Production in Polarized p+p Collisions at $\sqrt{s}=200$ GeV.
    PHENIX Collaboration, Phys.Rev. D82 (2010) 112008, Erratum-ibid. D86 (2012) 099904.
12. Event Structure and Double Helicity Asymmetry in Jet Production from Polarized p+p Collisions at $\sqrt{s}=200$ GeV.
    PHENIX Collaboration, Phys.Rev. D84 (2011) 012006.
13. Cross section and Parity Violating Spin Asymmetries of $W^{\pm}$ Boson Production in Polarized p+p Collisions at $\sqrt{s}=500$ GeV.
    PHENIX Collaboration, Phys.Rev.Lett. 106 (2011) 062001.
    **Citations: 61**
14. Double-Helicity Dependence of Jet Properties from Dihadrons in Longitudinally Polarized p+p Colli-

sions at √s=200 GeV.
PHENIX Collaboration, Phys.Rev. D81 (2010) 012002.
15. Inclusive cross section and double helicity asymmetry for $\pi^0$ production in p+p collisions at √s=62.4 GeV.
PHENIX Collaboration, Phys.Rev. D79 (2009) 012003.
**Citations: 85**
16. The Polarized gluon contribution to the proton spin from the double helicity asymmetry in inclusive $\pi^0$ production in polarized p+p collisions at √s=200 GeV.
PHENIX Collaboration, Phys.Rev.Lett. 103 (2009) 012003.
**Citations: 92**
17. Inclusive cross-Section and double helicity asymmetry for $\pi^0$ production in p+p collisions at √s=200 GeV: Implications for the polarized gluon distribution in the proton.
PHENIX Collaboration, Phys.Rev. D76 (2007) 051106.
**Citations: 210**
18. Improved measurement of double helicity asymmetry in inclusive midrapidity $\pi^0$ production for polarized p+p collisions at √s=200 GeV.
PHENIX Collaboration, Phys.Rev. D73 (2006) 091102.
19. Measurement of transverse single-spin asymmetries for mid-rapidity production of neutral pions and charged hadrons in polarized p+p collisions at √s=200 GeV.
PHENIX Collaboration, Phys.Rev.Lett. 95 (2005) 202001.
**Citations: 162**
20. Double helicity asymmetry in inclusive mid-rapidity $\pi^0$ production for polarized p+p collisions at √s=200 GeV.
PHENIX Collaboration , Phys.Rev.Lett. 93 (2004) 202002.
**Citations: 80**
19. Mid-rapidity neutral pion production in proton proton collisions at $\sqrt{s}$ = 200-GeV
PHENIX Collaboration, Phys.Rev.Lett. 91 (2003) 241803.
**Citations: 332**

## *STAR:*
1. Measurement of the transverse single-spin asymmetry in $p^\uparrow + p \to W^\pm/Z^0$ at RHIC,
STAR Collaboration, submitted to PRL, arXiv: 1511.06003
2. Observation of Transverse Spin-Dependent Azimuthal Correlations of Charged Pion Pairs in $p^\uparrow$+p at √s =200 GeV,
STAR Collaboration, Phys.Rev.Lett. 115 (2015) 242501
3. Measurement of longitudinal spin asymmetries for weak boson production in polarized proton-proton collisions at RHIC.
STAR Collaboration, Phys.Rev.Lett. 113 (2014) 072301.
4. Precision Measurement of the Longitudinal Double-spin Asymmetry for Inclusive Jet Production in Polarized Proton Collisions at √s =200 GeV,
STAR Collaboration, Phys. Rev. Lett. 115 (2015) 092002
5. Neutral pion cross section and spin asymmetries at intermediate pseudorapidity in polarized proton collisions at √s=200 GeV,
STAR Collaboration, Phys. Rev. D 89 (2014) 012001.
6. Single Spin Asymmetry $A_N$ in Polarized Proton-Proton Elastic Scattering at √s=200 GeV.
STAR Collaboration, Phys. Lett. B 719 (2013) 62.
7. Transverse Single-Spin Asymmetry and Cross-Section for $\pi^0$ and η Mesons at Large Feynman-x in Polarized p+p Collisions at √s=200 GeV.
STAR Collaboration, Phys. Rev. D 86 (2012) 51101.
**Citations: 50**
8. Longitudinal and transverse spin asymmetries for inclusive jet production at mid-rapidity in polarized p+p collisions at √s=200 GeV.
STAR Collaboration, Phys. Rev. D 86 (2012) 32006.
**Citations: 50**

9. Measurement of the W $\to$ e $\nu$ and Z/$\gamma^*\to$e$^+$e$^-$ Production Cross sections at Mid-rapidity in Proton-Proton Collisions at √s=500 GeV.
   STAR Collaboration, Phys. Rev. D 85 (2012) 92010.
10. Measurement of the parity-violating longitudinal single-spin asymmetry for W$^\pm$ boson production in polarized proton-proton collisions at √s=500 GeV,
    STAR Collaboration, Phys. Rev. Lett. 106 (2011) 62002.
    **Citations: 50**
11. Longitudinal double-spin asymmetry and cross section for inclusive neutral pion production at midrapidity in polarized proton collisions at √=200 GeV.
    STAR Collaboration, Phys.Rev. D80 (2009) 111108.
12. Longitudinal Spin Transfer to Lambda and anti-Lambda Hyperons in Polarized Proton-Proton Collisions at √s=200 GeV.
    STAR Collaboration, Phys.Rev. D80 (2009) 111102.
13. Forward Neutral Pion Transverse Single Spin Asymmetries in p+p Collisions at √s=200 GeV.
    STAR Collaboration, Phys.Rev.Lett. 101 (2008) 222001.
    **Citations: 150**
14. Longitudinal double-spin asymmetry for inclusive jet production in p+p collisions at √s=200 GeV.
    STAR Collaboration, Phys.Rev.Lett. 100 (2008) 232003.
    **Citations: 118**
15. Measurement of transverse single-spin asymmetries for di-jet production in proton-proton collisions at √s=200 GeV.
    STAR Collaboration, Phys.Rev.Lett. 99 (2007) 142003.
16. Longitudinal double-spin asymmetry and cross section for inclusive jet production in polarized proton collisions at √s=200 GeV.
    STAR Collaboration, Phys.Rev.Lett. 97 (2006) 252001.
    **Citations: 197**
17. Cross-Sections and transverse single spin asymmetries in forward neutral pion production from proton collisions at √s=200 GeV.
    STAR Collaboration, Phys.Rev.Lett. 92 (2004) 171801.
    **Citations: 303**

# 6.3 THE CHARGE

08/30/2015

## Charge for an Updated Plan for Physics with Polarized Protons at RHIC

During the DOE Nuclear Physics RHIC site visit on July 23, 2015, Associate Director of Science for Nuclear Physics, Tim Hallman, asked BNL to develop and submit an updated plan ("Spin Plan") of the key cold QCD measurements utilizing polarized p+p and p+A collisions at RHIC. This deadline for submission of this plan to DOE/NP is January 31, 2016. In order to meet this deadline and allow for in-depth review and feedback by a group of senior experts, we ask you to submit a preliminary draft of the updated Spin Plan to BNL no later than December 10, 2015.

Specifically, the document should address the following issues:

- What are the compelling physics questions the future polarized p+p and p+A program at RHIC can address? While the Plan should reconfirm the physics case for the 500 GeV polarized p+p run currently planned for 2017, its main focus should be on physics opportunities in "cold" QCD using polarized protons during the planned hard probes campaign after the installation of sPHENIX.

- What is the anticipated scientific impact of future RHIC data in view of the complementary measurements from LHC, COMPASS, and JLab 12 GeV. With respect to the physics program at a future electron-ion collider, the Spin Plan should discuss, which of the key measurements are critical for the planning of the EIC physics program or are necessary as sources of critical information for the interpretation of the expected EIC data.

- The Plan should describe possible modest detector upgrades that are required to perform the proposed measurements. Their cost should be estimated and a realization plan should be outlined.

- The required integrated luminosities, figures-of-merit, and the possible need for collision systems other than p+Au (e.g. an A-scan in p+A) should be listed. The luminosity requirements should be based on recent guidance by C-AD (http://www.rhichome.bnl.gov/RHIC/Runs/RhicProjections.pdf)

- Unique "must-do" measurements, which require running beyond the currently planned RHIC runs (~20 weeks of Au+Au and ~10 weeks each of p+p and p+Au, all at 200 GeV), should be briefly described, but should not form the sole justification for proposed experimental upgrades.

This plan should take full advantage of the unique opportunities provided by the flexibility and polarization of the RHIC beams.

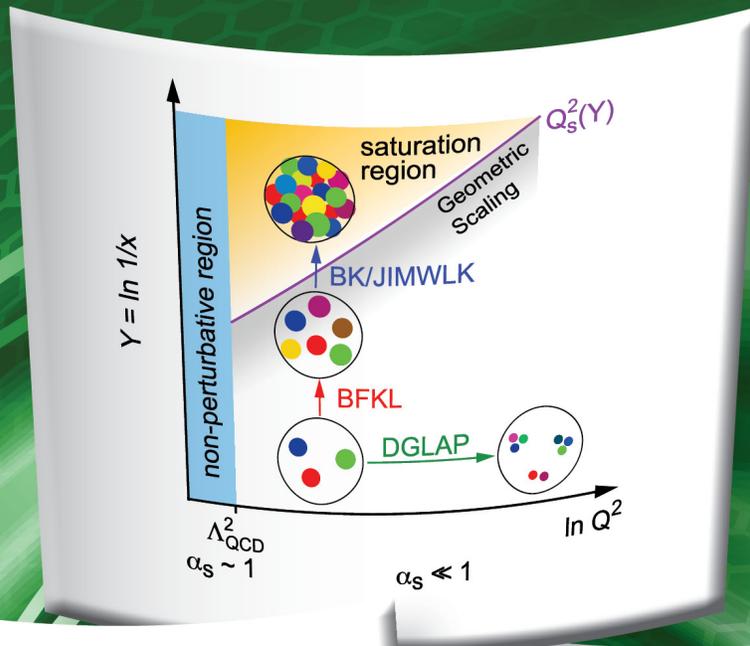
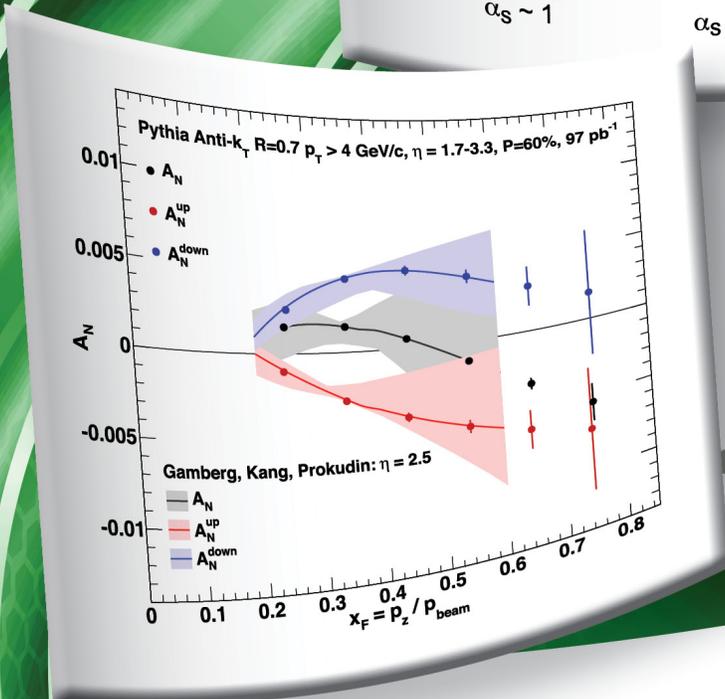
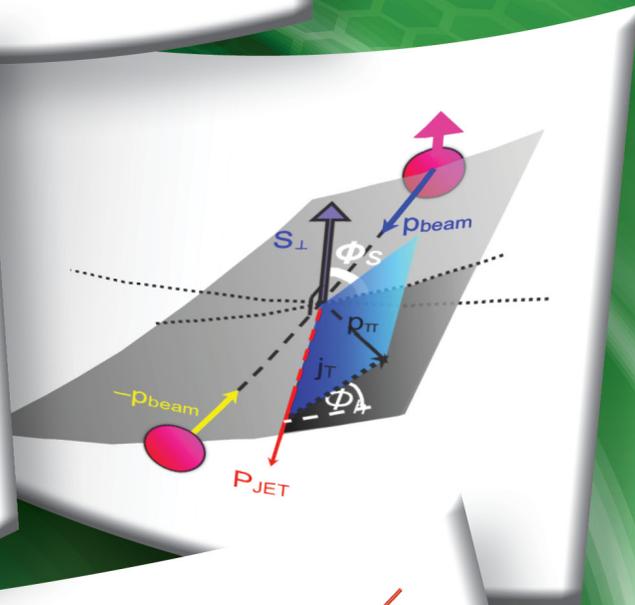
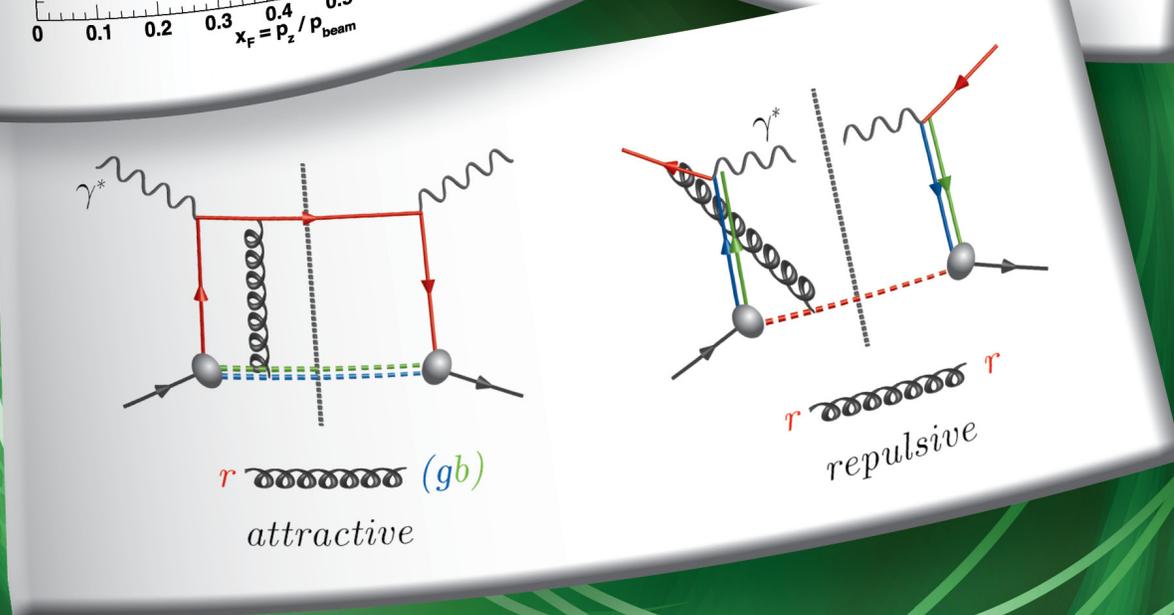